# Eight-order mosaic structure theory of the glass transition and macromolecular motion

Jia-Lin Wu

*(College of Material Science and Engineering, State Key Laboratory for Fibers and Polymer Materials, Donghua University, Shanghai 201620, China)*

**Abstract:** A universal theoretical framework has been proposed that the solid-to-liquid glass transition and macromolecular motion within the entire temperature range from $T_g$ to $T_m$ and the liquid flow are absolutely determined by the intrinsic 8 orders of transient 2-*D* mosaic geometric structures formed by exciting interfaces, without any presupposition and relevant parameter. Interface excitations are cross-coupled electron pairs, which attribute to the Van der Waals repulsion electron-electron pairs with 2-*D* self-avoiding closed loops existing in any random system at any temperature. An interface excitation state is a vector with 8 orders of additional restoring torque, 8 orders of relaxation times, quantized energies and extra volumes. Dynamics occurred by the slow inverse cascade generates the 8 orders of potential loop-flows while the fast cascade rearranges the structure. The delocalization mode is solitary wave along the 8 orders of geodesic. Solitary wave is derived from the balance between inverse cascade and cascade. Increasing temperature only increases the number of inverse cascade and cascade. This model provides a unified mechanism to interpret ripplon, reptation, Boson peak, Ising model, non-ergodic, free volume, hard-sphere, tunneling, cage, compacting cluster, jamming behaviors, breaking solid lattice, geometrical frustration, potential energy landscape, flow-percolation, thermo-disorder-induced localized energy, average energy of cooperative migration along one direction, critical entanglement chain length, and Reynolds number, solitary wave, heteroclinic orbit in hydrodynamics. This model also directly deduces the Lindemann ratio, the famous 3.4 power law of viscosity of entangled macromolecules and the well-known WLF equation and shows that the Clapeyron equation governing all first order phase transitions in thermodynamics only holds true in each subsystem, instead of the system in glass transition. The formation of the paradigm for solid-to-liquid transition has been proposed that the chemical potential is defined as Fermi energy on the occasion when the temperature is zero in solid physics; whereas the random motion energy, $kT_g$, is defined as the localized energy of glass transition on the occasion when the chemical potentials in subsystems are always zero in solid-to-liquid transition.

## Article Outline

## 1. Introduction

One of the most challenging problems in condensed matter physics is to understand the glass transition [1, 2] and the macromolecular entanglement motion in polymer physics. In this paper, the main theoretical concepts are proposed that the mode of the multi-macromolecular motion is just that of the solid-to-liquid glass transition. This groundwork will deal with the current differ theories of many-subjects, such as: quantum effect and many-body effect, ripplon and supercooled liquid, reptation model and Ising model, universal Lindemann ratio and solid-to-liquid transition, broken symmetry and order parameter, delocalization and disorder induced localization, topological analysis and non-integrable geometric phase, self-similar and critical phase transition, mosaic structure and entanglement structure, soft-matter and geometrical frustration, cage effect and jamming behaviors, ergodic and non-ergodic, flow-percolation and hydrodynamics, cascade and inverse cascade, self-avoiding walk and ideal random walk, geodesic and solitary wave, Mott disorder, Anderson disorder and Flory disorder and so on.

Recent theories of glass transition focus on microscopic dynamical mechanisms of the short-ranged attractive colloidal system [3-6] and the Boson Peak [7-11]. For various hard-sphere colloidal systems, mode-coupling theory, which is originally developed for atomic systems interacting through a spherically symmetric intermolecular potential, has played a leading role [12]. One of the various models within mode-coupling theory is the short-ranged square-well model, in which extensive molecular dynamics simulations have also been carried out [12]. For example, a colloidal gas-liquid transition is induced when the short-ranged attractive interactions attain sufficient magnitude [13]; the glass transition behavior is induced by the geometrical frustration which appears due to the existence of short loops in the generalized Bethe lattices [14]. Since Liu and Nagel proposed an idea of unifying the glass transition i.e. the jamming behavior [15, 16], the jamming behavior has been extensively studied [16]. De Gennes [17] pointed out that the possible weakness of the mode-coupling theories is the ignorance of geometrical frustration effect and he set up a cluster picture with mosaic structure. The mosaic structure is based on the compact primary clusters which roughly correspond in size to the Boson peak [18]. These clusters retain a finite size because of the size limited by frustration effects [17]. The pacing involved cannot propagate up to large distances. However, the detailed frustration process will vary from system to system. For flexible polymers, the nature of the frustrated packing is unclear [17]. De Gennes also pointed out that the clusters move rather than molecules [17]. Wolynes [9] proposed that quantized 'domain wall motions' connected with the mosaic structure predicted by the 'random first-order transition' theory [19] of glasses could explain the Boson peak. The random first-order transition theory of the glass transition suggests a dynamical mosaic structure which is directly manifested in the dynamical heterogeneity. Mean field calculations on supercooled liquids and glasses suggest a *discontinuous transition* without latent heat at a Kauzmann temperature $T_K$, as found in infinite range spin glasses that lack reflection symmetry, which is called 'random first-order' [20]. More recently Wolynes shows that the transition states in a statistical field theory are instantons possessing additional replica symmetry breaking in the entropic droplet and argues that this new replica symmetry breaking state arises from configurational entropy fluctuations [20].

One of the most striking is that a population of fast particles moves co-operatively in string-like paths and clusters into transient, fractal structures that grow rapidly in size as the glass transition is approaching [21]. Weeks [22] emphasizes that the existence and behavior of the clusters indicate that the relaxations are very inhomogeneous, both temporally and spatially. This correlated motion can play a critical role in the glass transition, and its consequences must be incorporated in any theoretical treatment.

This paper mainly aims to find out the intrinsic mosaic geometric structures in the ideal (flexible polymer) glass transition and in macromolecular motion, from which current various theories and explanations on glass transition and on reptation can be involuntarily deduced. The paper also tries to obtain a universal picture of cooperative migration of particle-clusters in a whole temperature range from $T_g$ to $T_m$. More complex jamming behavior of 200 chain-particle cooperative rearrangements will also be described in this paper. It will be proposed that the additional symmetry breaking in the solid→liquid glass transition deals with the additional Lindemann distance increment. The 8 orders of 2-*D* clusters, which are transient states possessing additional replica symmetry breaking in the glass transition, will be strictly defined and the 3-*D* clusters are affirmed to turn out to be the hard-spheres with the negative attractive-potential character. Before the hard-sphere model is modified, the current models and concepts in the glass transition and in the macromolecular motion have to be seriously considered as follows.

On the one hand, some experiments indicate a correlation between the nature of glass transition and the relative concentration of tunneling modes [23], which is a localized quantum mechanical two-level tunneling system. In particular, the tunneling model is based on the assumption that collective structural rearrangement, by means of tunneling of a local conformational state into another, is the main relaxation mechanism in glass. These two nearly degenerate conformational states form a two-level system. Although a broad range of experimental observations can be explained in terms of this idea, there are many open questions concerning the standard two-level model [24]. The simple 'two-state' models [25] have recently reappeared [26-28]. Merolle and co-workers [29] have proposed a modified 'two-state' model with Gaussian widths for the site energy of both ground and excited state. This model assumes the presence in condensed phase of degrees of freedom which, in real space, can exist in ground and excited states. The excitations of the two-state models correspond to mobility excitations leading to dynamic heterogeneity [27]. A site can only change its state if two of its neighbors are simultaneously in their excited state [29].

In this paper, the external degree of freedom for generating a new larger cluster is always equal to 1 in the glass transition as long as the two-state of 'site' is modified as that of the 'interface' between two sites. These modifications are in accordance with the result of Wolynes [30, 31], who with his co-workers have described a mosaic model in which the inter-domain *boundary energy* plays a vital role in the thermodynamics.

The so-called inter-domain boundary here will be defined as the evolution cluster (the thawing local zone from small to large), i.e., the instant symmetric loop surrounded by the interfaces in excitation states defined in this paper.

Wolynes [9] also have pointed out that the multilevel behavior of these domains can be pictured as involving the concomitant excitation of the *ripplon* modes of the glassy mosaic along with the transition between local minimum energy configurations of a mosaic cell. Thermally generated spontaneous density fluctuation gives rise to an interface tension. The interface tension acts as a restoring force on the fluctuation and it excites an interface tension wave, which is called a capillary wave or ripplon. These excitations occur also above $T_g$ so long as the system is below $T_A$ [9]. The capillary waves occurring on liquid-liquid interfaces have quantized energy [9, 32] and various wavelengths, and each frequency depends on each wavelength. Especially, one of the important results proposed by Wolynes, the $l$-th ($l = 2, \ldots 9$) mode of a sphere is $(2l + 1)$-fold degenerate and here are no harmonics of higher than 9th to 10th order, corresponding to an atomic scale half-wavelength. $l = 1$ corresponds to a domain translation and $l = 2$ is in agreement with the boson peak.

This result of existence of the limited domain wall vibration frequencies will be applied to form the interfaces of 8 orders of different transient 2-*D* clusters in size in this paper, however, this paper will point out that in 3-*D* space, forming a *compacted* $i$-th order of 3-*D* clusters (3-*D* domains) results from the cooperative contributions of all frequencies because of the anharmonic vibrations in ripplons.

On the other hand, in polymer physics, macromolecules can only move extremely slowly in the topological direction of chain-segment according to the reptation model (or the tube model) proposed by de Gennes [33]. The energy of reptation comes from the so-called defect, which is similar to one type of ideal gas. The reptation model in melting state must arise from the solid-to-liquid glass transition. Thereby, in glass-thawing, the cluster migration should be to a certain extent restricted by reptation of a macromolecular chain. This hypothesis is conceivable that the glass transition has the molecular weight dependency [34]. In polymer physics, when the increasing of molecular weight exceeds a certain value, the transition from liquid-state to 'solid-state' will occur in the system. Current hard-sphere models, however, have not been able to reflect such restrictions. It is noteworthy that Garrahan's [35] result (that the mobility of particle carries a direction to persist in a glass former for significant periods of time) and Wolynes's [30] picture of site-excitation (that one of the dynamical rules is in the limit of maximal directional persistence and the defect moves induce sites along its trajectory to relax) also have more or less reflected such restrictions. These pictures are reminiscent of the spatial fluctuations of an order parameter close to a continuous phase transition, such as the magnetization of an Ising model near criticality, and the order parameter is purely a dynamic object [9, 36].

A possible case absent in current hard-sphere models is that the interval in 'forming a dynamical hard-sphere' (which migrates one step in local z-axial direction) is so long (glass transition time scale) that hides the relevancy of all 2-*D* clusters *along one direction*, with different relaxation times, on a local x-y projection plane during the interval.

## 2. Basic concepts and model

### 2.1. Key concept of viscosity: number of particle-clusters moving along one direction

In flexible polymer system, there are two well-known empirical laws: one is Williams–Landel-Ferry (WLF) equation [44] to describe temperature and time dependent behaviour in the glass transition, the other is the 3.4 power law of viscosity for entangled macromolecules, i.e. the viscosity depends on the molecular weight as $\eta \sim M^{3.4}$, independent of temperature, solvent and molecular species [45]. Theorists have been trying to prove the two empirical laws directly from fundamental theories. All the two empirical laws deal with the microscopic descriptions of viscosity in Flory ideal random system, which includes the number of particles in cooperative migration, needful energy and time to excite particle-clusters *along one direction*. The microscopic description of viscosity is so complex that the migration along one direction will entangle with the particle-motions along other directions. If the excitation energy is quantized, as in the Ising model, a local z-axial direction can be selected as *the z-component* particle-cluster delocalizing direction in a referenced 3-*D* local field (chain-particle $a_0$ as center in Figs.1, 3 - 6). This paper will detailed discuss the thawing processes in the referenced local z-axial space, or say, on a local (2-*D* x-y) projection plane of referenced chain-particle $a_0$.

### 2.2. Mott disorder theory and interface excitation state

Mott disorder theory [46] is one of the fundamental theories in condensed matter physics. Mott deems that the outside electrons of all atoms (molecules) in random system are correlated and the electrons form a lattice to maximally avoid each other in order to minimize the totally dominant potential energy.

In solid-to-liquid transition, in the same way, all the outside Van der Waals repulsion electron-electron (two repulsion electrons maximally avoid each other in local z-space, see Fig.1, the red circle in (a) and the blue circle in (b)), appearing on the interfaces of two adjacent particles in random system, are also correlated and always *appeared one by one in time in 3-D space* to maximally avoid *simultaneity*. And these coupled repulsion electron pairs always tend to occur in *a certain space-direction* in a 3-*D* solid-domain, e.g. in local z-space in a reference domain, and are capable of forming local 2-*D* lattice in order to minimize the totally dominant delocalized energy for z-component particle-clusters migrating along z-axial. The local 2-*D* lattice in the solid-to-liquid glass transition is namely the 8 orders of instant 2-*D* mosaic geometric structures, and will be revealed in this paper.

Van der Waals interaction is the weakest of all intermolecular attractions between molecules. However, with a lot of *one-special repulsion states* in Van der Waals interaction between *two self-similar particle-clusters*, the interaction can play a *critical role* in the solid-to-liquid glass transition (soft-matter theory).

There are both states of attractive and repulsive in Lennard-Jones (L-J) potentials (van der Waals forces, see Fig. 7) that control the thermo-fluctuation distance between two chain-particles. Attractive state (on the right side of L-J potential curve) involves the interactions among induced dipoles that arise from fluctuations in the charge densities that occur between adjacent uncharged non-bonded atoms. Repulsive state (on the left side of potential curve) involves the interactions that occur when uncharged non-bonded atoms come very close together but do not induce dipoles. The repulsion is the result of the *electron-electron repulsion* that occurs as two 'clouds' of electrons begin to overlap.

The possible weakness of the 'clouds' is the ignorance of the *synchrony* with dipoles, especially in the case of the critical phase transition. When an instantaneous dipole atom approaches an adjacent atom; it can cause that atom to also produce dipoles. The adjacent atom is then considered to have an induced dipole moment. Even though these two atoms are interacting with each other, their dipoles may still fluctuate. However, *they must fluctuate in synchrony* in order to maintain their dipoles and stay interacted with each other. The attractive state of synchronizing fluctuation of two dipoles can cause the two repulsion electrons to also form a synchronizing coupled electron pair in the repulsive state.

Therefore, one special interface repulsion state in Fig.1 (a) and (b), the coupled repulsion electron pairs of Van der Waals interacting between two chain-particles $a_0$ and $b_0$ can be singled out in the critical fluctuation in critical phase transition.

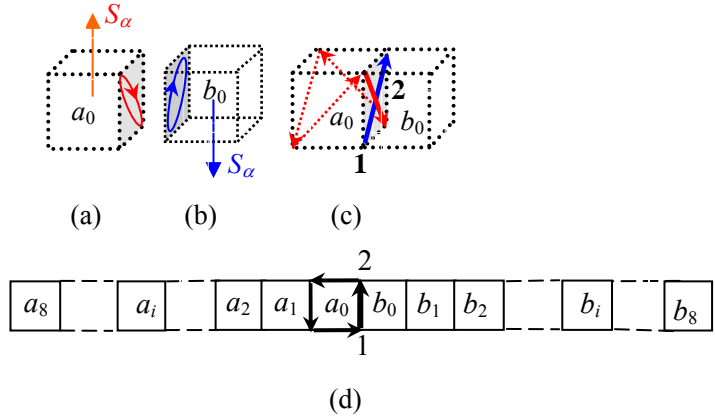

Fig.1. Only in the case of a coupled repulsion electron pair (the arrow 1-2) on the interface between particle $a_0$ in (a) and particle $b_0$ in (b) forms an interface of a closed loop in (d), the two coupled repulsion electrons (red arrow-blue arrow) simultaneously move an interface from sideline 1 to sideline 2 in (c), and $S_\alpha$ in (a) and $S_\alpha$ in (b) have suddenly the same instantaneous spin component $S_\alpha$ in (c). (d) The black thick arrow ↑ represents the cross-coupled electron pair between $a_0$ and $b_0$ on x-y projection plane in $a_0$ local field. Each arrow is scaled by the quantized interface excitation energy $\Delta\varepsilon(\tau_i)$ and has a same unit length for different relaxation times.

The singularity is that the two coupled repulsive electrons have respectively 'oscillated' on their own repulsive interfaces, seeing the electron on the interface of $a_0$ in (a) and the electron on the interface of $b_0$ in (b) in Fig.1, until a 'closed loop' (the 4 interface coupled repulsion electron pairs occur one by one) encircled $a_0$ is appeared.

Therefore, only in the case of closed loop appears, the two coupled repulsion electrons *simultaneously move* from the sideline 1 to the sideline 2 in local z-space in Fig.1 (c) (Notice the special manner of the two repulsion electrons, the strongly repulsion of the two repulsion electrons at one dot will supply the condition for compacting cluster and resonance energy transfer in the critical phase transition.), or say, *from the position 1 to 2* on the interface between particles $a_0$ and $b_0$ on x-y projection plane in Fig.1 (d). The 'moving an interface' for a cross-coupled electron pair is called *an excitation interface*, which is an instantaneous vector on x-y projection plane in particles $a_0$ local field. The appearance and the state of the cross-coupled electron pair on projection plane is respectively called as an *interface excitation* and an *interface excitation state* in solid-to-liquid glass transition.

It is clear, the presence and the occurrence of the interface excitation state have no use for any presupposition and relevant parameter. The cross-coupled electron pair is in fact the degenerate-oscillation state for a coupled repulsion electron pair on an interface. And the coupled repulsion electron pair is actually, *an especial electron-electron microscopic state*, one of the incalculability microscopic states within a classical Van der Waals repulsive 'clouds' between two neighboring z-component particles and exists *in any random system* (solid state, liquid state and gaseous state) *at any temperature*. Thus, the term the interface excitation states are in nature *singled out the especial state* in the normal Van der Waals interactions, which has the *maximal interface repulsive energy*, the ability of *compacting cluster* and *resonance energy transfer* in critical phase transition. In Section 4, we will proof that this especial state corresponds to the intrinsic stability fixed point of L-J potential.

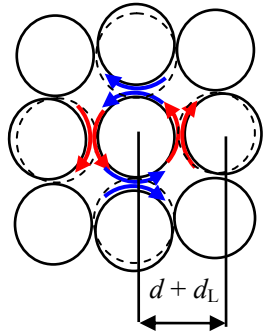

Fig.2. The additional Lindemann distance increment $d_L$ between two hard-spheres (dashed line) appears by thermally excitation interfaces (arrows) in solid-to-liquid glass transition. $d_L \approx 0.10\sigma$ is a universal constant for all materials. $\sigma$, the diameter of hard-sphere.

In other words, based on the Mott disorder theory, the 'local thermal motion energy' in the local z-space can more and more be higher than that in the local x- or y-space in a reference domain (compare the physical picture with magnets! the balance of random thermal motion energy in all domains) when the temperature or the time increases. Thus, in order to expediently refract the increasing number and energy of interface coupled electrons when the temperature and the time increases, we also call interface coupled repulsion electron pair as interface excitation state (the strictly definition for interface excitation is the cross-coupled electron pair in Fig.1 (c), however, all coupled repulsion electron pairs must change as cross-coupled electron pairs in critical phase transition based on Mott disorder), which possesses the added interface 'excitation energy' $\Delta\varepsilon$, relative to the average Van der Waals' interface energy absent coupled electrons.

Therefore, all energy to excite solid-to-liquid glass transition comes from the additional energy of these excitation interfaces excited by temperature and time. We only need to consider the entire interface coupled repulsion electron pairs (cross-coupled electron pairs) in a reference domain in the glass transition. The number and the distribution of these interfaces excitation states should meet the minimum energy requirement to excite the glass transition.

In the solid-to-liquid glass transition, when two interface repulsion electrons occupy an interface excitation state, there exists also an instantaneous Lindemann distance increment $d_L$ (or say an extra volume, $\Delta v$) on the interface, see Fig.2.

**2.3. An additional spin system with number of spin components $n = 0$**

In the $n$ vector model [33], the magnetic atoms are located on a periodic lattice. Each magnetic atom carries a spin $S$, this is a vector, with $n$ continuous spin components $S_{\alpha, \beta, \gamma...n}$, $\alpha, \beta, \gamma... \in (0, n)$; while $S_\alpha$ is only a spin component. De Gennes pointed out that the spin system with number of spin components $n = 0$ can correspond to the macromolecular system, which formed by all different spatial conformational macromolecules of self-avoiding walks of $N$ steps linking the two ends of chain-length $N$ on the lattice.

The link between the number of spin components $n = 0$ and the macromolecular chain results from that when $n = 0$, only one type of average is nonvanishing, i.e. the quadratic term $\langle S_\alpha \cdot S_\beta \rangle_0 = \delta_{\alpha\beta}$; where $S_\alpha$ is the spin component coming of one chain-particle with one covalent electron, and the two spin components $S_\alpha$, $S_\alpha$ of two chain-particles combined by a covalent bond involve the same spin component (the position-phase difference of the two neighboring chain-particles on a chain is $\pi$), satisfying $\langle S_\alpha \cdot S_\alpha \rangle_0 = 1$. This system for the number of spin components $n = 0$ contains all *spatial* (does not involve the temporal or mosaic) macromolecular chains linking the two ends of chain-length $N$ on the lattice. However, there is no additional energy to excite self-avoiding walks of a *spatial* long-chain in the system.

A new additional spin system with number of instantaneous spin components $n = 0$ has been found and proposed in this paper. This is a concomitancy spin system with macromolecular motion and provides its energy to rearrange conformation. When two coupled repulsion electrons are in the cross-coupled state in Fig.1 (c), the two instantaneous spin components $S_\alpha$ in Fig.1 (a) and $S_\alpha$ in fig.1 (b) being in the state of in Fig.1 (c), also involve the same component (the position-phase difference of the two chain-particles on two neighboring chains is also $\pi$), and also satisfying $\langle S_\alpha S_\alpha \rangle_0 = 1$. The 1 means one excitation interface state. Because all excitation interfaces occur one by one in time, there is no two excitation interfaces appearing simultaneously with a certain time interval based on Mott disorder. The two instantaneous spin components in an $n$-dimensional space is only a 'point', thus, the number of instantaneous spin components $n = 0$.

The special property is that the additional spin system contains *only one spatial reference excitation interface* with a certain time interval and the other *temporal excitation interfaces on all self-avoiding closed loops*, in which each loop must pass through the reference excitation interface in order to satisfy the instantaneous spin components $n = 0$ condition for the reference coupled electron pair in Fig.1 (c).

Using the property, we can find out the *entire spatial and temporal excitation interfaces* in a reference domain, which corresponds to the maximal 'Hamiltonian' in a reference domain.

The theoretical method in this paper is to firstly obtain the maximal 'Hamiltonian' in Z-space using geometric method and frustration, which is the activation energy to break solid lattice (Section 3.17), and to find out the maximal 2-*D* closed-loop with mosaic structure (Section 2.17), then to get the minimum energy to glass transition using the mean field method (Section 3.13).

**2.4. Method and notation to describe geometrical phases**

The excitation interface (from 1 to 2 in Fig.1 (d)) is a *vector*, denoted as $j(a_0\uparrow b_0)$. The particle on the left side of the arrow ↑ in the parentheses (in this case $a_0$) means that the succedent excitation interfaces (because of the one by one excitation in time) will *first* form a loop around $a_0$.

According to the reptation model [33], the thawed reference particle $a_0$ will migrate along, e.g. the local z-axial, which is also the z-component direction of covalent bond (or the direction of stronger bond, the unexcited Van der Waals' bond for small molecular system) connecting with $a_0$. The condition of migration or delocalization for $a_0$ is that the 4 interfaces which encircle $a_0$ should be first thawed, or say, the 4 interfaces of $a_0$ should be first excited on $a_0$ local x-y projection plane. Thus, the concept of 2-Dimension interface excitation energy loop-flow of an excited particle should be introduced, which is also a z-vector with *relaxation time* (namely, needful time exciting the 4 interfaces). As is shown in Fig.1 (d), the 4 excitation interfaces encircled $a_0$ will form an additional *interaction excitation energy loop-flow* on $a_0$ local x-y projection plane. In the glass transition, because only closed cycle-flow (that will generate non-integrable phase, an additional induced potential, see Section 2.16) corresponds to the abnormal heat capacity and Boson peak; we need only discuss all the interaction excitation energy loop-flows on the reference particle $a_0$ local filed. The discussion is further simplified to only consider the loops formed by several consecutive arrows on a 2-*D* projection plane, seeing Fig.3- 6.

**2.5. Eight orders of relaxation time spectrum**

Note that the excitation interface $j(a_0\uparrow b_0)$ also comes from the cooperative contribution of the neighboring particle-sites $a_i$ and $b_i$ ($i \in [0, 8]$), see Fig.1 (d), here comes the result of Wolynes: $j(a_0\uparrow b_0)$ only contains the contribution of 8 harmonic frequencies. The interface excitation thus is quantization of energy and has the relaxation (or creation) time spectrum $\tau_i$ ($i \in [0, 8]$), so the $\Delta v$ and the $\Delta\varepsilon$ should be respectively modified as $\Delta v(\tau_i)$ and $\Delta\varepsilon(\tau_i)$. As the contribution of different neighboring particle-pairs at different times to the interface excitation between sites $a_0$ and $b_0$ always keep the direction of projection wave vector component, i.e., the direction of the arrow ↑ between sites $a_0$ and $b_0$, the energy quantization of interface excitation is hidden.

That means *once the first excitation interface with the direction 1 → 2 shown in Fig.1 (d) appears* (at the local time $t_{0, 1}$) *in $a_0$ local field, the directions of all the succedent excitation interfaces* (appear respectively at the local time $t_{i, l}$, $i \in [0, 8]$, the number of order; $l \in [1, L_i]$, $L_i$, the number of excitation interfaces of *i*-th order, see Section 2.13.) *will be completely confirmed so as to form 8 orders of interface excitation energy loop-flows* as shown in Fig.3 - 6, otherwise, the total excitation energies in the system will increase. This property also comes from Mott disorder theory: all interface excitations in random system are correlated. This property is called as *the priority of the first interface excitation*, which is the primary reason to generate 'cage effect', and non-ergodic in the glass transition.

That is the reason and also the advantage to bring the signature of arrowhead into the standout mode of excitation interface so as to get the minimum energy in exciting glass transition.

Here the interface excitation represented by arrow is scaled by the quantized excitation energy $\Delta\varepsilon(\tau_i)$ and thus the arrow is of a *unit length* for different wave lengths or frequencies on the (energy) lattice model. The convenience of this description is that the contribution of different harmonic frequencies or the different relaxation times to interface excitation on $j(a_0\uparrow b_0)$ will be expressed by different sizes of 2-*D* symmetric loop-flow surrounding $a_0$ on x-y projection plane.

All interfaces in excitation state with the extra volumes are correlative in space-time. The geometric method will be used in this paper in order to refract 'the interface-interface correlation' and the geometric frustration in local space. Quantized excitation energy $\Delta\varepsilon$ does not depend on temperature in flexible polymer system. Increasing temperature only *increases the number of interface excitation per unit time* to accelerate the thawing of domain frozen in the glass transition.

It is convenient that the model is described in terms of a lattice scaled by $\Delta\varepsilon(\tau_i)$. Whole excitation interfaces can be represented on the lattice formed by square cells, which is the requirement of minimized energies to excite glass transition. In Fig.1 (d), the thick arrow 1-2 represents the projection direction of 'interface capillary wave' (so-called Ripplon) with the extra volume $\Delta v(\tau_i)$ and excitation energy $\Delta\varepsilon(\tau_i)$. Therefore, all energy (expressed by $kT$) to excite solid-to-liquid glass transition comes from the additive energy loop-flow, in manner of the *energy flow-percolation* (the percolation connected one by one by interface excitation energy loop-flows) formed by all excitation interfaces in 3-*D* space.

One of the singularities of glass transition is that the thermally and time excited interface energy loop-flows only occur on the interfaces of a few thawed domains, which are able to form an energy flow-percolation during observation time. That is consistent with the dynamic heterogeneity [27] and non-ergodic [37]. What is being cared is that

(i) How much the number of whole excitation interfaces is, i.e. the *number of whole interface excitation states*, to thaw a domain in the glass transition?

(ii) How much the flow-percolation energy is to form an interface excitation energy loop-flow in 3-*D* space?

**2.6. Pair of additional restoring torque**

The position of particle $a_0$ or $b_0$ in solid state in Fig.1 (d) is in equilibrium, thus the universal Lindemann displacement $d_L$ in Fig.2 generates a pair of additional restoring torque (comes of the cross-coupled state of two electrons in Fig.1 (c)), denoted as $M_0(a_0\uparrow b_0)$, which is the contribution coming from the interface tension between the *pair* of particles $a_0$ and $b_0$, or the torque with relaxation time $\tau_0$ between the two 'solid-state' cells of $a_0$ and $b_0$. $M_0(a_0\uparrow b_0)$ also gives rise to an additional *position-asymmetry* on the projection plane because $a_0$ firstly surrounded by coupled electron pairs. There are 8 orders of additional restoring torques with $\tau_i$ on $j(a_0\uparrow b_0)$, which come from the contributions of the pair of microscopic 'solid state' blocks of $(a_0 + a_1 \ldots + a_i)$ and $(b_0 + b_1 \ldots + b_i)$ in Fig.1 (d), denoted as $M_i(a_0\uparrow b_0)$, in which torque energies, i.e., $\Delta\varepsilon(\tau_i)$, are also stored in the excitation interface $j(a_0\uparrow b_0)$.

Here the theoretical treatment is made to the transient lattice model (mean filed method): the harmonics wavelength in ripplon corresponds to the length of such 'solid-state' lattice-block, whereas the homologous anharmonicity in ripplon is represented by both the behavior of *mosaic-blocks* with mosaic characters and the 8 orders of compacted 3-*D* clusters with more complex mosaic behaviors.

A region of space that can be identified by a single mean field solution is called a mosaic cell [38]. The details of additional replica symmetry breaking using geometric method are going to be discussed in the following part.

**2.7. Denotation of energy loop-flow and definition of acting particle**

In the denotation of $j(a_0\uparrow b_0)$, it is denoted that the particle on the left side of the arrow $\uparrow$ in the parentheses is the central particle in local filed and will be *first* (because of phase difference of $\pi$ between $j(a_0\uparrow b_0)$ in $a_0$ local field and $j(b_0\uparrow a_0)$ in $b_0$ local field) surrounded by 4 one after another excitation interfaces to form a minimum energy loop-flow, in the form of interface excitation energy loop-flow, as the 4 red-arrows in Fig.1 (c), or the 4 arrows in Fig.1 (d) or the 4 thick (a pair of blue and a pair of green) arrows in Fig.3 (b), denoted as $V_0(a_0\uparrow b_0)$.

Here $V_0(a_0\uparrow b_0)$ also defines an *acting particle* (or say, an excited particle, a contributing particle to the glass transition) $a_0$ surrounded by interface excitation energy loop-flow (or by 4 z-component-excitation interfaces) in solid-to-liquid transition. Thus, an acting particle has topological properties: cyclic direction, cyclic jumping-off point and the *first interacting interface* $j(a_0\uparrow b_0)$ in the 4 neighboring interfaces of $a_0$. Clearly, there is a phase difference of $\pi$ between $V_0(a_0\uparrow b_0)$ and $V_0(b_0\uparrow a_0)$ and the latter denotes the excited particle $b_0$ in $b_0$ (local) field.

The denotation of the excited particle $V_0(a_0\uparrow b_0)$ can be simplified as $V_0(a_0)$.

Notice, in Fig.3 (b), a pair of blue arrows and a pair of green arrows are all the coupling (exciting) interfaces of $\pi$ phase difference, but there is $\pi/2$ phase difference between the former and the latter (comes from cross-coupled electron pair). Thereupon, the interface excitation energy loop-flow of an acting particle can be also considered as the 'inner-mosaic' loop-flow of the two pairs of coupling excited interfaces of $\pi/2$ phase difference.

**2.8. First order of symmetry energy loop-flow and first order of transient 2-D cluster**

Within the loop of $V_0(a_0\uparrow b_0)$, the Lindemann displacements of the 4 excitation interfaces let particle $a_0$ replace to the *primal equilibrium position* in real space, while to the central of block $a_0$ in the (energy) lattice model.

As the interface tension with $\tau_0$ on $j(a_0\uparrow b_0)$ comes from the contribution of the torque between two particles $a_0$ and $b_0$, thus the condition to form symmetric cycle of $V_0(a_0\uparrow b_0)$ for eliminating additional position-asymmetry is that the 4 neighboring symmetric loop-flow of $a_0$: $V_0(b_0\uparrow a_0)$, $V_0(c_0\uparrow a_0)$, $V_0(d_0\uparrow a_0)$ and $V_0(e_0\uparrow a_0)$, should also be one by one finished (because of the Lindemann displacement $d_L$ being the *coupling value* of two particles). At the instant time $t_1$ in $a_0$ (local) field, once 4th cycle $V_0(e_0\uparrow a_0)$ is finished, a new loop-flow of $a_0$ surrounded by the excitation interfaces with $\tau_1$ timescale will occur as in Fig. 3 (b). The new symmetric loop-flow is defined as the interfaces of the *first order of transient 2-D cluster* in $a_0$ field, which is denoted as $V_1(a_0\uparrow b_0)$ and simplified as $V_1(a_0)$.

The cycle path of $V_1(a_0\uparrow b_0)$ is 1→ 5→ 6→ 4→ 7→ 8→ 3→ 9→ 10→ 2→ 11→ 12→1 ($\delta$ →0), whose cycle direction is negative, contrary to that of $V_0(a_0\uparrow b_0)$. The energy walking *one step and only one step* of the loop-flow surrounded by 12 one after another excitation interfaces is the energy of one excitation interface with relaxation time $\tau_1$, $\Delta\varepsilon(\tau_1)$.

$V_1(a_0\uparrow b_0)$ turns out to be the first order of concomitant $2\pi$ loop-flow with particle $V_0(a_0)$ at the instant time $t_1$ and at the moment the additional position-asymmetry of particle $V_0(a_0)$ has been eliminated within $V_1(a_0)$ and the harmonic interface tension relaxations of $\tau \le \tau_1$ have been first realized on the 4 excitation interfaces of $V_0(a_0)$.

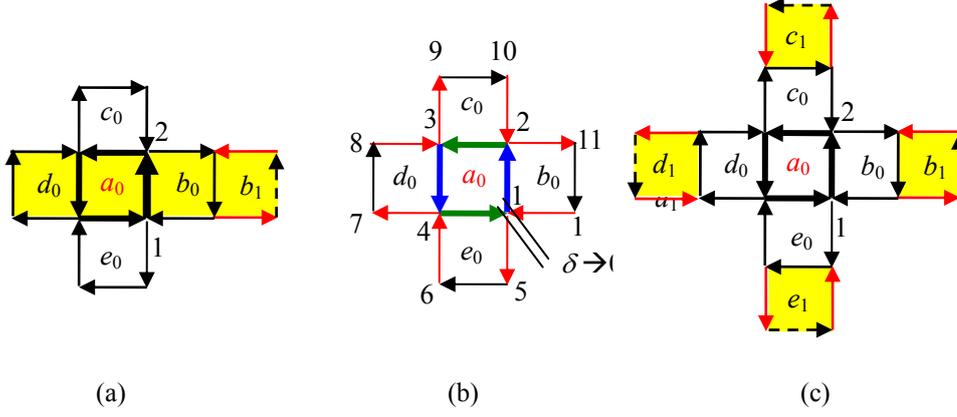

(a)           (b)           (c)

Fig.3. (a) The first order additional restoring torques $M_1(a_0\uparrow b_0)$ (thick arrow 1-2 with coupling two yellow blocks $(a_0 + d_0)$ and $(b_0 + b_1)$) breaks the position-symmetry in loop $V_0(a_0\uparrow b_0)$ of particle $a_0$ and new additional removing asymmetrical process should be adopted in the succedent evolution. (b) The symmetric $V_0(a_0)$ interacts one by one with its four symmetric neighboring excitation particles, $V_0(e_0)$, $V_0(d_0)$, $V_0(c_0)$ and $V_0(b_0)$, and the 12 excitation interfaces (red-black thin arrows) form $V_1(a_0\uparrow b_0)$, which is the first order of transient symmetrical loop of excited particle $V_0(a_0)$ in $a_0$ local field. Within $V_1(a_0\uparrow b_0)$, the interface tension relaxation of $\tau \le \tau_1$ has been realized on the 4 interfaces of $V_0(a_0)$. (c) 4 yellow blocks are concomitant mosaic cells with $V_1(a_0\uparrow b_0)$.

Note that the 'solid-block' $b_1$ in the torque $M_1(a_0\uparrow b_0)$, i.e., the thick arrow 1-2 with coupling blocks $(a_0 + d_0)$ and $(b_0 + b_1)$, on $j(a_0\uparrow b_0)$ in Fig. 3(a) is a *mosaic cell* that does not appear in the succedent $V_1(a_0\uparrow b_0)$. There are 4 such torques: $M_1(a_0\uparrow b_0)$; $M_1(a_0\uparrow c_0)$; $M_1(a_0\uparrow d_0)$; $M_1(a_0\uparrow e_0)$ on the 4 interfaces of particle $V_0(a_0)$, so, there are 4 mosaic cells, i.e. $b_1$, $c_1$, $d_1$ and $e_1$ when $V_1(a_0\uparrow b_0)$ appears in Fig.3(c).

**2.9. Topological properties of the first order of transient 2-*D* cluster**

First order transient symmetry loop-flow (also the first order transient 2-*D* cluster) has 12 excitation interfaces and is of the interface energy of $12\Delta\varepsilon$ ($\tau \le \tau_1$). The number of cooperative excitation particles in first order transient symmetry loop-flow is 5 (i.e. $a_0$, $b_0$, $c_0$, $d_0$ and $e_0$ in Fig.3 (a)). Forming a first order symmetry loop-flow deals with the *determinate path*, *cycle direction* and *cyclic jumping-off point* of the 16 'self-avoiding walks' of excitation interfaces, i.e., the given 16 interfaces (steps) to form $V_0(a_0\uparrow b_0) + V_1(a_0\uparrow b_0)$ illustrated in Fig.3 (b). (In local z-space, the 16 steps are in self-avoiding.)

The walk path of the given 16 self-avoiding walks of excitation interfaces is the elementary method to form the following orders of transient 2-*D* clusters to the *replica symmetry breaking* in larger areas.

It can be seen that the 16-self-avoiding-walk of interface cross-coupled electron pairs is provided with the characteristic of Boson (see 4.18).

The formation of 12 excitation interfaces of first order symmetry loop-flow is also realized as that, seeing Fig.3 (b), the 8 red arrows (i.e. 4 pairs of coupling interfaces) contributed from 4 neighboring local fields ($b_0$, $c_0$, $d_0$ and $e_0$) plus the 4 new interfaces (i.e. 4 black arrows in Fig.3 (b)) with $\Delta\varepsilon$ ($\tau \leq \tau_1$) transferred from the 4 interfaces (thick blue and green arrows) of particle $V_0(a_0)$.

### 2.10. The 2nd order of transient 2-D cluster

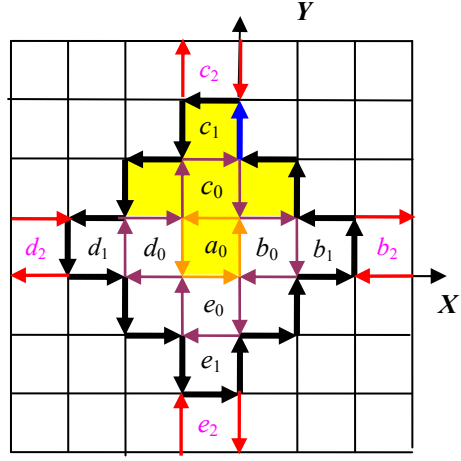

Fig.4. The 20 thick arrows denote 2nd order of transient 2-D cluster in $a_0$ local field, $V_2(a_0 \uparrow b_0)$ that only appears on the instant lattice graph at the instant time $t_2$. The yellow shadow represents symmetric $V_1(c_0 \uparrow a_0)$ loop in $c_0$ field. Neighboring 4 symmetric loops, $V_1(c_0 \uparrow a_0)$, $V_1(d_0 \uparrow a_0)$, $V_1(e_0 \uparrow a_0)$ and $V_1(b_0 \uparrow a_0)$ cooperatively generate $V_2(a_0 \uparrow b_0)$ in $a_0$ field. The 'solid-block' $e_2$ in the torque $M_2(a_0 \uparrow e_0)$ on $j(a_0 \uparrow e_0)$ is a mosaic cell that does not appear in the succedent $V_2(a_0 \uparrow b_0)$.

Once $2\pi$ cycle $V_1(a_0 \uparrow b_0)$ is finished, the reference particle $V_0(a_0)$ will replace the center of block $a_0$, whereas the additional restoring torques $M_2(a_0 \uparrow e_0)$ (i.e., torque between the two 'solid-state blocks' of ($a_0 + c_0 + c_1$) and ($e_0 + e_1 + e_2$)) with $\tau_2$ on $j(a_0 \uparrow e_0)$ will first and again break the position-symmetry of particle $V_0(a_0)$ in the block $a_0$, as in Fig. 4. Thus a new process to eliminate position-asymmetry should be again adopted in the succedent evolution.

In order that the reference particle $V_0(a_0)$ again replaces the center of the block $a_0$, 4 symmetric loops: $V_1(c_0 \uparrow a_0)$ in $c_0$ field (shadow in Fig.4), $V_1(d_0 \uparrow a_0)$ in $d_0$ field, $V_1(e_0 \uparrow a_0)$ in $e_0$ field and $V_1(b_0 \uparrow a_0)$ in $b_0$ field, should one by one be adopted in $a_0$ field. Therefore, the 2nd order of transient 2-D cluster with $\tau \leq \tau_2$, denoted as $V_2(a_0 \uparrow b_0)$, is formed at the instant local time $t_2$ when the 2nd order $2\pi$ loop-flow finished in $a_0$ (local) field, whose cycle direction is positive, contrary to that of $V_1(a_0 \uparrow b_0)$. The cyclic jumping-off point of 2nd order $2\pi$ loop-flow is the blue arrow in Fig.4. Note that the 'solid-block' $e_2$ in the torque $M_2(a_0 \uparrow e_0)$ on $j(a_0 \uparrow e_0)$ in Fig. 4 is a mosaic cell that does not appear in the succedent $V_2(a_0 \uparrow b_0)$. There are 4 mosaic cells, i.e. 'solid-blocks' $e_2$; $b_2$; $c_2$ and $d_2$ in Fig.4.

### 2.11. Topological properties of the second order of 2-D cluster

The number of interfaces of $V_2(a_0 \uparrow b_0)$ loop-flow is 20 (20 thick arrows in Fig.4). The number of interfaces of torque relaxed with $\tau \leq \tau_2$ inside $V_2(a_0 \uparrow b_0)$ loop-flow is 12, and equals to the number of the interfaces of $V_1(a_0 \uparrow b_0)$. This means that the evolution energy from $V_1(a_0 \uparrow b_0)$ to $V_2(a_0 \uparrow b_0)$ is $8\Delta\varepsilon\,(\tau \le \tau_2)$.

As same as generating $V_1(a_0\uparrow b_0)$, seeing Fig.4, forming $V_2(a_0\uparrow b_0)$ in $a_0$ field is the contribution of 4 $V_1$-clusters ($V_1(c_0\uparrow a_0)$ in $c_0$ filed, the yellow shadow; $V_1(d_0\uparrow a_0)$ in $d_0$ filed; $V_1(e_0\uparrow a_0)$ in $e_0$ filed and $V_1(b_0\uparrow a_0)$ in $b_0$ filed) in 4 neighboring local fields of $a_0$.

As same as forming $V_1(a_0\uparrow b_0)$ in Fig.3 (b), the 20 excitation interfaces on $V_2(a_0\uparrow b_0)$ in Fig.4 are also realized as the contributions of the 8 red arrows (4 pairs of coupling interfaces of $\pi$ phase difference in Fig.3(c)) in the 4 mosaic cells ($b_1$, $c_1$, $d_1$ and $e_1$ in Fig.3(c)) plus the 12 transferred interfaces from that on $V_1(a_0\uparrow b_0)$.

This evolution rule holds true for the concomitant 8 orders of transient 2-D clusters in $a_0$ field. The number of cooperative excitation particles in second order of cluster is 13 (13 instant cells (small squares) surrounded by 20 excitation interfaces in Fig.4).

The energy walking one step of the loop-flow surrounded by 20 one after another excitation interfaces is the energy of one excitation interface with relaxation time $\tau_2$, $\Delta\varepsilon(\tau_2)$.

**2.12. Topological properties of the third order of 2-D cluster**

In the same way, the concomitant third order of transient 2-D cluster at the instant time $t_3$ in $a_0$ field, $V_3(a_0\uparrow b_0)$ can be obtained as in Fig. 5.

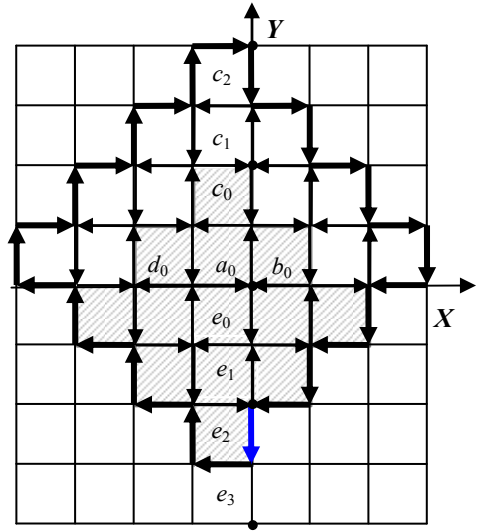

Fig.5. The third order transient 2-D cluster $V_3(a_0\uparrow b_0)$ with the interface tension relaxations of $V_0(a_0)$ in the range of $\tau \le \tau_3$. The shadow area represents the symmetric $V_2(e_0\uparrow a_0)$ loop in $e_0$ local field, whose symmetric center is $e_0$. The 'solid-block' $e_3$ in the torque $M_3(a_0\uparrow e_0)$ on $j(a_0\uparrow e_0)$ in $a_0$ local field is a mosaic cell of $V_3(a_0\uparrow b_0)$ and appears with $V_4(a_0\uparrow b_0)$.

The number of interfaces of $V_3(a_0\uparrow b_0)$ loop-flow is 28 (28 thick arrows in Fig.5). The number of interfaces of torque relaxed in the range of $\tau \le \tau_3$ inside $V_3(a_0\uparrow b_0)$ loop-flow is 20, which equals to the number of the interfaces of $V_2(a_0\uparrow b_0)$. This means that the evolving energy from $V_2(a_0\uparrow b_0)$ to $V_3(a_0\uparrow b_0)$ is $8\,\Delta\varepsilon\,(\tau \le \tau_3)$.

The energy walking one step of the loop-flow surrounded by 28 one after another excitation interfaces is $\Delta\varepsilon(\tau_3)$.

As same as forming the 20 excitation interfaces on $V_2(a_0\uparrow b_0)$, the 28 excitation interfaces on $V_3(a_0\uparrow b_0)$ in Fig.5 are also realized as the contributions of the 4 pairs of coupling interfaces of $\pi$ phase difference in Fig.4 in the 4 mosaic cells ($e_2$, $b_2$, $c_2$ and $d_2$ in Fig.4) plus the 20 transferred interfaces from that on $V_2(a_0\uparrow b_0)$.

The number of cooperative excitation particles in third order of cluster is 25 (25 instant cells surrounded by 28 excitation interfaces in Fig.5), or say, 28 excitation interfaces can excite 25 particles to cooperative thaw in third order of transient 2-D cluster along local z-axial direction.

From the results of $V_1$, $V_2$ and $V_3$, it is clear that the more the number of cooperatively thawed particles is, the less the average needful interface excitation energy for each particle is.

The key point is that what the minimum excited energy is and what the number of cooperatively thawed particles excited by the energy in the glass transition is.

**2.13. Topological properties of 8 orders of 2-D cluster**

(i) The number of interfaces of *i*-th order of transient 2-D cluster can be calculated as

$L_i = 4(2i+1) = 12, 20, 28, 36, 44, 52, 60, (68)$ for $i = 1, 2… 8$, and $L_0 = 4$.

Note that $L_8$ will be corrected as 60 by percolation.

(ii) The number of cooperative excited particles in *i*-th order of cluster is respectively:

$N_i = N_{i-1} + 4i = 5, 13, 25, 41, 61, 85, 113, (145)$ for $i = 1, 2… 8$, and $N_0 = 1$.

Note that $N_8$ also will be corrected as 136 by percolation.

(iii) For *i*-th ($i = 0, 1, 2… 7$) order of transient 2-D z-axial cluster, the number of mosaic cells is 4; the number of mosaic interfaces is 8, or say, the 4 pairs of coupling interfaces of $\pi$ phase difference, and the evolving energy from $V_i$ to $V_{i+1}$ is $8\Delta\varepsilon$ ($\tau \le \tau_{i+1}$), the numerical value of the evolving energy is $8\Delta\varepsilon$.

(iv) For *i*-th order of transient 2-D cluster, the energy of the loop-flow walking one step (and only one step) is the energy of one excitation interface, $\Delta\varepsilon(\tau_i)$. However, the numerical value of energy of loop-flow is still $\Delta\varepsilon = \Delta\varepsilon(\tau_i)$. The singularity is that the *i*-th order of interface excitation energy, $\Delta\varepsilon(\tau_i)$, is only one of the 8 components of one quantized energy $\Delta\varepsilon$.

(v) The phase difference between *i*-th order (interface excitation energy loop-flow) and ($i+1$)-th order is $\pi$.

**2.14. 5-particle cooperative excited field to break solid-lattice**

There are *4 neighboring concomitant excitation centers surrounding the referenced particle center*. Thus, the excited field by the 16-self-avoiding-walk can be also called 5-particle cooperative excited field to break solid-lattice. The forming of *i*-th order of cluster always results from the cooperative contributions of the 4 neighboring (*i*-1)-th order of clusters around the reference particle.

During the time of ($t_{i-1}$, $t_i$) in the referenced $a_0$ local field, the 4 neighboring local coordinates can take any direction (fragmentized and atactic lattices) because of fluctuation. However, *at the instant time $t_i$,* the direction of *i*-th order cluster always starts *sticking to the direction* of the referenced first order cluster (this characteristic in random system comes of the regression characteristic of the graph of Brownian motion) and forming the instant inerratic energy-lattices, such as Figs.3 – 6, to satisfy the *minimum energy requirement* of the glass transition. The energy to excite *i*-th order cluster orientable evolution thus is the energy of *one i*-th order *external degree* *of freedom* with $\tau \leq \tau_i$ of cluster.

**2.15. Energy of one external degree of freedom and potential well energy**

Each order of cluster has 8 more excitation interfaces on x-y projection plane or in local z-component space than the cluster of the sub order, which contributed by the 4 pairs of coupling interfaces of $\pi$ phase difference in 4 mosaic cells. This means that the induced potential energy of one *i*-th order of external degrees of freedom to excite *i*-th order of cluster migrating along z-axial is equal to the evolution energy of $8\Delta\varepsilon(\tau_i)$ on x-y projection plane, which has no option but to equal the potential well energy $\varepsilon_0(\tau_i)$ (seeing following Section 4.7) on z-axial of *i*-th order of cluster because generating a bigger cluster *always comes back to its equilibrium position on x-y projection plane* of the referenced $a_0$ particle. For ideal solid system, three potential well energies on x, y, z-axial always are equal. Therefore, the potential well energy in z-axial

$$\varepsilon_0(\tau_i) = 8\Delta\varepsilon(\tau_i) \tag{1}$$

**2.16. 8 orders of potential well energies $\varepsilon_0(\tau_i)$**

It should be admitted that the energy $\varepsilon_0(\tau_i)$ in Eq. (1) in fact is the non-integrable geometric-phase-induced potential in multiparticle system. All multiparticle systems feature an important topological geographic property revealed by Aharonov–Anandan [49] cyclic theory. This property states that the motion loop-flow generates a non-integrable phase factor, $\beta$, after it makes a $2\pi$ cycle movement in the potential field according to *parallel transport law* along a spatial curved surface closed cycle and back to the cyclic initial state. The nun-integrable phase factor is called A-A phase: $\beta = 1/2\Omega$. Here $\Omega$ is the solid angle for the induced charge on particle-cluster to a $2\pi$ cycle movement on potential curve-surface. The potential curve-surface of A-A cyclic in glass transition *degenerates into* the cylindrical surface (i.e. z-component space in Fig.3 - 6) parallel to the local z-axis, thus, the solid angle $\Omega$ is $2\pi$, the non-integrable geometric phase factor $\beta_i = \pi$.

The singularity in the glass transition is that the induced loop-flow kinetic energies have 8 degenerate states: $\Delta\varepsilon(\tau_i)$, and $\Delta\varepsilon(\tau_1) + \Delta\varepsilon(\tau_2) +\ldots+ \Delta\varepsilon(\tau_8) = 8\Delta\varepsilon = \varepsilon_0$ (see Section 4.7). Because kinetic energy always balances with potential energy and both phase difference is $\pi$ (see Section 5.4), the induced potential corresponding to the non-integrable phase factor $\pi$ also has 8 orders of components: $\varepsilon_0(\tau_i)$. $\varepsilon_0$ denoted as $\varepsilon_0(\tau_i)$, is a invariable value for flexible polymer (this result is as same as [39]), and $\varepsilon_i(\tau_{i,l}) = 8\Delta\varepsilon_{i,l}(\tau_{i,l})$, ($i \in [0, 8]$; $l \in [0, L_i]$), $\varepsilon_i \neq \varepsilon_{i+1}$ is of 8 orders of potential well energies for really complicated system. The latter may correspond to the so-called potential energy landscape [9, 39].

Each neighboring (*i*-1)-th order of cluster also has one external degrees of freedom of energy as $\varepsilon_0(\tau_{i-1})$, with $\tau \leq \tau_{i-1}$, so the *i*-th order of cluster in the (*i*+1)-th order is of 5 *inner degrees of freedom* with $\tau \leq \tau_i$ (called as 5 *i*-th order of inner degrees of freedom) and taking any directions during the time of $(t_i, t_{i+1})$ in referenced *local* excited field.

Each order of transient 2-D cluster connects with 4 deferred reaction mosaic cells that belong to the senior order of cluster.

Note that each excited particle on the x-y projection plane, in Fig.3 – 6, also connects with a z-component of covalent bond, and the direction of covalent bond can be in any direction. This also shows a picture of Flory ideal random distribution of covalent bonds inside *i*-th order of cluster.

**2.17. The 8th order of 2-D cluster and percolation**

In the same way, the 8th order of transient 2-D cluster, $V_8$ ($a_0\uparrow b_0$), can be obtained in $a_0$ local field, as in Fig.6. As there is no 9th cluster in system, percolation should be adopted at the instant time $t_8$, and the 8th order of cluster must be corrected for the anharmonicity in ripplon.

The uncorrected number of interfaces of $V_8(a_0\uparrow b_0)$ is 68, denoted by blue-black arrows in Fig. 6. When percolation appears in system, the 4 excited cells with $\tau \leq \tau_8$ in $V_8(a_0\uparrow b_0)$, by thick upward diagonal lines, are the mosaic cells of in the 4 neighboring $V_7$-clusters (Note: here the 4 $V_7$-clusters are respectively in the 4 neighboring 5-particle cooperative excited fields) and 4 cells with $\tau \leq \tau_7$ in the 4 neighboring $V_7$-clusters, by the color of dark green, are also the mosaic cells in $V_8(a_0\uparrow b_0)$. The 4 thick-black *inverted arrows* in Fig.6 are the *mutual interfaces* of $V_8(a_0\uparrow b_0)$ and its 4 neighboring $V_7$-clusters. Thus the number of interfaces of $V_8(a_0\uparrow b_0)$ should be corrected as 60, i.e. $L_8$ (corrected by percolation) = 60.

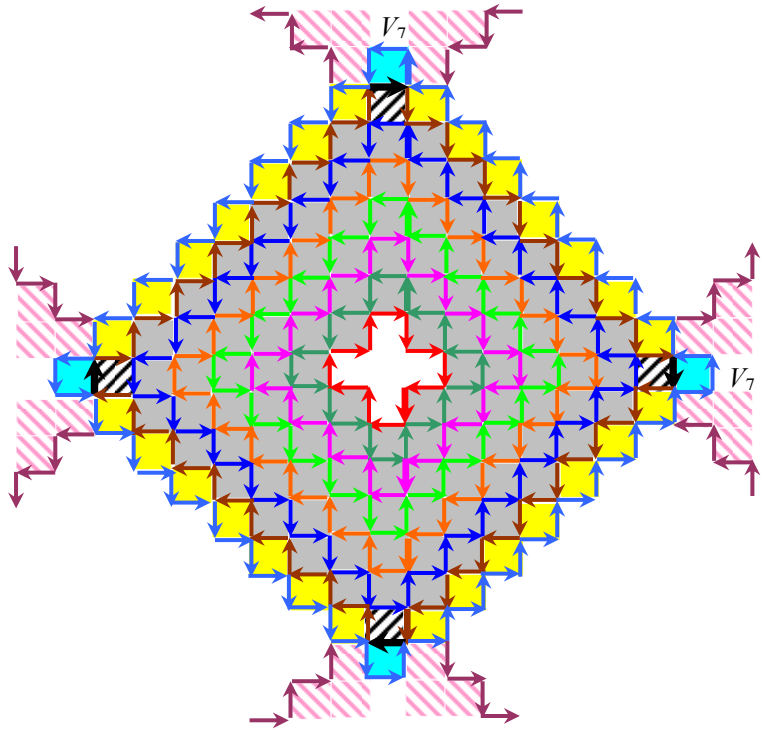

Fig.6. Sketch of the dynamical $V_8$-$V_7$ cluster percolation. The loop-flow encircled by 68 cambridge blue arrows is uncorrected $V_8$-cluster. The corrected number of interfaces of $V_8$-cluster is 60. Each of the 4 small squares in the reference $V_7$-cluster, marked by thick upward diagonal lines, respectively is also the mosaic cell of the corresponding neighboring $V_7$-cluster. The central 5 cavity cells representing the interface tensions of 5 particles will be first relaxed so as to first 'jump out' of the projection plane after time $t_8$.

Note that there are 4 *inverted* thick-black arrows in the corrected $V_8$ loop-flow in Fig. 6. Thus, the number of concomitant excited particles with $V_0(a_0)$ in $V_8(a_0\uparrow b_0)$ should be corrected as 141 because the 4 cells (4 mosaic cells of $V_8$), marked by the color of blue green, belong to that of the neighboring $V_7$-clusters.

In addition, the central 5 empty cells in $V_8(a_0\uparrow b_0)$ representing the harmonic interface tensions of 5 particles in $V_1(a_0\uparrow b_0)$ will be first fully relaxed (i.e. the 5 particles will be in non-interacting with other 136 in $a_0$ local coordinate system) when percolation appears.  The 5 particles will first 'jump out' ($a_0$ local coordinate system invalidates) of the projection plane of $V_8(a_0)$ after $a_0$ local time $t_8$.  Therefore, the number of cooperative excited particles in 8th order of cluster should finally be corrected as 136.

$N_8$ (corrected by percolation) = 136.

The 8 colorized thick arrows, in Fig.6, denoted the 8 cyclic jumping-off points, which are also the 8 degenerate states for the first excitation interface $j(a_0\uparrow b_0)$, the thick arrow in Fig.3 (a), in $a_0$ local field, and the 8th order of cyclic jumping-off interface will in fact evolve into a new first excitation interface in a new local field (a new local coordinate system) after fast vibrating cascades from $V_8$ to $V_0$ (see Fig.9).

### 2.18. Average energy of cooperative migration, $E_{mig}$

Each of the 136 excited particles also connects with a covalent bond z-component in *parallel transport state* encircling excited center $a_0$ with resonance transfer energy $\Delta\varepsilon$ between covalent bond z-component and 4 Van der Waals excitation interfaces. The mechanism of resonance energy transfer also attributes to that each one of the 4 Van der Waals interfaces of a chain-particle is exposed the compacting force of two cross-coupled electrons meeting on a dot; and the energy, $\Delta\varepsilon$, of walk-one-step of the 4-interface loop-flow transfers to the covalent bond. That means the resonance energy transfer only occurs in the minimum energy loop-flow surrounded by 4 excitation interfaces, and the resonance transfer energy $\Delta\varepsilon(\tau_i)$ of covalent bond z-component belongs to the induced potential of additional non-integrable phase.

As all energy to excite glass transition comes from the additional energy, the energy of cooperative migration along one direction *in a domain* is $136\Delta\varepsilon$, or $17\varepsilon_0$, which comes from the contribution of the specially selected all excitation interfaces at the instant time $t_8$ on the graph of Brownian motion, as shown in Fig.6, in Flory ideal random distribution.  As the migrating direction in a domain can be statistically selected as x-, y- z-axial, and the appearing of all thawing domains is one by one, the *average energy of cooperative migration*, $E_{mig}$, along one direction in a percolation field (formed by 2-D loop-flows in different direction) can be denoted by random thermal motion energy of $kT_2$, and

$$E_{mig} = kT_2 = 17/3\varepsilon_0 \qquad (2)$$

The random motion energy of $kT_2$ is similar to the energy of Curie temperature in magnetism.  It will be proved (seeing 6.11) that $kT_2$ here is also the energy of a 'critical temperature' existing in the glass transition presumed by Gibbs based on thermodynamics years ago [47]. The denotation of $kT_2$ is the same as that of Gibbs.

The average energy of cooperative migration along one direction, e.g., along the direction of external stress in the glass transition, is an *intrinsic attractive potential energy*, $E_{mig}$, that balances the random motion kinetic energy, $kT_2$ in 3-D space, for flexible polymer system, *independent of temperature and external stress and response time*. That is one of the key concepts to directly prove the WLF equation. It stands to reason that $E_{mig}$ is a transient attractive potential comes of the interface excitations in co-Brownian Motion existing in any random system.

### 2.19. Mosaic geometric structure and long-range correlation effect

The picture implies that there is a *long-range correlation effect* (with other interfaces in 4 neighboring 5-particle cooperative excited local fields) of a short-range interaction-interface through its 8 orders of relaxation times and mosaic geometric structure. Especially, the 4 inverted interfaces of the modified 8th loop-flow in Fig.6 denotes that each one of the 4 *inverted slow action torques* still interacts with, e.g., the 1-2 interface $j(a_0 \uparrow b_0)$ in Fig.1(d), in local 2-D projection plane. Thus, the reference particle $V_0(a_0)$ has not yet *fully* re-coupled with other clusters at local time $t_8$. *i*-th order of reference cluster is always covered with reversed (*i*+1)-th order of reference clusters. The 2-D picture is similar to that of the famous critical phase transition based on Kadanoff [43] study on isotropic ferromagnetism.

One important conclusion is that the generating of a $V_8$-cluster always comes from the contributions of its 4 neighboring $V_7$-clusters. That also implies that all neighboring 5-particle cooperative excited local fields in system are one after another excited with phase difference of $\pi$ *with* each other, which perhaps is from the period character of ripplons [30].

**2.20. Fractal dimension of excitation interfaces, 3/2**

In the discussion above, the 8 orders of closed loop-flows in Fig.6 are the *track records* of selected excitation interfaces, from many coupled repulsion electron pairs in fluctuation in 3-D local space, respectively at the instant of $t_1, t_2…t_8$ discrete time on a 2-D projection plane. In other words, in 3-D local space, the cyclic direction, selected from many random oriented excitation interfaces, returns to the direction of ± z-axial at the instant of $t_i$ (*i* = 1, 2, …8) time, which indicates that the distribution of coupled repulsion electron pairs in 3-D local space obeys the distribution of the Graph of Brownian Motion. Therefore, it can be guessed that the Hausdorff fractal dimension $d_h$, or the Box fractal dimension $d_c$ of the cross-coupled electron pairs in the glass transition is the fractal dimension of Graph of Brownian Motion [48], i.e.

$$d_h = d_c = 3/2 \tag{3}$$

**2.21. Symmetry braking, elementary excitation and order parameter**

The symmetry breaking in solid-to-liquid transition is that the *position-symmetry* of Lindemann distance increment $d_L$ to two particle position centers is broken by one particle that firstly forms its closed loop. We called the 8 orders of transient 2-D interface excitation energy loop-flows, (ripplon) as the *elementary excitation* to replica symmetry breaking in solid-to-liquid transition.

Edwards [33] propounded that the *order parameter* $\psi$ is an 'initiator' (or 'terminator') for long-chain $N$; and thought both ends of long-chain equivalent [33]. The order parameter $\psi$ is similar to a quantum mechanical creation (or destruction) operator [33].

We will prove that the motion of a z-component long-chain $N_z$ is the z-space solitary wave (Section 7.4.); the determinate particle-energy of solitary wave is $n_z \varepsilon_0(\tau_i)$ and the number of step of its traveling wave is $N$, where $n_z \varepsilon_0(\tau_i)$ is a small energy of the conformational rearrangement for each chain-particle on a long-chain; $n_z$ is the number of the 8th order of 2-D loop-flow and $n_z \varepsilon_0(\tau_i)$ connects with the two ends of a chain-length $N_z$ in each one-step-walk of z-space solitary wave.

We can understand the $n_z$ 8th order of 2-D loop-flows are connected with the 'initiator' and 'terminator' of $\psi$, which corresponds to the z-space solitary wave moving *from one end to other end*. Here both ends of chain-long are distinguished. It is useful to think of $\psi$ as the 'magnetization' of a spin system with a number of spin components $n = 0$, proposed by de Gennes [33]. De Gennes has pointed out that $n = 0$ corresponds to the self-avoiding walk of long-chain

*N*, *N* and *ε* are conjugate variables; energy *ε* is here a small 'reduced temperature' [33]. De Gennes emphasized that the temperature *T* of the polymer system is not related to the reduced temperature *ε* of the magnetic system [33]. Thus a more concrete statement is the following: in solid-to-liquid transition, the conjugate variables are $n_z\varepsilon_0(\tau_i)$ and *N*, the determinate 'quantized' particle-energy of z-space solitary wave and the number of step of traveling wave; the order parameter *ψ* is an 'initiator-terminator' for a z-space solitary wave of z-component long-chain *N*. The *generalized rigidity* is the 8th order of transient 2-*D* interface excitation loop-flow, and the *defect* is the extra vacancy volume in cross-coupled electron pair.

## 3. Eight orders of hard-spheres

### 3.1. Jamming behaviors and tube model

It is emphasized that the picture of the concomitant 8 orders of transient 2-*D* cluster with 4 interface relaxations of a reference particle only respectively appears at the discrete time $t_i$ in reference local field, the *i*-th order of cluster, being of 5 (*i*-1)-th order of, inner degrees of freedom and one *i*-th order of external degree of freedom, excites cluster possible tiny displacement along ±z-axial during the time of $(t_i, t_{i+1})$. For flexible polymer system, it can be noted that if $V_0(a_0)$ with the relaxation time $\tau_0$ takes the z-axial positive direction, the concomitant first order cluster $V_1(a_0)$ with $(\tau \leq \tau_1)$ will take the z-axial negative direction, and the concomitant cluster $V_2(a_0)$ with $(\tau \leq \tau_2)$ positive, and $V_3(a_0)$ with $(\tau \leq \tau_3)$ negative and so on. Therefore, the energy to excite reference particle $V_0(a_0)$ migration along z-axial repeatedly takes $\varepsilon_0(\tau_0)$, -$\varepsilon_0(\tau \leq \tau_1)$, $\varepsilon_0(\tau \leq \tau_2)$, -$\varepsilon_0(\tau \leq \tau_3)$… until $\varepsilon_0(\tau \leq \tau_8)$, (as the picture of ripplon modes, and according with the multilevel behavior of glassy mosaic domains pictured by Wolynes [9]). During (0, $t_8$), through the excitation of increasing concomitant orders of $V_0(a_0)$ with flipping signs of direction, the reference particle $V_1(a_0)$ migrates laggardly along the z-axial flipping the sign of direction. Since the z-axial excited energy $\varepsilon_0(\tau) = 8 \Delta\varepsilon$ ($\tau \leq \tau_i$) has *i* orders of relaxation times and the interface excitation energy quantized, all the migrations of $V_0(a_0)$ along z-axial in fact *cancel out* until the appearance of the largest $V_8(a_0)$, when the 4 interface tensions of $V_0(a_0)$ may be suddenly fully relaxed (see following Section) and $V_0(a_0)$ will jump out the bondages (in z-axial or on x-y projection plane) of its 4 neighboring particle fields.

This migrating manner of particles and clusters is similar to that of z-axial walking one step in 'tube model' in polymer physics. It can be seen that during (0, $t_8$), all particles in local field are in different degrees of relaxation state or jamming state, *none of particles is in fully relaxation state* and solely obtains the evolution energy $\varepsilon_0$ to migrate along z-axial. That further describes the details of jamming behavior in the glass transition and the details of least 'tube' within local filed in the multi-chain macromolecules motion. The fully depiction of jamming behavior is exactly the solitary wave, see Section 7.4.

The moving energy of a reference chain-particle in tube model here is affirmed to turn out to be the induced potential of non-integrable geometric phase, which numerical value equals to the potential well energy $\varepsilon_0$ and comes from the contribution of the 4-neighboring particle fields for the reference chain-particle field.

### 3.2. First order of hard-sphere

The conditions of particle $V_0(a_0)$ migrating along the z-axial is that $V_0(a_0)$ should be of fully relaxation interface tensions and 5 degrees of freedom. However, at the instant time $t_8$, the interface tensions of particle $V_0(a_0)$ has not yet been fully relaxed because of the percolation influence: 4 interface tensions with reverse direction, marked by 4 thick arrows in Fig. 6, inlaid on the $V_8$-loop-flow, thus the *reverse torques* with $\tau_7$ react on the interfaces of $V_0(a_0)$ and the particle $V_0(a_0)$ is still to a certain extent astricted from its 4 long-distance *neighboring excited fields* in Fig.6.

These anharmonic interface reverse torques are in nature the *long range attractive interactions* between reference $a_0$ 5-particle cooperative excited field and 4 neighboring 5-particle cooperative excited local fields in time and space, which result from the anharmonic vibrations in ripplons. Nevertheless, in $a_0$ 5-particle local field, at a certain time of $t \gg t_8$, 4 neighboring first order excited clusters of $V_1(a_0)$, i.e. $V_1(b_0)$, $V_1(c_0)$, $V_1(d_0)$ and $V_1(e_0)$, each containing their 5 particles in respectively their 2-D local space, can *dynamically take any directions* in 3-D local space, relative to that of $V_1(a_0)$ through fluctuation and one by one *fully eliminate the first order of* anharmonic interface torques between each other, thus the first order of 'hard-sphere' with diameter $\sigma_1$ can be statistically constituted.

### 3.3. Compacting cluster and density fluctuation

The first order of hard-sphere $\sigma_1$ contains 5 (first order clusters of $V_1(a_0)$) + 12 (the number of interfaces of $V_1$, each interface corresponds to a particle, e.g. the interface of arrow 9 → 10 in Fig.3 (b) corresponds to the particle $c_1$ in Fig.3 (c)) = 17 chain-particles that are *compacted* to form a hard-sphere or a 3-D cluster which has an internal density larger than average, because the interface cross-coupled electron pairs with extra volumes on $V_0(a_0)$ inside $V_1(a_0)$ have been compacted and transferred to that on $V_1(a_0)$. The ability of compacting cluster comes from the energy of the two cross-coupled electrons meeting one dot on an interface.

It is noted that the first order of hard-sphere $\sigma_1$ is a vector. The direction of $\sigma_1$ is negative, as same as the direction of the first order clusters of $V_1(a_0)$, if the direction of $V_0(a_0)$ is positive.

Note that a concept has been introduced that the interface torque is eliminated accompanied with the extra volume vanished on the interface. This 3-D cluster is a dynamical hard-sphere surrounded by *finite* acting facets that occur *in different local space-time coordinate systems*. The complexity of dynamical hard-sphere here comes down to the acting facets in mosaic structures.

In the same way, the second order of hard-sphere $\sigma_2$ contains 13 (the number of particles in $V_2$) + 20 (the number of interfaces of $V_2$) = 33 chain-particles. The direction of $\sigma_2$ is positive.

Therefore, the number, $S_i$, of particles in $i$-th order of hard-sphere $\sigma_i$ can be obtained as

$S_i = -17, +33, -53, +77, -105, +137, -173, (+200,$ see Section 3.7) for $i = 1, 2 \ldots 8$.   (4)

In Eq (4), the sign denotes the moving direction along z-axial of $i$-th order hard-sphere.

### 3.4. Complex frequencies implicated effects

By $i$-th order of hard-sphere is meant that the interaction-interface energy of its two-body is quantized interface excitation energy or transferred energy, $\Delta\varepsilon(\tau_i)$, on 2-D local projection plane, independent of the distance of two excited cents in 3-D space.

First order of hard-sphere size $\sigma_1 = 17^{1/3} \approx 2.57$ (chain-particle unit), is as same as [9], denoting the free motion zone size for a chain-particle, with 5 zero-th order of inner degrees of freedom and one first order of external degree of freedom. However, specially note that the characteristic size of $\sigma_1$ appears after time $t$ ($t \gg t_8$) is not only correlated with the relaxation times $\tau_1$ and $\sigma_1$ size results from the cooperative contributions of *all relaxation times* in the local zones from $V_1$ to $V_8$. That refracts a complex *frequencies implicated effects* of the characteristic size of $\sigma_1$.

**3.5. Statistical lengths of 8 orders of chain-segments**

In Fig.6, each of 200 (see Section 3.7) acting particles connects with a covalent bond z-component; the *topologically additional resonance energy state* of a covalent bond z-component (Section 2.18) is thus represented by its 4 surround excitation interface states on x-y local projection plane. Fig.6 shows there are 200 *different* topologically additional resonance energy states of covalent bond z-components that also categorize as 8 orders.

Accordingly, in flexible macromolecular system, if the migrating direction of a chain-particle $a_0$ is in +z-axial, the lengths and the moving directions of the 8 orders of statistical chain-segments $l_i$ (if chain-long $N \geq 200$) are

$$l_i = -17, +33, -53, +77, -105, +137, -173, +200; \quad \text{for } i = 1, 2 \ldots 8. \qquad (5)$$

The plus sign in Eq (5) denotes a topologically twisting chain-segment surrounded z-axis of chain $+2\pi$ cycle and the minus sign denotes a topologically twisting chain-segment surrounded z-axis of chain $-2\pi$ cycle on a long-chain. As this $2\pi$ cycle movement can be also regarded as the parallel transport surrounded z-component particle $a_0$ for the chain-particles connected with interfaces of hard-sphere, the $\pm 2\pi$ cycle movement will generate a $\pm \pi$ non-integrable geometric phase and a topologically additional potential that induce the centric chain-particle $a_0$ delocalizing. It must be emphasized that the 8 orders of topologically $2\pi$ - twisting directions of chain-segment (on a 'free' long-chain in 3-D space) is only along z-axial, instead of the chain figure direction in 3-D space, in order to induce the reference chain-particle $a_0$ delocalizing along +z-space.

**3.6. 8 orders of self-similar hard-spheres and chain-segments**

The remnant anharmonic interface tensions on first order of hard-sphere can be further relaxed in a larger hard-sphere. Therefore, the statistic model of 8 orders of self-similar hard-spheres (the self-similar 3-D clusters in small molecular system), and 8 orders of self-similar segment sizes (in macromolecular system), named as $\sigma_i$ (or $l_i$), can be one after another constituted; and each characteristic size of $\sigma_i$ (or $l_i$) has the frequencies implicated effects that give rise to the complicacy of the glass transition (or to the reptation of macromolecules, see Section 7).

**3.7. Number of structure rearrangements**

The number of particles in the 8th order of hard-sphere, $N(\sigma_8)$, (or the 8th order of chain-segment sizes, $l_8$, in macromolecular system), also is the number of particles $N_c$ in encompassing rearrangements and eliminating interface torques of a reference domain in 3-D space, can be easily found out from Fig.6: the number of cooperative excited particles *along one direction* in corrected $V_8$ is 136, the number of interfaces of $V_8$ is 60 before percolation is 68, from which subtracts 4 mutual interfaces (that belong to 4 neighboring local fields) of $V_8$ and its 4 neighboring $V_7$-clusters when percolation appears, and each of the 64 (= 68 − 4) interfaces *respectively* relates to an external excited particle in 3-D local space-time coordinate systems, which can eliminate the corresponding interface torque on $V_8(a_0)$ *along z-axial*. Thus, the certain number of acting particles in the 8th order of hard-sphere $\sigma_8$ is 136 + 64 = 200, a constant for flexible system, i.e.

$$N(\sigma_8) = l_8 = N_c = 200. \qquad (6)$$

The numerical value is consistent with the conjectural results of encompassing rearrangements in [40].

### 3.8. Critical chain length

A foremost result for polymer physics is that the mode of multi-macromolecular motion is that of the slow inverse cascade orientation and the fast cascade re-orientation (relaxation). The critical chain length that is able to precisely contain whole topologically *additional resonance energy states in z-component space*, instead of 3-*D* space, for flexible system is 200 (chain-particle units), independent of temperature, which is an inherent character reflecting the entangled motion (Section 7.4) of macromolecules in the range of the temperature from $T_g$ to $T_m$. The numerical value is in accordance with the experimentally determined critical entanglement chain length of ~ 200 [41].

### 3.9. Evolution direction of local cluster growth phase transition

From $N_c = 200$, $\sigma_8 \approx 5.8$ (chain-particle units), which is the size of 'cage' in the glass transition. According to the definition of 2-*D* clusters and 3-*D* clusters (hard-sphere) in this paper, an interaction on x-y projection plane of two *i*-th order of 3-*D* clusters is always equal to the surface exchanging interaction with relaxation time $\tau_i$, i.e., the quantized *exchange energy* $\Delta\varepsilon(\tau_i)$, *independent of the distance of two clusters in 3-D space*. That means that the cluster in [17] turns out to be *i*-th order of hard-sphere. During the time of ($t_i$, $t_{i+1}$), *i*-th order of 3-*D* clusters cannot be welded together, as same as [17], however, at the instant time $t_{i+1}$, all *i*-th order of 3-*D* clusters in local field will be compacted together to form a compacted (*i*+1)-th order of 3-*D* cluster with the evolution direction of the first order of 3-*D* cluster, corresponding to (*i*+1)-th order of local cluster growth phase transition in the glass transition.

### 3.10. Localization energy, mobility edge and critical flow-percolation energy

The steady excited energy is exactly the critical flow-percolation energy in percolation on a continuum, or say, the mobility edge of classical model in condensed matter physics [45]. The energy of steadily 'excited state energy flow' in the process of $V_8$ vanishing and reoccurring is defined as the *localization energy*, named as $E_c$ (the same denotation of $E_c$ as Zallen did [45]) in the glass transition. The 8 orders of mosaic geometric structures directly manifest that *the external degree of freedom of an* (inverse cascade) *excited state energy flow is 1 and the inner degree of freedom of a* (cascade) *i-th order cluster (i < 8) in free state is 5* in the solid-to-liquid glass transition.

### 3.11. Macroscopic melting state and renewed energy of microscopic cluster

It is a very important theoretic consequence and also foreshows that the external degrees of freedom of an excited state energy flow is 5 in the melting state phase transition and that for flexible polymer system, the completely renewed energy of a microscopic cluster, i.e. the energy for an *i*-th order cluster ($i < 8$) is in free motion within (*i*+1)-th order cluster zone, is *numerically* equal to the energy of the macroscopic melting state of $kT_m$, namely, $kT_m(\tau)$, ($\tau \leq \tau_i$, $i < 8$), $kT_m(\tau_i) = E_c(\tau_i) + 4\varepsilon_0(\tau_i)$ (corresponding to 5 external degrees of freedom), in which $E_c(\tau_i)$ is *i*-th order localization energy, see Eq. (9). And the energy of the macroscopic melting state is denoted as $kT_m(\tau_8)$.

Note that the 5 external degrees of freedom of an excited state energy-flow in the melting state phase transition are exactly evolved to the 5 independent variables in hydrodynamics, in which the *first* energy flow-percolation generated in glass transition corresponding to the entropy wave (the thermal wave) in hydrodynamics [63].

**3.12. Thermo-disorder-induced interface electron-electron pair localization**

Disorder-induced localization is the essential concept of Anderson disorder theorem in condensed matter physics [45]. It can be seen from Fig.6 that the localization object induced by thermo-disorder in glass transition and in macromolecular motion is namely the *maximal loop-scale* in the 8 orders of 2-*D* symmetric interface excitation energy loop-flows existing in a few thawed domains, in order to induce particle-clusters migrating along ± z axial.

**3.13. Localized energy $E_c$**

Now the geometric method is used to derive the localized energy.

The step (interface) number of the 8th order 2-*D* loop-flow is $L_8 = 60$. $E_c$, is less than $60\Delta\varepsilon(\tau_8)$ because of the quantized energy effect of excitation interface and the dynamical mosaic structure of excitation interface energy flow. The interface excitation energies on $V_8$-loop-flow in Fig.6 are *shared by* $V_0(\tau_0)$-$V_8(\tau_8)$ interfaces and $V_7(\tau_7)$-$V_8(\tau_8)$ interfaces.

The key concepts are that a few of interfaces (named as $L_{inverse}$) on the $V_8$-loop-flow will be, in manner of slow inverse cascade, excited by the interfaces with relaxation time of $\tau_i < \tau_8$ in the new local fields after $a_0$ (the effect of mosaic structure, or say, $L_{inverse}$ is the step number of mosaic), and the others (named as $L_{cascade}$) on the $V_8$-loop-flow will one by one vanish and their excited energy will rebuild many new $\tau_0$-interfaces, in a manner of *fast cascade vibration*, in the new local fields.

Thus, in the flow-percolation on a continuum, the 60 interfaces of a reference $V_8$-loop-flow in Fig.6 are *dynamically* divided into two parts: the $L_{cascade}$ interfaces that occur at the local time of $t_8$ in $a_0$ field and the $L_{inverse}$ interfaces that are the mosaic structure of energy flow and occur at a time after $t_8$. Formula (7) is obtained

$$L_{cascade} = L_8 - L_{inverse} \tag{7}$$

The energy of $L_{cascade}$ is also the fast-process cascade vibrant energy of rebuilding new $V_0$ loop-flows when the $L_{cascade}$ interfaces vanish. The energy of $L_{inverse}$ is the slow-process inverse cascade energy of all $V_i$ loop-flows from $V_0$ to $V_7$ in new local fields.

The balance excited energy of inverse cascade and cascade in flow-percolation (contained many local fields) is exactly the localize energy $E_c$. Therefore, in the fast-process cascade vibrant, $E_c/\Delta\varepsilon(\tau_8) = L_{cascade}$, in which $\tau_8$ is used as timescale of $V_8$ vanishing, $E_c$ here is the cascade energy in flow-percolation.

Since the excited energy in inverse cascade is not dissipated, the *evolution energy* of each order closed cycle is $8\Delta\varepsilon(\tau_i) = \varepsilon_0(\tau_i)$, namely, the 'singular point energy' of *any i*-th order (energy flow) closed cycle (Gauss theorem).

Each mosaic step (either the inverted arrow, or the shared interface by $V_7(\tau_7)$-$V_8(\tau_8)$) in Fig.6 connects a closed cycle $V_7(\tau_7)$ that does not belong to the reference $a_0$ local field. Thus, in the slow-process inverse cascade from $V_0$ to $V_7$ in flow-percolation, the number of the inverse cascade energy $E_c$ forming mosaic step is $E_c/\varepsilon_0(\tau_7) = L_{inverse}$. For flexible polymer, $\varepsilon_0(\tau_i) = \varepsilon_0(\tau_8) = \varepsilon_0$. Therefore, formula (8) is obtained

$$E_c = 60\Delta\varepsilon(\tau_8) - E_c/8 \tag{8}$$

Or $\quad E_c(\tau_i) = 20/3\,\varepsilon_0(\tau_i)$ (9)

Eq. (8) is a representative mean field formula. It can be seen that the physics meaning of the right term containing left term on equation is the contribution of mosaic structure.

Eq. (9) denotes that the localized energy $E_c$ (or say, the energy of mobility edge, the critical flow-percolation energy, the transfer energy from inverse cascade to cascade) in the glass transition has 8 components, $E_c(\tau_i)$, $i = 1, 2\ldots 8$ and the numerical value of each component energy is the same, namely, $20/3\,\varepsilon_0$. This is also one of the singularities in the glass transition.

$E_c(\tau_i)$, is a characteristic invariable with 8 order of relaxation time spectrum in the glass transition, independent of glass transition temperature $T_g$. The localized energy $E_c(\tau_i)$ is very important to interpret the macroscopic glass transition.

### 3.14. Intrinsic thermo random motion energy independent of glass transition temperature

If a *thermo random motion energy*, $kT_g°(\tau_i)$, of $i$-th order clusters with relaxation time $\tau_i$ is used to denote the energy of $E_c(\tau_i)$, that is

$$kT_g°(\tau_i) = E_c(\tau_i) = 20/3\,\varepsilon_0(\tau_i) \quad (10)$$

The numerical value of $T_g°(\tau_i) = T_g°$ that can be called the *fixed point* in renormalization of clusters from small to large, in the critical local cluster grow phase transition in glass transition, it is independent of glass transition temperature $T_g$. Furthermore, in following Section 5.7, 5.8, the statistical physics will be used to prove $T_g°(\tau_8) = T_g$, $T_g$ is traditionally accepted as glass transition temperature, which is obtained by slow heating rate. Therefore, $E_c$ is a measurable magnitude by experiments. From Eq. (2), numerical relationship is:

$$kT_g° = kT_2 + \varepsilon_0 \quad (11)$$

### 3.15. Geometric frustration effect

It can be seen that the Eq. (9) is based on the result of corrected $L_8$ (the number of interfaces) value and this correction from $L_7$ to $L_8$ in Fig.6 also can be regard as the geometric frustration effect, which appearance corresponds to the glass transition [14]. This flow-percolation also indicates that the percolation in glass transition belongs to the *high-density percolation* [42], see following (3.20) free volume.

### 3.16. Total interface excitation states

Fig.6 has given out the number of all excited states in the ideal glass transition: the 320 different interface tension relaxation states. The numerical value is from

$$\sum_{i=1}^{8} L_i + 8 = 320. \quad (12)$$

$\quad\quad\quad\quad = 8$ (spatial evolution) $\times$ 8 (temporal evolution) $\times$ 5 (5-particle cooperative excited)

The numerical value 8 in Eq. (12) is the 8 evolution states. The 8 evolution states together with the 4 fully relaxed interface states on a reference particle $a_0$ will evolve a new first order symmetry loop-flow being of 12 interfaces when 8th cluster appears.

It can be seen that a mode of the glass transition has been proposed, which is also the mode of breaking solid lattice and the mode of multi-macromolecular motion in the range of $T_g$ to $T_m$. For flexible polymer system, in brief, the mode to break a 3-D solid domain is the interface-torque progressive relaxation of the 8 orders of 2-D clusters in local z-space, which relates to the 320 different interface tension relaxation states.

### 3.17. Activation energy to break solid lattice

The energy summation of the 320 different interface tension relaxation states is named the *cooperative* (*orientation*) *activation energy* to break solid lattice, denoted as $\Delta E_{co}$. The reason to call it activation energy is that although in macroscopical, the energy to break solid lattice is seemly only $kT_g$, $< \Delta E_{co}$, in microcosmic, 320 interface energy (320 interface space-time states) of every solid domain is needed. Thus,

$$\Delta E_{co} = 320\Delta\varepsilon = 40\varepsilon_0 \qquad (13)$$

And

$$E_c = kT_g^\circ = 1/6\Delta E_{co} \qquad (14)$$

Using $T_g^\circ(\tau_8) = T_g$, thus,

$$kT_g = 1/6\Delta E_{co} \qquad (15)$$

(By the way, the dispute concerning the concept of activation energy has been in existence all the time in physical chemistry. An exact definition, the energy of all interface space-time states, for activation energy is also proposed in this paper, the correctness of the concept of activation energy in Eq. (13) will be validated in theoretical proof of the standard WLF equation in the glass transition in Section 6: comparing with the conventionally theoretical proof of the free volume, where the activation energy is used to WLF equation.)

### 3.18. Percolation threshold in the glass transition

The ratio of $E_c/\Delta E_{co} \equiv \phi_c(E_c) \equiv 1/6$, its physical meaning is that $\phi_c(E_c)$ specifies the occupied fraction of 320 interface-excited states that allow flow of energy $E_c$ to occupy. The ratio is consistent with the result of Zallen [45], who suggests $\phi_c(E_c) \approx 0.16$ for the percolation on a continuum in 3-*D* space. Here, the occupied fraction of all interface-excited states allowed to flow of energy $E_c$, takes the place of the occupied fraction of space allowed to particles of energy $E_c$ proposed by Zallen.

The invariant ratio of 1/6 is exactly the percolation threshold $p_c$ in the glass transition. The singularity of percolation in the glass transition is that the *percolation threshold is an invariable value*, independent of the glass transition temperature, and equals to the occupied fraction $\phi_c(E_c)$, i.e. $p_c \equiv \phi_c(E_c) \equiv 1/6$, because the filling factor [45] for the lattice in Fig.6 is right 1 (high-density percolation, the filling factor in Fig.1 (c) is also 1).

If the average thermo random energy $kT = kT_g^\circ(\tau_8) = kT_g$, the glass transition would occur after time $t > \tau_8$, while if $kT < kT_g^\circ(\tau_8)$, the lower temperature glass transition also can occur, takes long time to obtain the occupied fraction $\phi_c(E_c)$ in a domain.

The occupied fraction is also an invariant, independent of the glass transition temperature, which is also one of the singularities in the glass transition: *the increase of transition temperature only increases the number of excitation interface energy loop-flow, or the number of thawing domains to expedite glass transition.*

### 3.19. Free volume, free volume fraction, non-ergodic and ergodic

The classical free volume ideas [41] have been questioned because of the misfits with pressure effects, but here the pressure effects should be indeed minor [17]. It is interesting in our model that there is no so-called classical free volume with an atom migrating in system unless percolation (subsystem) appears. Except the 8th order of 2-*D* clusters (hard-spheres), in the *i*-th order of local phase transition, all (*i*-1)-th order of 2-*D* clusters *are compacted* to form *i*-th order of 2-*D* clusters, and the extra volumes of the surfaces of all (*i*-1)-th order of 2-*D* clusters vanish and reappear on the surfaces of *i*-th order of clusters. So it is proposed that when the 8th order of local cluster growth phase transition (i.e., percolation) appears, the extra volumes in the surfaces of the 5 7-th order of 2-*D* clusters suddenly form vacancy volumes of 5 cavity sites in Fig.6. Here the definition of free volume is modified by the cooperatively appearing 5 cavities, i.e., using clusters rather than atoms, as same as that proposed by de Gennes [17]. The modified free volume ideas is still true, because the action of *pressure should be only through the same percolation field generating cavity volume* and the cavity volume imparted by extra volumes in ripplons (here, the 8 orders of interface excitation loop-flows in Fig.6 has been imagined as a thermo-excited ripplon of acting particle $a_0$) is of interface tension energy, which should be balanced with the external pressure or tension (see Section 6). Each cavity is an orientation vector (Section 2.16.). Since the reference 5 cavities maintaining their own *orientations* need for long latency time (see Section 7.2 (x), non-ergodic in the long latency time) in the 5 orientation 2-*D* excited fields on a percolation field (a reference subsystem) connected by $N$ domains ($N \geq N_c$, or $N \to \infty$) in system. However, taking the space-time scale of a reference local excited field as statistical unit and calculating the average value of the 5 cavities for all innumerability local excited fields in all subsystems (ergodic!) that appear at different time in system, 5 statistically isotropic 3-*D* cavities (so-called classical free volume) thus are obtained per 200 statistically isotropic particles. In other words, as the average occupying cavity volume per particle (involved its all 200 acting particle states in cooperative rearrangement in Fig.6) is as the free volume fraction, and the 5 statistically cavity volumes only occur after many cycles in structure rearrangements, each cycle also need 200 statistically cooperative rearrangement particles in local 3-*D* space, thus the free volume fraction: 5/200 = 0.025 can be directly and explicitly obtained, which is in accordance with the experimental results for flexible polymer (so-called the free volume theory in the glass transition).

### 3.20. Experimental value of interface excitation energy, $\Delta\varepsilon$

As the mode of breaking solid lattice is also that of the cooperative migration to excite particle-clusters along one direction, it is presupposed that $\Delta E_{co}$ the activation energy to break solid lattice is also the activation energy of *cooperative orientation* of macromolecules in elongational flow. Here, the orientation activation energy in polymer physics is assured to come of all the interface excitation states along one direction in system. As the method of selected 320 interface states along one direction in ideal random system obeys the track records of the graph of Brownian motion, the relationship between stretch-orientation viscosity and $\Delta E_{co}$ obeys

$$\eta_{co} \sim \exp(\Delta E_{co}/T)$$

Therefore, $\Delta E_{co}$, $\Delta\varepsilon$ and $\varepsilon_0$ can also be measured by the stretch-orientation experiment with melt high-speed spinning along one direction. It should be noted there that the relation between the spinning parameters and the development of structure and mechanical properties of the as-spun fibers formed at high speed of winding has been investigated rather thoroughly [50]. There is a stretch-orientation zone *along one direction* on melt high-speed spinning-line. The structure of the as-spun yarn, formed at a relatively low speed (below 1000m/min), is quite unstable, and the glass transition of the yarn has not been completed. Whereas, when the work of the stress on spinning line reaches the cooperative orientation activation energy, $\Delta E_{co}$, the structure of the yarn formed at a super high-speed spinning (up to 5000m/min) is stable and quickly full orientation, so-called FOY (Full Orientation Yarn) in current polyester fibre industry. The glass transition of the yarn at spinning of 5000m/min, called as stress-induced glass transition, has been completed on spinning-line. The rate of change of the stress-induced glass transition is $10^5$ times of the general. Therefore, the orientation activation energy can be measured on melt high-speed spinning-line. In fact, the author [50] had obtained the experimental data of orientation activation energy $\Delta E_{co}$ for polyethylene terephthalate (PET), by using the on-line measuring on the stretch orientation zone during melt high-speed spinning at 2200–4200M/min. The experimental result shows $kT_g$ (for PET) $\approx 1/6\Delta E_{co}$ (for PET), and $\Delta E_{co}$ (for PET) = $2035k$. Thus, for PET, the added interface excitation energy $\Delta\varepsilon$ of Van der Waals coupled repulsion electron pairs

$$\Delta\varepsilon = 6.4\ k\ (= 5.5\times 10^{-4}\ \text{eV}) \quad (16)$$

And $\quad \varepsilon_0 = 8\Delta\varepsilon = 51k$

Comparing the energy $5.5\times 10^{-4}$ eV of $\Delta\varepsilon$ with the energy $\sim 10^{-2}$ eV of phonon, we see the energy of interface excitation is too small to be directly observed. However, in the theory of the solid-to-liquid transition the transient interface excitation 2-D loop-flows induce particle-cluster delocalizing is considered as of primary importance.

The experimental result of Eq. (16) will be further proved by the WLF experimental law, seeing Eq. (83).

### 3.21. Universal picture of cooperative migration of particle-clusters

A universal picture of cooperative migration of particle-clusters has been proposed in the entire temperature range from $T_g$ to $T_m$. The picture can be simply described as that: in z-component space of a reference particle $a_0$ local filed, the 320 interface excitations (on the x-y projection plane) appearing one by one, on the 8 orders of 2-D mosaic lattices formed by 200 z-component particles, step by step and from fast to slow induce 136 particles to move back and forth along local ±z-axial. Once 320-th interface excitation appears, the 8th orders of 2-D interface excitation energy loop-flow cascades and compacts 200 particles, and the +z-axial $a_0$ particle firstly delocalizes and moves '*one step*' (the step-scale is in fact very small because the appearing probability of the 8th orders loop-flow is very small, see Section 7.4) away from, one by one, its 4 neighboring -z-axial particles and gives rise to 5 z-axial particle-cavities. However, this migration of the $a_0$ particle will still be entangled by the 200 particles in z-space because they need share the localized energy $E_c$. After that, the 199 particles one after another in their own z-component local fields, the same process will occur. Only after a long time and repetitious rearrangements of 200 particles has the reference $a_0$ particle full re-coupled with its primal 4 neighboring particles.

Two intrinsic attractive potential energies independent of temperature: the average energy of cooperative migration along one direction $E_{mig} = 17/3\varepsilon_0(\tau_i)$ and the localized energy $E_c(\tau_i) = 20/3\varepsilon_0(\tau_i)$ respectively balance two intrinsic random motion energies, the Gibbs' critical random energy $kT_2$ existing in the glass transition and $kT_g^\circ(\tau_i) = 20/3\varepsilon_0(\tau_i)$ with $\tau \leq \tau_i$, $\varepsilon_0(\tau_i)$ is here the potential well energy of *i*-th order cluster with relaxation time $\tau_i$. For flexible polymer, $kT_g^\circ(\tau_8) =$

$kT_g$.

In flexible polymer, all 320 different interface excitation states in turn occur on 8 orders of 2-*D* mosaic lattices and all the excited states possess the same numerical value of quantized interface excitation energy $\Delta\varepsilon \approx 6.4\ k \approx 5.5\times 10^{-4}$ eV, but have different interacting times, relaxation times and different time-space phases in loop-flows. Furthermore, the dynamical rule of these excited states should, from fast to slow and small to large, form 8 orders of transient $2\pi$ interface excitation loop-flows in inverse energy cascade, with the even number cycles along one direction and the odd number cycles, the opposite direction.

## 4. Fixed point for self-similar Lennard-Jones potentials in the glass transition

### 4.1. Essentiality of the theoretical proof for 8 orders of cluster growth phase transitions

The central idea of the intrinsic 8 orders of instant 2-*D* mosaic geometric structures is mainly based on the existence of the 8 orders of domain wall vibration frequencies, proposed by Wolynes and co-workers [9]. In this section, the strictly independent testification for fixed point of self-similar Lennard-Jones (L-J) potentials will directly educe that there are only 8 orders of self-similar hard-spheres in the glass transition and explain the origin of quantized interface excitation energy.

The existence of fixed point of L-J potentials will be a very important theoretical result to out and away simplify discussions on pairwise interactions, especially in computer simulations.

L-J potentials are widely used to describe atomic, molecular, cluster, even nano-particle [51] interactions and phase transitions. However, the fundamental theory to unify the mechanisms for the *truncated* L-J potential in simulation [52], the presence of long-ranged interparticle potentials [53], the density or cluster size fluctuation stability in mode-coupling theory [54] and the limited perturbation resulting narrow colloid size distribution [55] are still not simply and fully set up.

The relationship between glass transition and general displacive phase transitions in condensed matter physics has remained obscure.

In addition, one of the most prominent open questions in glass transition concerns the so-called tunneling in two-level [15], which will be unambiguously explained by using the theory of fixed point of L-J potentials in this section.

The other problem is to understand the origin of Boson peak [17] in glass transition, which will be also touched upon in this section.

### 4.2. Self-similar L-J potential curves

The classical L-J potential that accords with the scaling theory in critical phase transition is used to discuss the glass transition and the macromolecular motion

In Fig.7, due to the left-right asymmetry of potential curve (17) with regard to the balance position $q_{1,0}$, two clusters with $\sigma_1$ have unequal amplitudes on the two sides in *q*-axial (i.e. in z-axial in Fig.6). The two *q* values on the two sides of *U* are respectively denoted as $q_{1,L}$ and $q_{1,R}$.

In solid-to-liquid transition, the fractal vibration appears as is shown in Fig.8, in which the two $\sigma_1$ clusters (black sphere) are apart $q_{1,R}$ and in attractive state. On the other hand, these two clusters also respectively belong to two larger $\sigma_2$ clusters at the instant time of the 2nd order of local cluster growth phase transitions. As long as these two $\sigma_2$ clusters are also apart $q_{1,R}$ but on the right side of the 2nd order L-J potential in Fig.7, i.e. $q_{1,R} = q_{2,L}$, these two $\sigma_1$ in $\sigma_2$ clusters are also in repulsive state. The potential that the attraction and repulsion happens to be in equilibrium is denoted as $U_c$.

$$U_{1,2}(\sigma,q) = q^x f(\sigma/q) = 4\varepsilon_0[(\frac{\sigma}{q})^{12} - (\frac{\sigma}{q})^6] \tag{17}$$

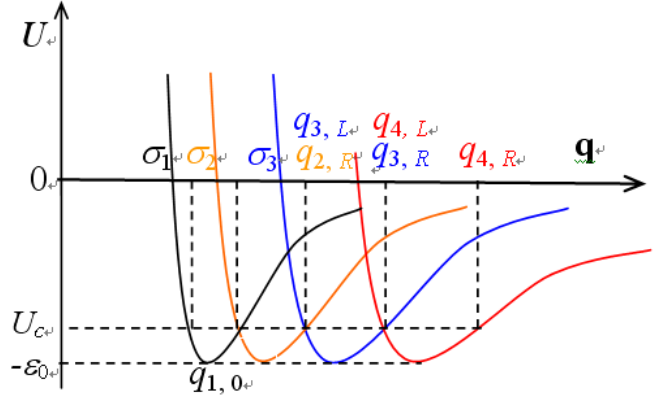

Fig.7. At the equilibrant point of attraction and repulsion, $U_c$, ($U_c \in (0, \varepsilon_0)$), there are no end of $U_c$ points and self-similar L-J potential curves satisfying $U_c = U(q_{i, R}) = U(q_{i+1, L})$. They are all unstable except for curves via the fixed point, $U_c^*$, of L-J potentials in glass transition.

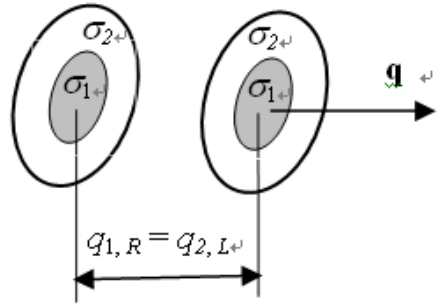

Fig.8. Self-similar clusters of two different sizes satisfy $U_c = U(q_{1, R}) = U(q_{2, L})$ along $q$-axis (i.e. z-axial in Fig.6), the direction of cluster growing. The two $\sigma_1$ in $\sigma_2$ clusters are in 'balance state' of attraction and repulsion during the 2nd order of cluster, when the quantized interface excitation energy transfer from inside $\sigma_2$ to the surface of $\sigma_2$.

### 4.3. Reduced geometric phase potential

By recursively applying this procedure, $M$ orders of clusters can be obtained from small to large in size along the q-axis seeing Fig.7, i.e. there are infinite self-similar L-J potential curves satisfying $U_c = U(q_{i, R}) = U(q_{i+1, L})$, where $U_c$ is in $(0, \varepsilon_0)$. These equilibrium points of attraction and repulsion at $q_{i R}$, the sharp-angled points in Fig.7, are unstable in general. The critical phase transition always occurs on an unstable critical point. Among all the possible values of $U_c$, there is, however, a point which is not only an unstable critical point in local cluster growth phase transitions but also a stable bifurcation point. We denote this point as $U_c^*$ which corresponds to the *intrinsic stability condition* in solid-to-liquid transition.

Let $q_{i,R}^*$ and $q_{i+1,L}^*$ satisfy: $U_c^* = U(q_{i,R}^*) = U(q_{i+1,L}^*)$.

Based on the scaling theory, the $(\sigma/q)$ in Eq. (17) has to be constant. Hereby a characteristic parameter is introduced

$$\chi = (\sigma/q)^6 \tag{18}$$

It will be interpreted later (see Section 4.5.) on that $\chi$ represents the induced potential formed by the non-integrable geometric phase factor between local (two-body interaction) and global ($2\pi$ cycle interaction of reference cluster surrounded by neighboring clusters), and is called *reduced geometric-phase-induced potential* in this paper, which is independent of temperature.
Equation (17) is rewritten as the non-dimensional form

$$U_{1,2}(\chi)/\varepsilon_0 = f(\chi) = -4\chi(1-\chi) \tag{19}$$

As $q$ increases, the tow-body interaction potential *does not necessarily increase* in glass transition, see Fig.7. Due to the left-right asymmetry of potential curve (17), $\chi$ in Eq. (19) must have two solutions, which correspond to two different constants in Eq. (18), denoted respectively as $\chi_{+1}$ and $\chi_{-1}$ (subscript represents slope; see the solutions of Eq. (25) and Fig.10).

**4.4. Recursion equation of potential fluctuation**

The density fluctuation stability in mode-coupling theory can be here regarded as the stability of compacting $i$-th order clusters with lower density and re-building $(i+1)$-th order with higher density in the 8 order of local cluster growth phase transitions. At the unstable critical point of $U_c = U(q_{i,R}) = U(q_{i+1,L})$ in Fig.7, suppose there is a disturbance of the reduced geometric-phase-induced potential arising from the displacive fluctuation (distance-fluctuation) in compacting $i$-th order clusters denoted as $\Delta\chi_i$, there is a corresponding disturbance in re-building $(i+1)$-th order clusters, denoted as $\Delta\chi_{i+1}$, $\Delta\chi_{i+1}$ then vice versa adds to its interaction potential $f(\chi_i)$ to excite disturbed $i$-th order cluster as an additional excited fluctuation potential. The recursion equation is given by

$$f(\chi_i) + \Delta\chi_{i+1} = f(\chi_i + \Delta\chi_i), \quad i=1,2,\ldots \tag{20}$$

Eq. (20) comes of: $\Delta f(\chi_i) = (df(\chi_i)/d\chi_i) \cdot \Delta\chi_i$, or a series of self-similar equations:

$\Delta f(\chi_i)/\Delta\chi_i = df(\chi_i)/d\chi_i;\ \Delta f(\chi_{i+1})/\Delta\chi_{i+1} = df(\chi_{i+1})/d\chi_{i+1}$ and so on.

As these self-similar equations and the two functions of $f(\chi_i)$ and $\chi_i(\sigma_i/q_i)$ are all non-dimensional functions of unit as 1, the recursive rule $\Delta\chi_{i+1} = \Delta f(\chi_i)$ then is adopted in Eq. (20). Notice that the type of the recursion equation is in nature different from that in current literature [56]. The recursively defined variable is the potential fluctuation $\Delta\chi$ in reduced geometric-phase-induced potential$\chi$, whose recurrent motion of 'first reinforce – after restraint' let $\chi$ finally recursive from $\chi_{-1}$ to$\chi_{+1}$, see Fig.(10), corresponding the cluster evolving from small to large in the glass transition.

**4.5. Parallel transport in topological analysis**

Here the key idea is as that of the picture of additional replica symmetry breaking in glass transition described in the Section 2.8: on projection plane perpendicular to $q$-axial (this means the interface excitations on a $2\pi$ closed loop-flow belong to *parallel transport* surround $q$-axial in topology analysis [57]), only by fluctuations of excitation interfaces may a $2\pi$ interface excitation loop-flow of ($i$+1)-th order 2-D cluster ($\sigma_{i+1}$) appear at the instant time $t_{i+1}$ in reference $a_0$ local field to eliminate the additional position-asymmetry of any $i$-th order 2-D cluster ($\sigma_i$) during ($t_i$, $t_{i+1}$), however, because of the jamming effect (Section 3.1), this procedure is in fact retarded until the forming condition of 8th order loop-flow is satisfied, hereafter, the 8 orders of $2\pi$ closed loop-flows will one after another appear and let the mass centers of the 8 orders of 2-D clusters one by one return to the vibrant balance position of reference $a_0$ particle on projection plane.

Because of a $2\pi$ instant closed loop-flow on projection plane corresponding a singularity on $q$-axial, the singularity is exactly the non-integrable geometric phase factor [57] (being of geometric-phase-induced potential), here comes the restricted relationship of geometric phase factors between local and global (global here means $2\pi$ cyclic in parallel transport) in each order loop-flow, as also, the $i$-th order loop-flow and ($i$+1)-th order loop-flow, or in other words, the relationship between mean field $U(\sigma_i)$ and mean field $U(\sigma_{i+1})$ in Eq. (17), or say, *there is the non-integrable geometric-phase-induced potential between mean field $U(\sigma_i)$ coordinate system and mean field $U(\sigma_{i+1})$ coordinate system.*

**4.6. First fluctuation-reinforce and after fluctuation-restraint**

The recursion equation (20) is a mean field (renormalization) equation of mean fields in different size, which represents the restricted relationship. That is, Eq. (20) is a recursion equation containing reduced geometric-phase-induced potential *fluctuation stability* in local cluster growth phase transitions. The recursive procedure of Eq. (20) is that a disturbance $\Delta\chi$ on the bifurcation point, $\chi_{+1}$, via first fluctuation-reinforce and after fluctuation-restraint, finally corresponds to the recurrent point $\chi_{-1}$, and a disturbance $\Delta\chi$ on the bifurcation point $\chi_{-1}$ to the recurrent point $\chi_{+1}$, which will be explained in Fig.10.

Notice again that in random system, all transient 2-D local excited fields are in random orientations, thus, using statistical 3-D clusters instead of 2-D clusters in Eq. (17) does not influence our discussions, whereas, the evolution characteristics of hard-spheres with various sizes along a same direction and the non-integrable geometric-phase-induced potential of $2\pi$ closed loop-flow surrounding $q$-axial should be still reserved.

Using the linear approximation:

$$f(\chi_i + \Delta\chi_i) = f(\chi_i) + \frac{\partial f(\chi_i)}{\partial \chi_i} \Delta\chi_i \quad (21)$$

If there exists the fixed point of $\chi_i$, denoted as $\chi^*$, Eqs. (19), (20) and (21) may be rewritten as

$$\Delta\chi_{i+1} = \left.\frac{\partial f(\chi)}{\partial \chi}\right|_{\chi=\chi^*} \cdot \Delta\chi_i \quad (22)$$

$$f(\chi^*) = -4\chi^*(1-\chi^*) \quad (23)$$

Equation (23) is the fixed point equation of self-similar L-J potential functions. From (22):

$$\frac{\Delta\chi_{i+1}}{\Delta\chi_i} = \left.\frac{\partial f(\chi)}{\partial \chi}\right|_{\chi=\chi^*} \quad (24)$$

For stable fixed point, the absolute value of $\Delta\chi_{i+1}$ must be less than that of $\Delta\chi_i$. Thus, the stability condition of fixed point can be similarly derived as literature [56]

$$s \equiv \left|\frac{\Delta\chi_{i+1}}{\Delta\chi_i}\right| = \left|\frac{\partial f(\chi)}{\partial \chi}\right|_{\chi=\chi^*} \leq 1 \tag{25}$$

From (23) and (25), it can be identified that $3/8 \leq \chi \leq 5/8$. In local cluster growth phase transitions, all reduced geometric-phase-induced potential fluctuate between the two bifurcation points of $\chi$, respectively the minimum value 3/8, corresponding to compacting into $i$-th order clusters on sharp-angled points in Fig.7, and the maximum value 5/8, corresponding to forming ($i$+1)-th order clusters. According to the recursive rule, for the solutions of Eq. (25),

$$\chi_{+1} = 3/8 \quad \text{and} \quad \chi_{-1} = 5/8$$

(The solution for slope −1 in Eq. (25) recursive in 5/8, see Fig.10) are adopted.

Thus, there are only two types of stable reduced geometric-phase-induced potentials: $\chi_{+1}$ =3/8 stands for the faster reduced geometric-phase-induced potential, which is contributed by the attraction of the $i$-th order clusters in ($i$+1)-th order, and $\chi_{-1}$ = 5/8 stands for the slower reduced geometric-phase-induced potential, which is contributed by the repulsion of the ($i$+1)-th order clusters containing $i$-th order.

Explain: on the critical point $U_c$ crossed by $i$-th order L-J potential and ($i$+1)-th order in Fig.7, two attracting $i$-th order clusters in ($i$+1)-th order, lie on the right, $U(q_{i,R})$, of $i$-th order L-J potential, which fluctuation along positive slope at point $U(q_{i,R})$, so, the faster reduced geometric-phase-induced potential (first compacted $i$-th order cluster) as $\chi_{+1}$ in solutions of Eq. (25); and two repulsiving ($i$+1)-th order clusters lie on the left, $U(q_{i+1,L})$, of ($i$+1)-th order L-J potential, which fluctuation along negative slope at point $U(q_{i+1,L})$, thus, the slower reduced geometric-phase-induced potential as $\chi_{-1}$.

### 4.7. Fixed point of L-J potentials

By substituting $\chi_{+1}$ or $\chi_{-1}$ into (23), the sole fixed point of L-J potentials can be obtained

$$U_c^* = -15/16\varepsilon_0 \tag{26}$$

In addition, $\chi_{+1} + \chi_{-1} = 3/8 + 5/8 = 1$, which means that (because of $f(\chi_{+1} + \chi_{-1}) = 0$), on each sharp-angled point (the crossover point of $i$-th order L-J potential and ($i$+1)-th order L-J potential) in Fig.9, the sum of the two fast-slow geometric-phase-induced potentials exactly equals to the cluster evolution energy $\varepsilon_0$ of one external degree of freedom, which is consistent with the result of Section 2.14. By substituting $\chi_{+1}$, $\chi_{-1}$ into (19), it can be concluded that there are only two stable reduced geometric-phase-induced potentials during the local cluster growth phase transitions, though there can be $M$ solutions for degenerate state:

$$\chi_{+1} = (\sigma_i / q_{i,R})^6 \quad \text{and} \quad \chi_{-1} = (\sigma_{i+1} / q_{i+1,L})^6, \quad i = 1, 2 \ldots M \tag{27}$$

Satisfying $\chi_{+1} = 3/8$, $\chi_{-1} = 5/8$.

In solid-to-liquid transition, the fractal vibration appears as is shown in Fig.8, in which the two $\sigma_1$ clusters (black sphere) are apart $q_{1,R}$ and in attractive state.

On the other hand, these two clusters also respectively belong to two larger $\sigma_2$ clusters at the instant time of the 2nd order of local cluster growth phase transitions. As long as these two $\sigma_2$ clusters are also apart $q_{1,R}$ but on the right side of the 2nd order L-J potential in Fig.7, i.e. $q_{1,R} = q_{2,L}$, these two $\sigma_1$ in $\sigma_2$ clusters are also in repulsive state.

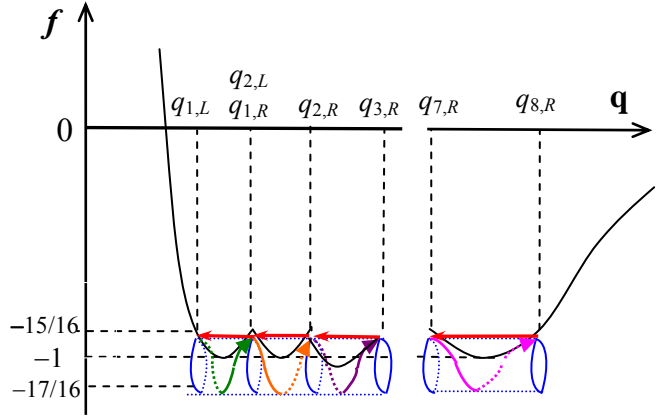

Fig.9. At the fixed point in glass transition, $U_c^*$, there are 8 sharp-angled points, $q_{i,R}$, forming the apart (re-coupling) paths of two $\sigma_1$ clusters along 8 orders of geodesic. $i$-th order geodesic is the shortest line of $2\pi$ cycle between $q_{i,R}$ and $q_{i+1,R}$ on $i$-th order of cylindric potential surface. 8 orders of cylindric potential surfaces are formed by $U_c^*$ surround axis $U = -\varepsilon_0$ one $+2\pi$ cycle of $\Delta\varepsilon(\tau_0)$, $-2\pi$ cycle of $\Delta\varepsilon(\tau_1)$, $+2\pi$ cycle of $\Delta\varepsilon(\tau_2)$, $-2\pi$ cycle of $\Delta\varepsilon(\tau_3)$,…until to the 8th $+2\pi$ cycle of $\Delta\varepsilon(\tau_8)$. The red arrows denote the paths of cascade vibrations: from $q_{8,R} \to q_{7,R} \to q_{6,R} \to \ldots \to q_{2,R} \ldots \to q_{1,R} \to q_{0,R}$.

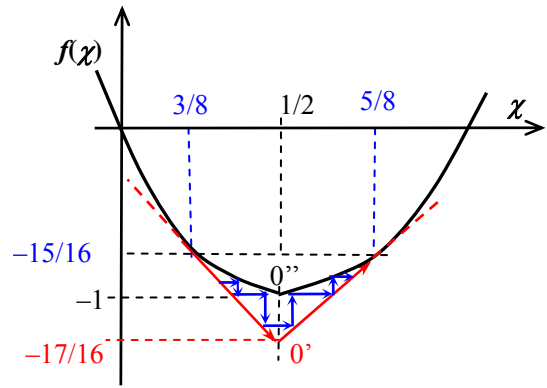

Fig.10. During glass transition, the path of compacting cluster starts from a disturbance $\Delta\chi$ on bifurcation point $(3/8, -15/16)$, via fluctuation-reinforce losing one interface excitation energy, to the point 0' along slope $-1$, then via fluctuation- restraint gaining one interface excitation energy, to the bifurcation point $(5/8, -15/16)$ along slope 1, in which one interface excitation energy of $1/8\varepsilon_0$ transfers from the surface on $i$-th order cluster to that on $(i+1)$-th order.

The potential that the attraction and repulsion happens to be in equilibrium is denoted as $U_c$.

**4.8. Stability condition of recursion equation**

The stability condition of recursion equation (20) can be also deduced from the simple self-similar condition of Eq. (28) in the new coordinate system (transformation: $\chi' = \chi - 1/2$, $f' = f + 17/16$) of point 0' (1/2, -17/16) as coordinate origin in Fig.10.  Eq. (28) and the difference of 17/16 −15/16 = 1/2 − 3/8 = 5/8 − 1/2 = 1/8 (the reduced quantized interface excitation energy) in Fig.10 clearly give voice to that the fixed point $U_c^*$ in Fig.7 and the origin of interface excitation and the transfer of excitation interface all result from the balance effect between the two-body interaction potential fluctuation, $\Delta f(\chi_i)$, and the cluster geometric-phase-induced potential fluctuation, $\Delta \chi_{i+1}$, in glass transition.  This is also the bifurcation stability condition in Eq. (20).  Fixed point $U_c^*$ is easy to be proved since in fig.7, none of the potential curves, except for those who have passed, correspond with the stability condition of Eq. (28).

$$\frac{\partial f'}{\partial \chi'} \equiv \frac{f'}{\chi'} \tag{28}$$

A series of important results thus are found as follow.

### 4.9. Two attraction-repulsion balance points in glass transition

There are two 'attraction-repulsion' balance points in glass transition, one is the point 0"(1/2, –1) in Fig.10, corresponding to the points ($q_{i,0}$, $\varepsilon_0$) in Fig.9, that of the solid state general syntonic vibrations of clusters, and the other is the point (1/2, –17/16) in Fig.10, corresponding to the balance both the reduced geometric-phase-induced potential fluctuation and two-body interaction potential fluctuation at point $\chi = 1/2$.  Note that the second balance points 0' (1/2, –17/16) only occur on a few thawed domains, which corresponding to the *abnormal heat capacity* and can be called the abnormal or *additional balance points* in the glass transition.  The additional attractive potential on point 0' comes of the self-similar fluctuation equation (28).

### 4.10. Apart path of clusters: 8 orders of geodesic

The apart path of two $\sigma_1$ clusters from $q_{i,R}$ to $q_{i+1,R}$ in Fig.9 rounds (do not pass!, refracts the parallel transport surround $q$-axial in topology analysis) the vibrate balance point $q_{i,0}$, i.e. the point 0" in Fig.10, and the sum of the two orthogonal geometric-phase-induced potentials on sharp-angled point $q_{i,R} = q_{i+1,L}$ equals to the potential well energy, which clearly indicate the arrow from $q_{i,R}$ to $q_{i+1,R}$ in Fig.9 is exactly the non-integrable geometric phase factor in multiparticle system [58] or the geodesics [59], the minimum energy path, in topological analysis.

In Fig.9, $i$-th order geodesic is the shortest line of $2\pi$ cycle of $\Delta \varepsilon (\tau_i)$ between $q_{i,R}$ and $q_{i+1,R}$ on the $i$-th order of cylindric potential surface.  The 8 order of cylindric potential surfaces are formed by $U_c^*$ surround axis $U = -\varepsilon_0$, $+2\pi$ cycle of $\Delta \varepsilon (\tau_i)$, $-2\pi$ cycle of $\Delta \varepsilon (\tau_1)$, $+2\pi$ cycle of $\Delta \varepsilon (\tau_2)$, $-2\pi$ cycle of $\Delta \varepsilon (\tau_3)$,…until to the 8th $+2\pi$ cycle of $\Delta \varepsilon (\tau_8)$.  In Fig.9, the red arrows denote the paths of cascade fractal vibrations: from $q_{8,R} \to q_{7,R} \to q_{6,R} \to \ldots \to q_{2,R} \ldots \to q_{1,R} \to q_{0,R}$.

It shows that the delocalization path or the re-coupling path of particle-pairs is along 8 orders of geodesics and the delocalization energy only as $1/8\varepsilon_0$ of L-J potential well energy, which is exactly the quantized interface excitation energy, $\Delta \varepsilon = 1/8\varepsilon_0$.

### 4.11. Increasing distance only increasing the interacting time of two-body interaction

The two-body interactions of two $\sigma_1$-clusters have 8 orders interacting (relaxation) times and the same numerical value of interaction potential of $1/8\varepsilon_0$, independent of their distance, increasing distance only increasing the interacting time through their $i$-th order cluster-interfaces.   .

### 4.12. Additional attractive potential lower than potential well energy

The potential energy $-17/16\varepsilon_0$ is a newly arisen additional attractive potential in the self-similar mean field (Eq. (28)) of mean fields of different sizes (Fig.9), which further resolves the puzzle why colloidal hard-sphere system has stronger attractive potential. The singularity of the mean field of mean fields of different sizes can be seen in $q_{i,R} = q_{i+1,L}$ (verified by Esq. (29)- (32)) that the right displacive amplitude of $q_{i,R}$ in $i$-th order mean field balances the left displacive amplitude of $q_{i+1,L}$ in ($i$+1)-th order, instead of the general left-right asymmetric vibrant in potential curve (Eq. 17), thus this displacive amplitude energy of particle-cluster is also the *additional potential* in the glass transition, which comes from the quantized interface excitation and transfer.

**4.13. The origin of quantized interface excitation**

'Self-similar fluctuation' in Eq. (28) itself is also a kind of 'ordering' relative to random fluctuation. It is interesting that the positions of two attraction potentials of $-15/16\varepsilon_0$ and $-17/16\varepsilon_0$ are in reflection symmetry relative to that of the potential well energy in Fig.10. This further explains that the origin of quantized interface excitation: *interface excitation not only balances both the reduced geometric-phase-induced potential fluctuation and two-body interaction potential fluctuation, but also balances both the additional self-similar attractive potential and the additional displacive amplitude energy* in the self-similar mean fields of different sizes in the glass transition, instead of so-called 'two-level'. *The abnormal additional displacive amplitude energy in self-similar system may be the causation of abnormal heat capacity and abnormal heat expansion in the glass transition.*

**4.14. Proof for 8 orders of intrinsic mosaic geometric structures**

From $\chi_{+1} = 3/8$, $\chi_{-1} = 5/8$, $\sigma_i$ and $q_{i,R}$ in Fig.9 can respectively be expressed as

$$q_{i,R} = q_{i+1,L}, \tag{29}$$

$$\sigma_i = \sigma_1 (5/3)^{(i-1)/6}, \qquad i = 1, 2, 3\ldots \tag{30}$$

$$q_{i,R} = \sigma_1 (5/3)^{(i-1)/6} (8/3)^{1/6} \tag{31}$$

$$\Delta q_{i+1} = q_{i+1,R} - q_{i+1,L} = q_{i+1,R} - q_{i,R} = (8/3)^{1/6} \left[(5/3)^{1/6} - 1\right]\sigma_i \approx 0.1047\sigma_i \tag{32}$$

In Eq. (31), when $i = 8$, $q_{8,R} \approx 2.137\sigma_1 > 2\sigma_1$, ($q_{7,R} < 2\sigma_1$), Therefore, only at the 8th order L-J potential field is the 8th order 'distance-increment vacancy' with excluded volume of $\sigma_1$ able to be put in between two $\sigma_1$ clusters with same excluded volume. In the meantime, when $i = 8$, the system also satisfies the classical condition ($v_c = 3b$, $b$ here is $\sigma_1$) of critical phase transition of the Van der Waals long range interaction equation (($P + a/v^2$) ($v - b$) = $kT$). However, *the critical energy to form the 8th order cluster is the localized energy, $E_c = kT_g^* = 20/3\varepsilon_0$ (Eq. (10)), less than the critical phase transition energy, $kT_c = 8\varepsilon_0$,* in Van der Waals long range interaction equation. This is clearly also the contribution of the long-range correlation of short-range interaction of excitation interfaces in local cluster growth phase transitions.

The result of $M = 8$ further confirms that *there are only 8 orders of intrinsic mosaic geometric structures and 8 orders of local cluster growth phase transitions in glass transition*.

It can be seen that the 8 orders of distance-increment vacancies in glass transition respectively offer one external degree of freedom for 8 orders of cluster motions in inverse cascade, which come from the 8 orders of geometric phase potentials formed by 8 orders of $2\pi$ interface excitation closed loop-flows, independent of temperature and frequency. This characteristic may be used to understand the origin of Boson peak.

**4.15. So-called 'tunneling'**

There are two degenerate states for reduced geometric-phase-induced potential on each sharp-angled point in Fig.9. The apart paths of two $\sigma_1$ clusters show in Fig.10: inside $\sigma_{i+1}$, $\sigma_i$ cluster is of 5 inner degrees of freedom (Section 3.11) and can migrate in the range of $\sigma_{i+1}$; thus, the slow process re-coupling or delocalization of two $\sigma_1$-clusters should be adopted accompanied with $\sigma_8$ cluster repetitious rearrangements. The delocalization energy is only $1/8\varepsilon_0$, the quantized transfer energy of interface excitation. Therefore, the 8 orders of apart (delocalization) paths of two-body and the local structure repetitious rearrangements and the attractive potential $-17/16\varepsilon_0$ reveal the mechanisms of the so-called 'tunneling'[23, 59] in glass transition.

The fixed point also clarifies that the so-called tunneling turns out to be the generating and transferring of all quantized interface excitations should pass through the additional attractive potential center of $-17/16\varepsilon_0$, which is lower than potential well energy and comes from the balance effect of reinforce-restraint of self-similar potential fluctuation.

It is interest that there are 8 sharp-angled saddle points, each with the same barrier $\Delta\varepsilon_h = 1/16\ \varepsilon_0$, and $g_s = \Delta\varepsilon / \Delta\varepsilon_b = 2$. $\Delta\varepsilon$ here is the energy of loop-flow. The $g_s$ here may be called the *additional classical spin g factor* of the interface cross-coupled electron pairs in random system, with the same value of the spin factor in one-electron theory.

**4.16. Universal Lindemann ratio $d_L = 0.1047…$**

Eq. (32) also shows that the stable ratio of $d_L = \Delta q_{i+1}/\sigma_i = 0.1047…$, a universal value, which is exactly the Lindemann ratio. Lindemann criterion has been applied to study solid-versus-liquid behavior [60]: distance-fluctuations cannot increase without destroying the lattice structure [61]. Here it has been found that the Lindemann ratio contained in Eq. (32) is equivalent to the stability condition of recursion equation (20), which also indicates the local cluster growth phase transitions in the glass transition is only determined by the intrinsic geometric structure, without any presupposition and relevant parameter. The numerical value of $d_L$ directly deduced from this paper accords with the approximation value of 0.10 in [9] and 0.1- 0.15 in [61]. This also validates the theory of fixed point for self-similar L-J potentials and the Eq. (29).

**4.17. Universal fixed point in non-ideal glass transition**

For general solid-to-liquid glass transition, there are 8 orders of different potential well energies $\varepsilon_i(\tau_i)$ and 8 orders of different quantized interface excitation energy $\Delta\varepsilon_i(\tau_i)$. However, if the non-dimensional reduced L-J potential $U_{1,2}(\chi) / \varepsilon_i(\tau_i) = f_i(\chi)$ and the non-dimensional reduced interface excitation energy $\Delta\varepsilon_{i,l}(\tau_{i,l})/\varepsilon_i(\tau_{i,l})$ are adopted, the equations (19) – (32) are still all hold true. The Fig. 6 is only modified as an aberrance lattice. Thereupon, $M = 8$, and the universal fixed point of L-J potentials obtained: $f_c^* \equiv -15/16$ and $\chi_{+1} \equiv 3/8$, $\chi_{-1} \equiv 5/8$, in general glass transition.

**4.18. Two orthogonal degenerate states of fast-slow geometric phase factors**

One of the most striking is that the existence of fixed point for self-similar L-J potentials can be actually proved without other assumption and complicated mathematical analysis. However, the stable fluctuation recursion equation in solid-to-liquid transition reveals the relationship between 'reinforce - restraint' of potential-fluctuation and Lindemann ratio.

This theory provides a unified mechanism to interpret the Lindemann ratio, the origin of quantized interface excitation, and the 8 orders of hard-spheres in glass transition, the hard-sphere long-range attractive potential and the tunneling.

Second universal behavior in glass transition is found that two orthogonal degenerate states of fast-slow geometric phase factors, the fast reduced geometric phase factor 3/8 and the slow reduced geometric phase factor 5/8, are accompanied with the appearance of each order of local cluster growth phase transitions.

The existence of the fixed point of L-J potentials moreover has proved that (a) the origin and transfer of interface excitation in the glass transition come of the balance effect between self-similar L-J potential fluctuation and geometric-phase-induced potential fluctuation; (b) a universal behavior: two orthogonal degenerate states, the fast reduced geometric phase factor 3/8 and the slow reduced geometric phase factor 5/8, is accompanied with the appearance of each order 2-*D symmetrical interface-excited energy loop-flow* in the glass transition. The sum of the two degenerate reduced phase factors is exactly equal to 1, corresponding to the origin of Boson peak.

**4.19. Five kinds of ordered potential motions balance with disordered kinetic motions**

The geometric phase factor here is exactly the non-integrable phase-induced potential to induce cluster migration. Thereby, the further cognition to the relationship between the random system and the glass transition can be obtained: in the thermal random motion system during glass thaw, there are at least five kinds of ordered (potential energy) motions which are balanced with disordered (kinetic energy) motions.

*The first motion is fluctuation self-similarity*. The self-similarity potential fluctuations of local and global, specially selected from random potentials fluctuations, leads to the lower attractive potential of $-17/16\varepsilon_0$ ($\varepsilon_0$, potential well energy) and the interface excitation and excited-interface transfer in the glass transition.

*The second motion is local transient 2-D symmetrical interface-excitation energy loop-flow and localized energy.* Some excitation interfaces, selected from fluctuating excitation interfaces (that is also an idea of soft matter!), form the instantaneous $i$-th order 2-*D* interface-excitation energy loop-flow in local z-component space (or on x-y projection plane, similar as Ising model), which defines $i$-th order cluster. In random motion, once a transient 2-*D* interface-excited energy loop-flow forms, the surrounded cluster will gain non-integrable phase-induced potential to induce the cluster migration along the z-axial. According to the topological analysis, the cluster is always in slow movement, the $2\pi$ energy flow is always in comparatively fast movement and the former is induced by $2\pi$ energy flow of symmetrical interface-excited energy loop-flow on the cluster interfaces. The localized energy $E_c$ independent of temperature, $E_c = kT_g^* = 20/3\varepsilon_0$ (Eq. (10)), shows that in the ideal (flexible polymer) random motion, the symmetrical 2-*D* closes loop is neither optional nor arbitrary large. In the temperature range from $T_g$ to $T_m$, there are only 8 orders of symmetrical 2-*D* closes loops with 8 orders of relaxation times in the thermal random motion. These loops form inverse energy cascade from small to large, whose energy is localized energy $E_c$.

*The third motion is cooperative orient migration in local field*, in which the 8 orders of 2-*D* interface-excitation energy loop-flows appear from small to large, from fast to slowly, in the manner of inverse cascade. This is also the manner of breaking a solid domain. When the 8th order loop-flow completes, its energy will transfer to random vibration kinetic energy in form of fast cascade vibration and the re-built excitation interfaces will begin new inverse cascade.

*The fourth motion is the degree of freedom of flow-percolation.* The increase of system temperature nothing but increase the number of inverse cascade-cascade in local zone, it can not increase the size and energy of 2-*D* interface-excitation loop-flow. That is the physical meaning of intrinsic localized energy in the thermo random system. The glass transition corresponds to the critical transition in which all the local inverse cascade motions are precisely connected (i.e. flow-percolation, which is a sub-system in system). At that time, each *i*-th order loop-flow gets the evolution energy $\varepsilon_0$ of one external degree of freedom of the *i*-th order cluster and the free motion energy $5\varepsilon_0$ of five inner degree of freedom of the centric (*i*-1)-th order cluster in the *i*-th order loop-flow. Thus, the external degree of freedom of flow-percolation in the glass transition is 1. The melt transition means that the external degree of freedom of flow-percolation is 5.

*The fifth motion is dynamical mosaic structure of flow-percolation*, which makes the evolvement of each reference local field consist of delaying contribution of 4 neighboring local fields to the evolvement of central local field. Thereupon, each excited domain does not return to the solid state for the disappearance of excitation interfaces in the initial reference local field, instead, they always consist of excitation interfaces of delaying contribution to the excited domain by the 4 local fields neighboring update reference local field.

## 5. Statistical thermodynamics in the glass transition
### 5.1. The fluctuation physical picture - the most effective way in the glass transition

Thermo disorder-induced localization is also one of the most important properties from $T_g$ to $T_m$ in the solid-to-liquid transition. In the Section 3.13, the localized energy $E_c = 20/3\varepsilon_0$ in the glass transition has been deduced by geometry method of the model. In the Section 5, $E_c \approx 20/3\varepsilon_0$ will be directly proved by statistic thermodynamics in the glass transition.

On the other way, although the instant 8 orders of 2-*D* mosaic geometric structures have been found by geometry method, all the other loop-flows will not appear as long as the 8th order loop-flow has not appeared, as emphasized in Section 2.

The problem is that in the critical state of the 8th order loop-flow being about to appear, how the 8 orders of symmetrical interface excitation loop-flow in the system realize it by the most effective way and that since the 8th orders of loop-flow is the event which occurs suddenly, how the fluctuation physical picture of the system is described before the 8th order loop-flow appears.

The two questions above will be also answered by the geometry method mentioned in Section 2.3, namely, the standout mode of an arrow on an excitation interface, in one solution. In the critical 8 orders of local cluster growth phase transitions of glass transition, the fluctuating 8 orders of *self-similar two-body and three-body coupling clusters* have appeared and their fluctuation interfaces will one by one abrupt change in the most effective way in order to realize 8 orders of 2-*D* mosaic geometric structures and complete the glass transition. In other words, *there are 8 orders of self-similar and fluctuating two-body and three-body coupling clusters* that excited by thermo random motion in the glass transition.

If this idea is correct, it will be expected that the sum of two potential energies of two-body and three-bod*y* should balance with the thermal random motion energy *kT* (current hard sphere model can not do as this [17]). This will be the sixth motion that potential balances kinetic energy. This motion will be discussed by way of the reduced second, third Virial coefficients in this section. In the percolation field, all the excited locals are one by one connected as a whole in time. While in the critical state, each excited local can also be seen as 8 orders of self-similar two-body and three-body clusters in the fluctuation-evolution states. Therefore, in a percolation field, all the *i*-th order two-body and three-body coupling clusters are connected with each other, namely, the *i*-th order inverse cascade energy flow in the percolation field, named as $V_i$-percolation field and $V_i$ is the volume of the *i*-th order of loop-flow. The second and the third Virial coefficients for *i*-th order two-body and three-body will be discussed respectively.

The Section 5 in this paper mainly aims to prove that the second Virial coefficients of clusters of different size exists fixed point, by using the classical thermodynamic theory and the scaling theory. It will be further proved that the potentials of the self-similar two-body and three-body coupling clusters balance the kinetic energy in the glass transition.

**5.2. Self-similar 2body-3body coupling clusters**

In the local cluster growth phase transitions of glass transition, if each 2-*D* interface excitation appears in the way of one arrow after another according to the appearance probability as mentioned in figures 3-6, the probability is rather low. In fact, the more probable situation is that each in the 8 2-*D* interface-excited energy loop-flows first form *two fluctuating three-body* (named as 2body-3body) clusters in the space and when the 8th loop-flow appears, 8 orders of 2body-3body coupling clusters cooperatively reduce the total excited energy to, one by one, form 8 order 2-*D* interface excitation loop-flows and eliminate interfaces in the loop-flow.

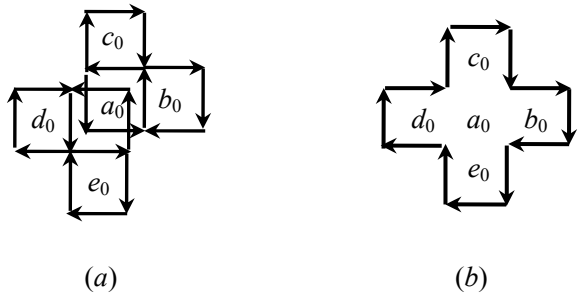

(*a*)          (*b*)

Fig.11. Schematic diagram of 2body-3body coupling cluster in the critical local cluster growth phase transitions of glass transition. When the two centers of $V_0(a_0)$ of two fluctuation three-body clusters, $V_0(a_0)+V_0(b_0)+V_0(c_0)$ and $V_0(a_0)+V_0(d_0)+V_0(e_0)$, coincide, first order interface excitation loop is formed (b).

As shown in Fig.11 (*a*), in a reference $a_0$ particle $V_0(a_0)$, at different times and in different spaces and positions, respectively form two 3body clusters: $V_0(a_0)+V_0(b_0)+V_0(c_0)$ and $V_0(a_0)+V_0(d_0)+V_0(e_0)$, with neighboring particle $V_0(b_0)$, $V_0(c_0)$ and particle $V_0(d_0)$, $V_0(e_0)$.

Each 3body cluster only needs the excitation energy of 10 excitation interfaces, which is less than that of 12 excitation interfaces in Fig.11 (*b*). The two particle $V_0(a_0)$ in two 3body clusters, excited at different times, move randomly and they are not in superposition. Once they are in superposition, Fig.11 (*a*) immediately mutates into Fig.11 (*b*) to form the first order interface excitation energy loop-flow and the excitation interfaces inside the loop-flow disappear. In the critical state of the 8th loop-flow's appearance, 8 orders of clusters of local field appear in the form of two fluctuation-3body clusters as shown in Fig.11 (*a*), named as *2body-3body coupling clusters* in the glass transition and solid-to-liquid transition.

### 5.3. Virial expansion in the glass transition

The Virial expansion is generally used to study the many-body problem. Eq. (33) is the expression of Virial expansion in classical gas system.

$$\frac{PV}{kT} = 1 + \frac{1}{b_0} B_2^0 + \frac{1}{b_0^2} B_3^0 + \cdots = 1 + B_2 + B_3 + \cdots \tag{33}$$

Where $B_2 = B_2^0/b_0$ and $B_3 = B_3^0/b_0^2$ are the reduced second and third Virial coefficient respectively. The numerical value '1' in Eq. (33) comes from the contribution of repulsive potential to surface transformed from the kinetic energy of one gas quasi-particle itself. The physical picture in classical gas system is described as follows. One quasi-particle *moves* along *one direction* and its motion is instantly *arrested* by surface in collision. The kinetic energy transforms into the potential. All other items in Eq. (33) are contributions of the kinetic energies from two-body, three-body and so on, transforming into *PV/kT*. *P* in Eq. (33) is a positive pressure. The sign in front of particle volume is positive, which represents that the *outward kinetic energy contributing to interface repulsive potential* in classical thermodynamic theory.

The theoretical picture of ideal (flexible polymer) solid-to-liquid glass transition is exactly *reversed* to that of classical ideal gas. 2body-3body clusters (from 'static' to 'moving' along one direction to thaw a solid domain) should 'catch' outside (i.e., neighboring 5-particle excited field) kinetic energy to be excited and have 'attraction' potential. A reference *i*-th order cluster (hard-sphere) is defined by the *i*-th order interface excitation loop-flow contributed by its 4 neighboring (*i*-1)-th order loop energies and the volume sign of *i*-th order cluster is the same with that of *i*-th order loop-flow. Here, one of the key proofs is that the direction of the *i*-th order loop-flow (the direction along which the *i*-th order cluster is driven to move) is always opposite to that of the (*i*+1)-th loop-flow, in other words, in critical local cluster growth phase transitions, the *free motion*, being of 5 degrees of freedom, of *i*-th order cluster always appears in the (*i*+1)-th induced potential field whose sign of induced direction is opposite to that of the *i*-th cluster.

### 5.4. Phase difference of $\pi$ between kinetic energy and potential energy in glass transition

Thus, in order to refract 'the *phase difference of $\pi$* between the cluster in free motion state and the induced potential', (i.e., the phase difference of $\pi$ between *kinetic energy and potential energy*, which is also one of the singularities of glass transition), a 'negative sign' should be attached to the front of the hard-sphere volume in the picture of critical local cluster growth phase transitions. That is, glass transition is a phenomenon of transformation from *inward kinetic energy contributing to interface attraction potential*. The role reversal between 'outward kinetic-repulsive potential' and 'inward kinetic- attraction potential' in physics shows that the classical thermodynamic theory still holds true for glass transition in percolation field.

In Eq. (33), the contribution of item '1' to *PV/kT* is replaced by that of a 2body-3body coupling cluster. This is because that the interface of hard-spheres is formed by interface excitation and each interface excitation always connects with two clusters. Therefore, there is no *single* cluster whose attraction potential alone contributes to *PV/kT*. Any items after the third item in Eq. (33) are also replaced by self-similar 2body-3body coupling clusters, which do not conflict with the mode-coupling theory [62].

In Eq. (33) the positive pressure is changed to negative attraction tensile stress which keeps the balance to the external stress on percolation field in the glass transition. In order to conveniently apply the classical thermodynamic, a 'negative sign' should be respectively attached to the front of the stress *P* and the volume *V* in this paper. The sign of *PV/kT* is still positive.

Eq. (34) is a Virial expansion around the volume variable $V_i$ of connected $v_i$-clusters on $v_i$-percolation field in the glass transition, where $V_i$ is only the *volume of sub-system* (percolation field), instead of the volume of the whole system, and $v_i$ is the volume of *i*-th order of cluster, $kT(v_i)$ is random motion energy of *i*-th order of cluster.

$$PV_i(v_i)/kT(v_i) = B_2(v_i) + B_3(v_i) \tag{34}$$

Note that Eq. (34) is only a Virial expansion of single $v_i$-cluster, and it will, as follows, be modified as that of self-similar clusters using average field approximate method in critical local cluster growth phase transition.

### 5.5. Self-similar equations of two-body clusters in the glass transition

From Enthalpy $H = E + PV$, the definition of Joule-Thomson coefficient $\mu_J$ [63] is

$$\mu_J = (\partial T / \partial P)_{H,N} = C_P^{-1}\left[T(\partial V / \partial T)_{P,N} - V\right] \tag{35}$$

Assume the state of $\mu_J \equiv 0$ corresponds to glass transition. When $\mu_J \equiv 0$

$$(\partial V / \partial T)_{P,N} \equiv V/T \tag{36}$$

This is the volume change of sub-system (also of cluster) as a function of temperature *T* (here *kT*, as the random motion energy, also is a function of cluster volume) when outside pressure (stress) remains constant. As long as the condition $\mu_J \equiv 0$ in Eq. (35) is satisfying, Eq. (36) holds true. In addition, it can be seen that $C_P$ in Eq. (35) may show an *abnormal change* during glass transition.

It can also be strictly proved that the reduced third Virial coefficient for hard-sphere system is *constant* 5/8; independent of temperature [64], i.e. $B_3$ in Eq. (34) is independent of temperature. Besides, from Fig.10, the non-dimensional interaction potential forming $B_2(v_i)$ in Eq. (34) can be thought as Eq. (37), because of the interface excitation energy, $\Delta\varepsilon_i(\tau_i) = 1/8\varepsilon_i(\tau_i)$, transferred away from $v_i$-percolation field to $v_{i+1}$-field and the random motion bound of $v_i$ cluster as $v_{i+1}$-field, the thermo- random motion energy as $kT(v_{i+1})$. So,

$$B_2(v_i) = \frac{PV_i(v_i)}{kT(v_{i+1})}, \quad (i = 1, 2...8) \tag{37}$$

On the other hand, using mean field method to Eq. (34): i.e. substituting $B_3 = PV/kT - B_2$ into Eq. (34), and modifying $B_3(v_i)$ with $B_3(v_{i-1})$ in Eq. (34). Because during the evolution from $v_{i-1}$-percolation field to $v_i$-field, the $B_3(v_{i-1})$ contributed by 3 $v_{i-1}$-body interaction also contains the $B_2(v_{i-1})$ of 2 $v_{i-1}$-body fluctuation interaction on $v_{i-1}$-filed, see Fig.11:

$$B_3(v_{i-1}) = \frac{PV_{i-1}(v_{i-1})}{kT(v_i)} - B_2(v_{i-1}) \tag{38}$$

The substitution of Eq. (38) into Eq. (34) yields

$$\partial B_2(v_i) = B_2(v_i) - B_2(v_{i-1}) = \frac{P(V_i(v_i) - V_{i-1}(v_{i-1}))}{kT(v_i)} = \frac{P \cdot \partial V_i(v_i)}{kT(v_i)} \tag{39}$$

From Eqs (37) and (39), $\dfrac{\partial B_2(v_i)}{B_2(v_i)} = \dfrac{\partial V_i(v_i) \cdot T(v_{i+1})}{V_i(v_i) \cdot T(v_i)}$.

Then modify Eq. (36) as follows, because of the $\pi$ phase difference between fast-moving kinetic energy $kT$ and slow-motion potential energy $PV$ when $i$-th order of cluster evolved to ($i$+1)-th order.

$$\frac{\partial V_i(v_i)}{V_i(v_i)} = \frac{\partial T(v_{i+1})}{T(v_{i+1})} \text{, and yields } \frac{\partial B_2(v_i)}{B_2(v_i)} = \frac{\partial T(v_{i+1})}{T(v_i)}.$$

Therewith, an important self-similar Eq. (40) in the glass transition has been obtained

$$\frac{\partial B_2}{\partial T} = \frac{B_2}{T} \tag{40}$$

Note that the variables of $B_2$ and $T$ are all the volumes of clusters and outside pressure remains constant. Eqs (36) and (40) show that the physical quantities of $\partial B_2/\partial T$ and $\partial V/\partial T$ on small scale are respectively equal to those of $B_2/T$ and $V/T$ on large scale, which represents the self-similarity between the small scale and the large scale in critical phase transition.

Eqs (36) and (40) thus are the self-similar phase transition equations of *sub-systems* in glass transition. Note that the enthalpy $H$ in Eq. (35) is *invariant* in any sub-systems *but not in the whole system*. These self-similar phase transition equations of sub-systems, together with the self-similar equation $\partial f'/\partial \chi \equiv f'/\chi$ mentioned in Eq. (28), are in fact the condition equations for occurrence of self-similar 2body-3body clusters in the glass transition.

## 5.6. Approximate solution for self-similar equations

In the Section 4, one of the important results is that the definition or origin of hard-sphere can be also deduced from the self-similar Lennard-Jones (L-J) potentials in ideal random system. Here the key point is the self-similar in system.

Now the approximate solution is deduced from the self-similar Eq. (40) also by L-J potential. Take [65]

$$B_2(T^*)_{L-J} \equiv B_2^0(T^*)_{L-J}/b_0 = \frac{4}{T^*}\int_0^\infty dx \cdot x^2 \left[\frac{12}{x^{12}} - \frac{6}{x^6}\right] \exp\left\{-\frac{4}{T^*}\left[\left(\frac{1}{x}\right)^{12} - \left(\frac{1}{x}\right)^6\right]\right\} = \sum_{n=0}^\infty \alpha_n \left(\frac{1}{T^*}\right)^{\frac{2n+1}{4}} \tag{41}$$

In this equation $\alpha_n$ can be represented by the $\Gamma$ function: $\alpha_n = -\dfrac{\sqrt{2}\Gamma\left(\dfrac{2n-1}{4}\right)}{2^{(2-n)}n!}$, in Eq. (41), $x = q/\sigma$.

$q$, $\sigma$ are all parameters in the L-J potential. $T^* = kT/\varepsilon_0$ is the reduced temperature, $\varepsilon_0$, the L-J potential-well energy.

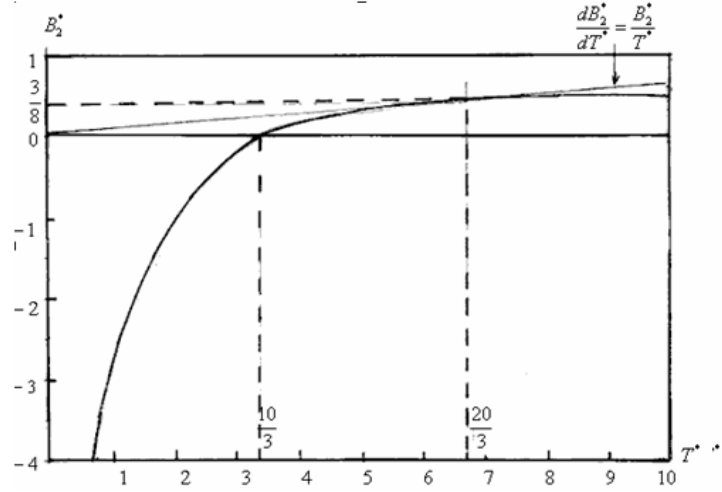

Fig.12. Diagrammatize to Eqs (40) and (41). In the figure, $T_g^* \approx 20/3$ denotes the reduced localized random kinetic energy point during glass transition when $B_2^*(20/3) \approx 3/8$. $T_g^*$ is a fixed point independent of temperature.

Note that Eq. (41) is a strict and proper solution of the second Virial coefficient to single cluster in statistical physics. However, if we regard the $kT$ in Eq. (41) as the thermo-random kinetic energy of self-similar clusters and find out the point of intersection of Eq. (40) and Eq. (41), the Virial coefficient in the glass transition may be obtained. There is no analytical solution after substituting Eq. (41) into Eq. (40). Graphical method gives the only set of approximate solution, see Fig.12.

$$\begin{cases} T_g^* \approx 20/3 \\ B_2^*(T_g^*) \approx 3/8 \end{cases} \quad (42)$$

**5.7. Intrinsic and invariant random kinetic energy**

The numerical solution in Fig.12 refers to the result of [63]. There are two reduced temperature points on the curve of $B_2(T)$, respectively corresponding to the two special *reduced random motion energy* and denoted as $T_g^* \approx 20/3$ and $T_0^* > 10/3$. Only the point $T_g^*$ will be discussed in this paper. Fig.12 shows that in the common expression of the second Virial coefficient in the ideal random system, only the point $(T_g^*, B_2^*)$, which corresponds to the local cluster growth phase transition in the glass transition, meets the condition of self-similarity. The 'reduced temperature' point $T_g^*$ seemingly corresponds to that of the glass transition, but does not generally do unless the 'slow process' nondimensional heat motion energy $kT/\varepsilon_0$ supplied from environment is exactly equal to $T_g^*$. Due to the fact that Eqs (34) – (42) only come into existence on sub-system, $T_g^*$ has no option but to be the local cluster growth phase transition temperature in sub-system, that is, $T_g^*$ is the *reduced critical random energy* in sub-system (flow-percolation field) for the occurrence of 8 orders 2body-3body cluster*s*.

$T_g^*$ is called as the *intrinsic* and *invariant* random kinetic energy which is an important

physical concept to characterize the intrinsic property in the glass transition. According to Anderson [46] theory of *disorder-induced localization*, $T_g^*$ is defined as the (nondimensional) *localized energy* of thermo random motion (kinetic energy) induced by thermo-disorder in glass transition.

From Fig.12, the localized energy, $E_c = T_g^* \varepsilon_0 \approx 20/3\ \varepsilon_0$, approximately equals to the result in Eq. (10): $E_c = kT_g^\circ(\tau_i) = 20/3\ \varepsilon_0 = T_g^* \varepsilon_0$.

## 5.8. Stable fixed point of second Virial coefficients in solid-liquid transition

There are 8 orders of localized energies $T_g^* \varepsilon_0(\tau_i)$ in glass transition, in which $\tau_i$ denotes the relaxation time of $i$-th order hard-sphere. The fixed point $T_g^*$ in Fig.12 shows that the numerical value of $T_g^*$ is invariable, but it contributes to different order of clusters in inverse energy cascade. Eqs (34) to (42) will always be valid and $B_2^*$ will always equal to 3/8, regardless if the temperature $T_g$ ($T_g$ is the apparent glass transition temperature) will further increase.

Point $(T_g^*, B_2^*) = (20/3, 3/8)$ is a stable fixed point. *When $T_g$ further increases, the increase of randomness in system cannot be otherwise than the increase of the number* of *inverse cascade-clusters*.

This foreshows that, on the fixed point of $T_g^*$ in the glass transition, there is an identical second Virial coefficient $B_2^*(T_g^*)$ for all difference clusters in size. Thus, by first determining the general second Virial coefficient expression $B_2(V_i)$, which depends on the order number, $i$, of clusters, and then the scaling equation for different clusters in size, $B_2^*$ can be finally deduced from the fixed point of the scaling equation.

Diagrammatic of Fig. 12 indicates:

$$B_2^*(T_g^*) \approx B_2^*(20/3) \approx 3/8$$

## 5.9. Fixed point theorem prove for $B_2^* \equiv 3/8$ of second Virial coefficients in glass transition

$B_2^* \approx 3/8$ in Eq. (42) is an approximate solution using L-J potential. Now $B_2^* \equiv 3/8$ is directly proved by scaling theoretical approach.

*The first step of testification is to find out the physical quantity- Fugacity,* which can reflect the contribution of clusters of different size to Virial coefficient expression in the classical Virial coefficient expression. Following the derivation for $B_2$ in classical statistical physics see as in [63].

Pressure (stress)

$$p = -\frac{\Omega(V,T,\mu)}{V} = kT \sum_{l=1}^{\infty} \frac{b_l(T,V)e^{\beta l \mu}}{\lambda^{3l}} \tag{43}$$

Particle density

$$\frac{\langle N \rangle}{V} = -\frac{1}{V}\left(\frac{\partial \Omega}{\partial \mu}\right)_{V,T} = \sum_{l=1}^{\infty} \frac{l b_l(T,V)e^{\beta l \mu}}{\lambda^{3l}} \tag{44}$$

Power series of Virial expansion

$$\frac{pV}{\langle N \rangle kT} = \sum_{l=1}^{\infty} B_l(T)\left(\frac{\langle N \rangle}{V}\right)^{l-1} \tag{45}$$

The symbols are the same as in [63], where $\beta = 1/kT$, $\mu$: the chemical potential, $B_l$: the $l$ order of Virial coefficient, $b_l$: the $l$ – cluster integral. Eqs (43) – (45) hold true for any sub-system in glass transition. Take the thermodynamic limits for all Equations of (43) – (45), i.e., $V \to \infty$, and notice that the limit condition corresponds to percolation field, which allows strange shapes [65] and here a fractal shape. $<N> \to \infty$. $<N>/V$ is constant. $b_l(T, V) \to \overline{b}_l$. Combine (44) – (45), and yield:

$$\left(\sum_{l=1}^{\infty} \frac{\overline{b}_l(T) e^{\beta l \mu}}{\lambda^{3l}}\right) \left(\sum_{n=1}^{\infty} \frac{n \overline{b}_n(T) e^{\beta n \mu}}{\lambda^{3n}}\right)^{-1} = \sum_{l'=1}^{\infty} B_{l'} \left(\sum_{n'=1}^{\infty} \frac{n' \overline{b}_{n'}(T) e^{\beta n' \mu}}{\lambda^{3n'}}\right)^{l'-1} \quad (46)$$

Expand both sides of Eq. (46) and make the two second power coefficients of $\lambda^{-3} e^{\beta \mu}$ equal, and the second Virial coefficient is obtained. Notice that the *factors* of $e^{\beta \mu}$ on both sides of Eq. (46) *can be canceled out*.

$$B_2(T) = -\overline{b}_2(T) \quad (47)$$

Eq. (47) is the second Virial coefficient expression of two-particle cluster for a general system. As is stated in the Section 5.8, it is necessary to respectively discuss the second Virial coefficient for the two-body clusters of different size at $T_g$ temperature in the critical local cluster growth phase transition to the $i$-th order cluster with cluster volume $v_i$ on $V_i$-percolation field.

It should be noted that the result of Eq. $B_3(v_i) \equiv 5/8$ is still correct, though it is the referenced result from [64], while the testification of [64] is just according to the two preconditions of hard-sphere with positive repulsive force and the random distribution of distance between hard-spheres. The hard-sphere with negative attraction potential and two-body interaction potential in glass transition just independent on the distance-increment between two-body, while the random distribution of distance-increment can be obtained from two random distributions of distance between hard-spheres, and the distance-increment between two-body also corresponds to the higher order hard-spheres in Fig.9, therefore, Eq. $B_3(v_i) \equiv 5/8$ can be obtained from each of the latter.

**5.10. Key role in scaling transformation: Fugacity**

Since in general gaseous system, only the second Virial coefficient for two-body cluster of a certain size is considered, an important physical quantity – Fugacity, $e^{\beta \mu} \equiv K$, is canceled out in the deduction of Eq. (47) from Eq. (46). There are 8 orders of different (relaxation time and size) two-body clusters in glass transition. Fugacity takes a key role in scaling transformation. From Eq. (43) it is found that $P/kT$ has the factor of $e^{\beta \mu}$, therefore, Eq. (47) may be rewritten as

$$B_2(v_i, K_i) = -\overline{b}_2(v_i) e^{\beta \mu_i} = -\overline{b}_2(v_i) K_i \quad (48)$$

Note that $V \to \infty$ as Eq. (47) is derived. This corresponds to the condition of $i$-th order inverse cascade energy flow in a percolation field. Percolation is a phase transition of geometric connection at a certain probability. Due to fluctuation effects, the variation of $K_i$, arose from the number change of the $i$-th order clusters in $V_i$-percolation field, is denoted as $\Delta K_i$ and in a similar way, the variation of $K_{i+1}$ is denoted as $\Delta K_{i+1}$. From Eq. (48)

$$\begin{cases} \Delta B_2(v_i, \Delta K_i) = -\overline{b_2}(v_i)\Delta K_i \\ \Delta B_2(v_{i+1}, \Delta K_{i+1}) = -\overline{b_2}(v_{i+1})\Delta K_{i+1} \end{cases} \quad (49)$$

From Eqs (37), (39), Eq. (40) may be rewritten as

$$B_2(v_i) = \frac{\partial B_2(v_i) \cdot T(v_i)}{\partial T(v_{i+1})}. \quad (50)$$

From Eq. (39)

$$\partial B_2(v_{i+1}) = \frac{P \cdot \partial V_{i+1}(v_{i+1})}{kT(v_{i+1})} \quad (50\text{-}1)$$

In Eq. (50-1), $P \cdot \partial V_{i+1}(v_{i+1})$ represents the *total extra work* made by the ($i+1$)-th order of interface-excited energy loop-flows, and $\partial V_{i+1}(v_{i+1})$ is the total extra volume contributed by the ($i+1$)-th order excitation interfaces. Which should be balanced with the ($i+1$)-th order of thermo-excited motion energy $k\partial T(v_{i+1})$ on $v_{i+1}$-percolation field, i.e. $P \cdot \partial V_{i+1}(v_{i+1}) = k\partial T(v_{i+1})$. And at the critical state before compacting clusters, $v_i$ and $v_{i+1}$ coexist, $kT(v_i) = kT(v_{i+1})$. So, from Eqs (50), (50-1), before compacting clusters, $B_2(v_i)$ has the form

$$B_2(v_i) = \frac{\partial B_2(v_i)}{\partial B_2(v_{i+1})}. \quad (51)$$

From Esq. (49), (51), $B_2(v_i)$ takes the form

$$B_2(v_i) = \frac{\overline{b_2}(v_i)\Delta K_i}{\overline{b_2}(v_{i+1})\Delta K_{i+1}}$$

If there exists fixed point of second Virial coefficient in above equation, denoted as $B_2^*$, according to the definition of fixed point: $\overline{b}_2(v_i) = \overline{b}_2(v_{i+1})$; so, at the critical state before compacting clusters, $B_2^*(v_i)$ is of the form

$$B_2^*(v_i) = \frac{\partial K_i}{\partial K_{i+1}}\bigg|_{K=K_c} \quad (52)$$

### 5.11. Scaling equation of fugacity

*The second step of testification is to deduce the $K_i$ scaling equation.* It has been proved that the ratio of the distance increment $\Delta q_{i+1}$ and the volume $\sigma_i$ between two $i$-th order clusters is constant, controlled by Lindemann ratio $\Delta q_{i+1}/\sigma_i = 0.1047$ and the direction of cluster growth is always along the local $q$-axial.

When the distance between the two $i$-th order clusters on $q$-axial is $q_i$ and the distance fluctuation is $\Delta q_{i+1}$, the two $i$-th order clusters are also the components of two ($i+1$)-th order clusters respectively. This picture should be equally represented by that of randomly distance fluctuation in Fig.13, in which an $i$-th order cluster on origin moves *randomly fluctuation* along a square with the side length of $\Delta q_{i+1}$ and the whole course is just the square on z-axial, i.e., the cluster randomly "walks out" the square alone + z-axial.

In Fig.13, the positive z-axial denotes the $q$-axial, which is also the increment direction of the cluster. The cluster can not evolve reversely midway, but evolve to the highest order (the 8th order) from small to large. That means the cluster can only move randomly along a side of square in the first quadrant. When it reaches the point of z =1 and continues to move along z-axial till outside the square, that means the cluster is moving in the (i+1)-th order cluster and is the component of it, or in other words, the (i+1)-th order cluster has formed.

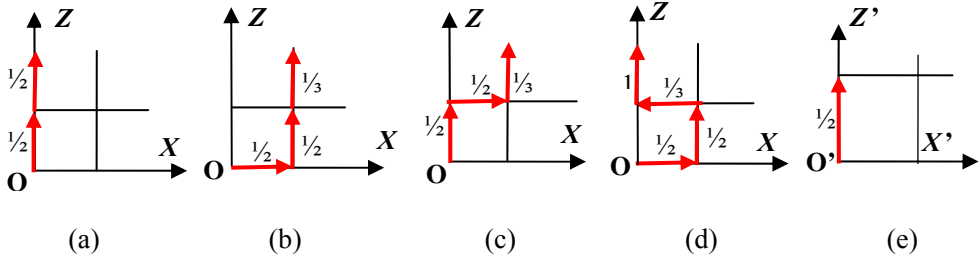

(a)        (b)        (c)        (d)        (e)

Fig.13. Schematic diagram of $K_i$ scaling transformation. Along z-axial, an $i$-th order cluster located at origin O has the 4 kinds of tracks (a), (b), (c), (d) evolving into an ($i$+1)-th order cluster (e) in thermo random motion.

The $K_i$ scaling transformation depends on all (4 kinds of) possible tracks for a reference $i$-th order cluster (located at origin O) evolving into an ($i$+1)-th order cluster (along the positive z-axis, walking out the square). The 4 kinds of tracks are shown in Fig.13 (a-d). The weighting of $K_i$ is shown at the edge of each square in Fig.13. The scaling equation of $K_i$ from Fig.13 (a-d) is in the following form

$$\frac{1}{2}K_{i+1} = \left(\frac{1}{2}K_i\right)^2 + \left(\frac{1}{2}K_i\right)^2\left(\frac{1}{3}K_i\right) + \left(\frac{1}{2}K_i\right)^2\left(\frac{1}{3}K_i\right)K_i$$

$$= \frac{1}{4}K_i^2 + \frac{1}{6}K_i^3 + \frac{1}{12}K_i^4 \qquad (53)$$

The 4 terms on the right side of Eq. (53) respectively represent the contributions from Fig.13 (a), (b), (c), (d) to (e) in scaling transformation. Denoting $K_c$ as the fixed point, from Eq. (53), the fixed point equation is

$$K_C = \frac{1}{2}K_C^2 + \frac{1}{3}K_C^3 + \frac{1}{6}K_C^4 \qquad (54)$$

The solutions of the fixed point from Eq. (54) give: 0, 1, ∞. Because the critical point is always an instability fixed point, thus taking $K_c = 1$.

From $K_c = 1 = e^{\beta\mu}$, we got the chemical potential in subsystem, $\mu \equiv 0$. Substituting Eq. (53) into Eq. (52) and taking $K_c = 1$, the result is

$$B_2^*(v_i) \equiv 3/8 \qquad (i =1, 2…8) \qquad (55)$$

### 5.12. Chemical potentials always zero in all sub-systems

This proof is self-consistent. If the macroscopic random heat energy $kT_g \geq T_g^* \varepsilon_0$, the increased random kinetic energy, $k\Delta T_g$, would increase the numbers of inverse cascade in system. If the mean random heat energy $kT$ satisfies: $1/8\varepsilon_0 < kT < T_g^*\varepsilon_0$, $1/8\varepsilon_0$ here is the interface excitation energy, due to the thermo fluctuation effects, it is possible for the local-excited energy-flows to form the geometric connected filed in a few of local fields within a long time, and low temperature glass transition thus occurs.

*The condition for fluctuation stability is that the chemical potentials are always zero in all sub-systems.* Fugacity $K_i$ is a fixed point, $K_i \equiv 1$. Eq. (44) is still derived from Eq. (48). The number of *i*-th order hard-spheres, $N_i$, in each order sub-system is invariant. Eqs (35) – (36) still hold true in all sub-systems.

Here, it is an interesting and profound theoretical comparison that the chemical potential is defined Fermi energy on the occasion when the temperature is zero in solid physics; whereas the random motion energy, $kT_g$, corresponds to that of the glass transition on the occasion when the chemical potentials in sub-systems are always zero in solid-to-liquid transition.

### 5.13. Kinetic energy always balance with potential energy

From Eqs (34), (37), (38) and (55), an impotent relationship equation has been deduced

$$\frac{PV(v_{i-1})}{kT(v_i)} = B_2(v_i) + B_3(v_{i-1}) \equiv 3/8 + 5/8 \equiv 1, \quad (i = 1,2,...8) \tag{56}$$

Eq. (56) holds true on all sub-systems, which means that the kinetic energy always keeps balances with the potential energy, in the manner of phase difference of $\pi$, in the mode of 8 orders of coupling 2body-3body clusters in the density fluctuation of glass transition.

Eq. (56) also shows that tow-body interaction always slower than that of three-body, in other words, three-body is always firstly compacted in order to minimize the totally interface excitation energy.

### 5.14. The balance picture between inner field and outer field in the density fluctuation

One of the interest balance pictures can be seen that the reduced slower two-body interaction $B_2 \equiv 3/8$ and the reduced faster three-body interaction $B_3 \equiv 5/8$ intrinsically and respectively connect with the fast reduced geometric phase factor $\chi_{+1} \equiv 3/8$ and the slower reduced geometric phase factor $\chi_{-1} \equiv 5/8$ in the glass transition. The reduced geometric phase factor is in fact the outer field contribution to inner field (corresponding to the critical states after compacted clusters), and two-body, three-body interactions are the contributions from inner field (corresponding to the critical states before compacting clusters). Thereupon, in the slow-process cluster fluctuation of inner-outer field, $B_2$ (= 3/8) + $\chi_{-1}$ (=5/8) = 1; and in the fast-process cluster fluctuation of inner-outer field, $B_3$ (= 5/8) + $\chi_{+1}$ (= 3/8) =1, which also refract the two kinds of balances between inner-outer field and fast-slow fluctuation.

### 5.15. The percolation in glass transition is only an infinity sub-system

The percolation in glass transition is only an infinity sub-system, which is connected, one by one, by local fields in excited state. Classical thermodynamics still holds true in each sub-system, but not in the whole system. The conditions for the self-similar Esq. (35), (36) in sub-systems permit heat capacity abnormality. From the solution of the self-similar equations we deduce the reduced localized energy of thermo random motion, $T_g^*$, which is a fixed point and independent of temperature, to characterize the universal intrinsic property in the temperature range from $T_g$ to $T_m$ in thermo random system.

# 6. Theoretical proof of the standard WLF equation in the glass transition

## 6.1. Semi-empirical WLF equation

Glass transition theory is an important subject in condensed-matter physics. There is the well-known semi-empirical Willams-Landel-Ferry (WLF) equation [66]

$$\log \frac{\eta(T)}{\eta_g} = -\frac{C_1(T-T_g)}{T-T_g+C_2} = -\frac{17.44(T-T_g)}{T-T_g+51.6} \tag{57}$$

In which $c_1$ and $c_2$ are two constants for most flexible polymer. Theorists have been trying to prove this time-temperature equivalent equation directly from fundamental theories of the glass transition and to understand the physical meaning of the two constants.

On the other hand, in the model of the intrinsic 8 orders of 2-*D* mosaic geometric structures, the two numerical values obtained directly by geometry method: the 320 interface excitation states and the 136 cooperatively oriented-migration particles, have not been experimentally proved. The aim of this section is to directly deduce the form of standard WLF experimental equation, based on the picture of particle-clusters cooperative migration along one direction, and to validate the two numerical values in the glass transition.

## 6.2. Relaxation expression of fast process stress work

The testification is carried out in three steps: the relaxation expression of fast process stress work; the relaxation expression of slow process stress work and Clapeyron equation in the glass transition.

A series of experiments to measure the relationship between viscosity and temperature can be illustrated as: in the temperature range from $T_g$ to $T_g+100C°$, temperature begins rising at constant speed from the point of $T_g$. In other words, interval of $\Delta T$ a response time $\tau_{res}$, the tensile viscosity of experimental sample is surveyed at constant strain rate. The constant rate can be described as: within the response time $\tau_{res}$, the volume increment of sample is $\Delta V(T)$ under the effect of external stress $\sigma(T)$ and the sample is still in random state without any orientation after it is stretched. Thus, viscosity (Note: the sign $\sigma$ in this section denotes as stress, according with general viscosity relationship equation)

$$\eta(T) = \sigma(T)\,\tau_{res} \tag{58}$$

Disorder-induced localization [60] is one of the fundamental concepts in condensed matter physics. The localized energy induced by thermo-disorder, $E_c(\tau_i)$, is an intrinsically invariable energy with 8 orders of relaxation times $\tau_i$ in the glass transition, independent of temperature $T$. The localized energy $E_c(\tau_i)$, reflecting the intrinsic characteristic in the glass transition, is defined the inverse cascade-cascade energy of excitation interface energy flow. For flexible polymer system, from Eq. (9)

$$E_c(\tau_i) = kT_g°(\tau_i) = 20/3\varepsilon_0(\tau_i) = kT_g°, \qquad (i=1, 2\ldots 8) \tag{59}$$

The inverse cascade has 8 orders, which have the same energy of $20/3\varepsilon_0$, however, since the creation time $\tau_i$ (or relaxation time) is different, the localized energy is denoted as $E_c(\tau_i)$, which shows the potential energy of *i*-th order interface excitation energy loop-flows in inverse cascade and also indicates the kinetic energy of *i*-th order clusters on a $V_i$-percolation ($V_i$-percolation is a field created by the connection of *i*-th order loop-flows, or *i*-th order (self-similar 2body-3body) clusters). Therefore, $E_c(\tau_i)$ can also be denoted as thermo random motion energy, $kT_g°(\tau_i)$, of $i$-th order clusters with relaxation time $\tau_i$.

The convenience of introducing $kT_g°(\tau_i)$ is that: (i) the induced potential of an $i$-th order interface excitation energy loop-flow is a relatively fast process, and the migration motion of $i$-th order clusters surrounded by the $i$-th order interface excitation energy loop-flow is a relatively slow process in topological analysis. It is unnecessary to care about the complex phase difference caused by mosaic structure since kinetic energy and potential energy always keep balance (Section 5.13). (ii) No matter in a local zone or a percolation field and also no matter for the 1-st clusters or the 8th order clusters, the numerical value of localized energy is always $kT_g°$.

In the experiment of tension deformation, only after the 8th order loop-flow appears, can the vacancy volume $\Delta V(T)$ needed in offering slow process cluster-migration or volume deformation appear. The slow induced potential of 8th order loop-flows always equals to that of the fast first order loop-flows on percolation fields.

In Eq. (59), $\varepsilon_0(\tau_i)$ is the potential well energy of $i$-th order cluster, also the relaxation energy of one external degree of freedom of $i$-th order cluster. Note that when the numerical value of $\varepsilon_0(\tau_i)$ is denoted by a certain temperature, $\varepsilon_0$ can be conveniently applied to the whole sample.

The other intrinsic invariable energy for flexible polymer system in the glass transition is the average cooperative migration energy in one direction $E_{mig}$ (Section 2.17):

$$E_{mig} = kT_2 = 17/3\varepsilon_0 \tag{60}$$

The physical meaning of $E_{mig}$ is that, in flexible system, $E_{mig}$ is the average attractive potential of collective motion in one direction of 136 neighboring particles in *an* inverse cascade-cascade to thaw a domain, whose energy numerical value is invariable (= $17/3\varepsilon_0$), independent of temperature $T$, external stress $\sigma(T)$ and the response time $\tau_{res}$. The 8 orders of attractive potentials of $E_{mig}(\tau_i)$ can balance the external stress work, thus, the external stress also has 8 orders of relaxation time $\tau_i$, denoted as $\sigma_i(T)$.

When the average random heat energy provided by outside temperature field is equal to the localized energy $E_c(\tau_8)$, such random motion energy, denoted as $kT_g$, is 'traditionally accepted as glass transition temperature'. From Eqs (59) and (60), the numerical relationship between $kT_g$ and $kT_2$ has the form

$$kT_g = kT_2 + \varepsilon_0 \tag{61}$$

### 6.3. Number of $V_8$-loop-flows to measure random heat energy $kT$

Since the average cooperative migration energy $E_{mig}$ and the localized energy $E_c(\tau_i)$ are all independent of temperature and *increase of temperature only increases the number of $V_8$-loops* (or the number of inverse cascade and cascade) that taking other directions in a reference local zone, so the key idea is that *the number of $V_8$-loop-flows can be used to measure the random heat energy $kT$ in system*:

$$kT = kT_2 + f\varepsilon_0(\tau_8) \tag{62}$$

$f$ here is the average number of degree of freedom with $\tau_8$ in a reference $V_8$ local zone, $1 \leq f \leq 5$.

The key idea here is that in a reference $a_0$ local field, seeing Fig.6, when $kT = kT_g$, $f = 1$, the reference $V_8(a_0)$-loop-flow is the only one that can be excited and provide one external degree of freedom energy $\varepsilon_0(\tau_8)$ to relax the 1-st order 2-D particle-cluster $V_1(a_0)$ in a local z-axial and its 4 neighboring loop-flows are in $V_7$-loop-flow states which can also be excited and provide $f$ external degrees of freedom energy $f\varepsilon_0(\tau_7)$ to relax the $V_1(a_0)$ in $f$ local directions when $kT > kT_g$.

At the instant when an external stress acts upon the test specimen, the specimen's instantaneous response can only come from the interactions among atoms (particles). These interactions include the extra volumes of interface excitation, which are the fractal vacancies with relaxation time $\tau_1$. Total extra volumes are vacancy volumes $\Delta V_1(T)$ formed by total excitation interfaces, of which the suffix 1 denotes the first order in inverse cascade. $\Delta V_1(T)$ excited by external stress $\sigma_1(T)$ are oriented, the fast process stress work, $W_1$, of first order loop-flows has the form

$$W_1 = \sigma_1(T) \cdot \Delta V_1(T) \qquad \text{(On } \tau_1\text{-percolation field)} \tag{63}$$

In Eq. (63), the fast process stress work of $\sigma_1(T) \cdot \Delta V_1(T)$ will fully change to the deformation energy $E_{defor}$ after cascade:

$$E_{defor} = \sigma_1(T) \cdot \Delta V_1(T) \tag{64}$$

In fast cascade process, together with structural rearrangement, the stable relaxation behavior appear, and the relaxation energy, or say, the deformation energy is just the $f\varepsilon_0$ in Eq. (62), thus, the deformation energy $E_{defor}$ at temperature $T$ also has the form balancing with kinetic energy:

$$E_{defor} = f\varepsilon_0 = k(T - T_2) \tag{65}$$

Eq. (65) not only shows the deformation energy needed to return to the random state during an inverse cascade-cascade in local field, but also indicates the deformation energy in delayed stable extensional deformation of the sample. Thus

$$\sigma_1(T) \cdot \Delta V_1(T) = k(T - T_2) \tag{66}$$

On the other hand, a flow-percolation can be considered as a very long 'chain (or a 'net') connected by one after another excited local fields, as same as a macromolecular chain, to bear external stress, each excited local field, as a 'chain' unit, randomly taking a direction of inverse cascade with the invariant attractive potential $E_{mig}$, thus, *flow-percolation is of the configuration entropy stress to balance external stress*.

Any small extensional deformation first needs an increment of orientation entropy stress of 'chain' and then relaxes the orientation work. In the stable extensional deformation, the 'length' of 'chain', i.e., the number of excited local fields with relaxation time $\tau_1$ at temperature $T$ is invariable; thus $\Delta V_1(T)$ is also invariable ($\Delta V_1(T)$ only determined by the total extra volumes in excitation interfaces), the increment work $\Delta W_1$ of extensional deformation that comes from the contribution of configuration entropy stress in flow-percolation may be obtained from Eq. (63)

$$\Delta W_1 = \Delta\sigma_1(T) \cdot \Delta V_1(T) \qquad \text{(On } \tau_1\text{-percolation field)} \tag{67}$$

From (66), (67), the relaxation expression of fast process stress work has the form

$$\Delta W_1 = \frac{\Delta\sigma_1(T) \cdot k(T - T_2)}{\sigma_1(T)} \tag{68}$$

**6.4. Relaxation expression of slow process stress work**

In the stable inverse cascade-cascade, at each inspecting time $\tau_{res}$, by the cooperation of external stress and temperature, the sample returns to the random state after experiencing the processes of orientation and re-orientation (relaxation). The stress work in slow process $W_{slow}(\tau_i)$, $i = 2, \ldots 8$, consists of two parts, of which one is consumed in thawing solid-domains as local orientation work, $E_{mig}(\tau_i) = kT_2(\tau_i)$, and the other is the relaxing work needed for pushing sample deformation in structural rearrangement, $E_{defor}(\tau_i) = f\varepsilon_0(\tau_i)$. In the flow-percolation, the average potential energy of all the loop-flows in percolation field, which contribute to the stress work $W_{slow}(\tau_i)$ is $kT$. Note that Eq. (62) is fit for any $\tau_i$. From Eq. (62), on $\tau_i$ flow-percolation fields, the relationship between $W_{slow}(\tau_i)$ and $E_{defor}(\tau_i)$ has the form

$$W_{slow}(\tau_i) = E_{defor}(\tau_i) + kT_2(\tau_i) \quad \text{(On } \tau_i > \tau_1 \text{ flow-percolation fields)} \tag{69}$$

Eq. (62) is used again on $\tau_{i+1}$ flow-percolation fields, and taking $kT_2(\tau_i) = kT_2(\tau_{i+1}) = kT_2$, $\varepsilon_0(\tau_i) = \varepsilon_0(\tau_{i+1}) = \varepsilon_0$ for flexible polymer, the expression of slow process stress work can be deduced as

$$\frac{W_{slow}}{kT} = \frac{E_{defor}}{k(T - T_2)} = 1 \tag{70}$$

Eq. (70) refracts the two kinds of balances between potential energy and kinetic energy. Since there is a non-integrable phase-induced potential between $i$-th order and $(i+1)$-th order percolation fields, the relationship between $\triangle W_{slow}(\tau_i)$ and $\triangle E_{defor}(\tau_i)$ can not be deduced from Eq. (70). Then the self-similar equation (36), on the $i$-th order of stable extensional stress $\sigma = P$ field (i.e. $i$-th order of flow-percolation field), in the glass transition at $T$ is adopted.

In Eq. (36), $V$ is the volume of $i$-th order of flow-percolation field, in which only $\Delta V$ is the effective volume to work, the total extra volumes of $i$-th order of excitation interfaces; $\partial V$ is the total extra volume-increment transferred from $i$-th order to $(i+1)$-th order. Using $E_{defor} = (\sigma \cdot V)_{defor} = \sigma \Delta V$, $W_{slow} = kT$, $\triangle E_{defor} = \sigma \partial V$, from Eq. (36), Eq. (71) obtained:

$$\frac{\Delta W_{slow}}{W_{slow}} = \frac{\Delta E_{defor}}{E_{defor}} \tag{71}$$

From Eqs (70) and (71), the relaxation expression of slow process stress work may be rewritten as

$$\frac{\Delta W_{slow}}{T} = \frac{\Delta E_{defor}}{T - T_2} \tag{72}$$

**6.5. Evolvement from solid-liquid coexist state to particle-flow**

Now it is time to discuss the first order phase transition of the so-called 'solid-liquid coexist' in the slow inverse-cascade and fast cascade process in a thawing domain, in which temperature changes by $\Delta T$, total extra volumes changes by $\Delta_{s-l}(T)$ and external stress changes by $\Delta\sigma_{s-l}(T)$. It is a microscopic 'solid-liquid coexist state' that once the 8th loop-flow appears in a reference $a_0$ local zone, the $a_0$ particle immediately migrates one step (one particle-distance or a tiny distance, see Section 3.21, 7.4.) along the local z-axial direction. Once $a_0$ particle migrates to a new position, the $a_0$ particle field in the new position will *partially* (because of mosaic structure) lost its 8 orders 2-*D* interface excitation energy loop-flows in local z-axial and change into a '2-*D* fast-slow liquid-solid state'. The re-activation of new interface excitation loop-flows is among the other arbitrary local z'-axial direction, thus, the 'isotropy orient migration' of liquid state in the local zones that follow a reference $a_0$ particle field, will come from the contributions of time after time flow-percolation fields. By the point of view of the rebuilt structure and relaxed orientation, a reference local $V_8(a_0)$-zone in a flow-percolation field (subsystem) first reaches the locally critical 'solid-liquid coexist' conditions to break the reference $a_0$ solid-lattice and migrate $a_0$ particle one step, then other solid-lattices connected with $a_0$ also satisfy the locally critical conditions one by one and follow $a_0$ particle migrating one step to form 'particle-flow with mosaic structure' (solitary wave, see Section 7.4). Therefore, the 'solid-liquid coexist' conditions in the glass transition evolve into that of the 8 orders of continuous fast-slow percolations of particle-cluster flow.

*The singularity of 'solid-liquid coexist' in the glass transition* is that ($i$-1)-th order relative fast-process percolation corresponds to the 'liquid phase' and $i$-th order relative (to ($i$-1)-th order) slow-process percolation corresponds to the 'solid phase', but the $i$-th order is also corresponding to the 'liquid phase' because it is the relative fast-process to ($i$+1)-th order. Especially, there is an unalloyed liquid phase state and an unalloyed solid phase state; the former corresponds to break the reference $a_0$ solid-lattice and the latter to a new 0-th order local cluster of rebuilding structure when 8th order of flow-percolation appears. Upon that, there are 8 orders 'solid-liquid coexist' phases, corresponding to the 8 orders of percolations in the glass transition. The law of first order phase transition of 'solid-liquid coexist' is still valid on the flow-percolation field (this is a subsystem in system).

Applying the Clapeyron equation in a flow-percolation field, that is

$$\Delta\sigma_{s-l}(T) \cdot \Delta V_{s-l}(T) = -\Delta T \cdot \Delta S_{s-l}(T) \tag{73}$$

A negative sign is assigned on the right hand side of Eq. (73) because here $\sigma$ is a tensile stress, not compression stress in liquid-gas phase transition. In Eq. (73), $\Delta S_{s-l}(T)$ is the entropy change of subsystem during solid-lattice broken.

The key point is that the procedure to break solid-lattices is non-ergodic in position and direction; instead, it occurs in the form of an energy-flow as a $V_8$-loop-flow (2-D local lattice) first on one flow-percolation subsystem, in one direction, then on the other flow-percolation subsystem, in other direction, and so on.

Note that the dimension of stress is [energy] / [volume], if all the energy is denoted by temperature, it is not important whether $\Delta\sigma_{s-l}$, $\Delta V_{s-l}$ and $\Delta S_{s-l}$ in Eq. (73) is the value in a $V_8$-zone or the value on a flow-percolation field. What is important is to get an entropy change of 2-D local lattice, so that $\Delta S_{s-l}$ can denote the entropy change of the sample in the tension deformation.

## 6.6. Entropy change of broken solid lattice

On a flow-percolation field, according to the law of critical phase transition of broken solid lattice, $\Delta S_{s-l}$ is of the form

$$\Delta S_{s-l} = \frac{\Delta E_{co}}{T} \tag{74}$$

$\Delta E_{co}$ is just the cooperative orientation activation energy for all particles in a thawed domain, and also the activation energy to break a solid lattice (Section 3.18) that is surrounded by a 2-D $V_8$-loop-flow. $\Delta E_{co}$ is equal to the energy of all 8 orders of interface excitations in a reference particle field, which is

$$\Delta E_{co} = 320\Delta\varepsilon_0 = 40\varepsilon_0 \tag{75}$$

The critical 'solid-liquid coexist' conditions also correspond to the deformation formed by generating vacancies and migrating particles, thus

$$\Delta E_{defor} = \Delta \sigma_{s-l}(T) \cdot \Delta V_{s-l}(T) \quad \text{(On flow-percolation fields)} \tag{76}$$

Combining Eqs (72), (73), (74), (76), $\Delta W_{slow}$ is of the form

$$\Delta W_{slow} = -\frac{\Delta T \cdot \Delta E_{co}}{T - T_2} \quad \text{(On flow-percolation fields)} \tag{77}$$

Each of flow-percolation fields is a subsystem in system, which bears *in turn* external work. Only part of the flow-percolation fields contain a reference 2-D $a_0$ local zone that orients in a certain direction and after considering all flow-percolation fields, all local zones are statistically in random orientation and Eq. (77) is correct.

In the comparison of Eq. (68) and Eq. (77), Eq. (68) means that in the local field, if we track the full inverse cascade and cascade of a fast process stress work, when the sample returns to the random state, the relaxation expression of fast process stress work will be Eq. (68); while Eq. (77) means that when the average is obtained both from time and space, the relaxation expression of slow process stress work is Eq. (77). When the left and right sides of the two equations are denoted by temperature, Eq. (68) and Eq. (77) should be equal to each other. Substitute Eq. (61) into Eq. (68) and Eq. (77) and make right sides of two equations equal, then yield

$$\frac{\Delta \sigma_1(T)}{\sigma_1(T)} = -\frac{\Delta E_{co} \cdot \Delta T}{k\left(T - \dfrac{kT_g - \varepsilon_0}{k}\right)^2} \tag{78}$$

In Eq. (78), $\sigma_1$ is immeasurable, because inverse cascade does not dissipate energy, if $\tau_{res} \geq \tau_8$, in the stable inverse cascade-cascade, we have $\Delta \sigma_1/\sigma_1 = \Delta \sigma \cdot \tau_{res}/\sigma \cdot \tau_{res} = \Delta \eta/\eta$, here $\sigma$ is average external stress, which is measurable, thus, Eq. (78) may be written as

$$\frac{\Delta \eta(T)}{\eta(T)} = -\frac{\Delta E_{co} \cdot \Delta T}{k\left(T - \dfrac{kT_g - \varepsilon_0}{k}\right)^2} \tag{79}$$

$$\int \frac{d\eta(T)}{\eta(T)} = -\frac{\Delta E_{co}}{k} \int_{T_g}^{T} \frac{dT}{\left(T - \dfrac{kT_g - \varepsilon_0}{k}\right)^2} \tag{80}$$

From Eq. (75), thus

$$\ln \frac{\eta(T)}{\eta_g} = -\frac{\Delta E_{co}}{\varepsilon_0} \cdot \frac{T - T_g}{T - T_g + \dfrac{\varepsilon_0}{k}} = -\frac{40(T - T_g)}{T - T_g + \dfrac{\varepsilon_0}{k}} \tag{81}$$

Or

$$\log \frac{\eta(T)}{\eta_g} = -\frac{17.37(T - T_g)}{T - T_g + \dfrac{\varepsilon_0}{k}} \tag{82}$$

Eq. (82) is in the form of the standard WLF equation.

### 6.7. Physical meanings of the constants in the WLF equation

This derivation gives clear physical meanings for both $C_1$ and $C_2$. For flexible-chain system, $kC_2$ turns out to be the potential well energy $\varepsilon_0$, therefore, $\varepsilon_0 = 51.6k$. $C_1$, taking logarithm, is the non-dimension activity energy $\Delta E_{co}/\varepsilon_0$ of broken solid-lattice. The 320 interface excitation states obtained by geometry method and Eq. (61) deduced from the numerical relationship between $kT_g = 20/3\varepsilon_0$ and $kT_2 = 17/3\varepsilon_0$ are also proved by the experimental result of WLF equation.

### 6.8. Nature of the dynamic glass transition

The theoretical proof of WLF equation gives some hints on the nature of the dynamic glass transition. In the subsystems of the glass transition, the Clapeyron equation governing the first order phase transition in thermodynamics still holds true, however, it only holds true in each subsystem that is also an infinite and a slow migration particle-cluster flow on percolation field.

The characteristic of the mean field in the glass transition is that: at a certain temperature $T$, when a representative subsystem that bears external stress is in relaxation state, other subsystem is forming and bearing external stress, and the relaxation by cascade balances that of configuration by $\Delta T$. Clapeyron equation should be *time after time* applied to all the subsystems so as to get Eq. (79). The integration of Eq. (80) implies the dynamics of particle-flow in subsystems in the glass transition.

### 6.9. Glass transition phenomenon is an emergent behavior of the subsystems of system

The multiply repeated applications of Clapeyron equation on the subsystems will result in glass transition. This is the cause of abnormality of the glass transition and reminiscent of the idea of Kadanoff [67] 'all the richness of structure observed in the natural world is not a consequence of the complexity of physical law, but instead arises from the many-times repeated application of quite simple laws'. The theoretical proof of the WLF equation directly validates the 'random first-order transition' theory [19], however, this proof furtherer shows that the dynamic glass transition phenomenon is an emergent behavior or *emergent property* of the subsystems of system.

If we consider the time-temperature dependency of viscosity in approaching glass transition, we would encounter the complexity of viscosity. The complexity of the glass transition is that the energies of interface excitation states represented by 2-$D$ projection plane are in nature implicative of each other in 3-$D$ space. It is convenient, however, to use the concept of 'average cooperative migration energy in one direction' that has been averaged in 3-$D$ space, in some cases, for example, in the WLF equation, in the discussions about 'viscosity of corresponding states'.

### 6.10. Theoretical explanation for applying ranges of WLF equation

If $f = 3$, i.e., $kT = kT_g + 2\varepsilon_0$, in a reference local zone, the minimum energy manner of cooperative migration of particle-cluster may will change as that: since among the three coordinate directions of local x, y, z coordinate system, $E_{mig}$ appears in turns, the isotropy orient migration in local zone can realize by several rearrangements and it is difficult to form long-distance 'chains' connected by many oriented local fields to bear external stress, Eq. (79) will lose effect. Therefore, the applying ranges of WLF equation are $T_g \sim T_g + 2\varepsilon_0 \approx T_g + 100°C$.

### 6.11. Average cooperative migration energy and Gibbs critical temperature

The energy of $kT_2$, not only can denote the average random kinetic energy in space and in time, but also can denote the average cooperative migration potential energy in local zone, which is similar to the energy of Curie temperature in magnetism. $kT_2$ here is also the energy of a 'critical temperature' existing in the glass transition presumed by Gibbs based on thermodynamics years ago [47]. The same denotation of $kT_2$ is used as Gibbs did. However, from our discussion, the energy of $kT_2$ is also equal to that of average ordering attractive potential $E_{mig}$, which can be exactly obtained through the geometric method in the Section 3 in this paper, without additional assumptions.

**6.12. Average energy of interface excitation state for flexible polymer**

For flexible polymer system, the average energy of interface excitation state in glass transition, $\Delta\varepsilon$, can also be obtained from WLF equation,

$$\Delta\varepsilon = 1/8c_2 k \approx 6.45k \ (\approx 5.6 \times 10^{-4} \text{ eV}). \tag{83}$$

This value also accords with the result of the Eq.(16) on line-measurement in high speed spinning.

# 7. Theoretical proof for the 3.4 power law of viscosity of macromolecules
## 7.1. Modification for Reptation model using the mode of multi-chain motion

The de Gennes' reptation model is a single chain model. The reptation model predicts that the viscosity of an entangled polymer melt is proportional to the cube of the chain-length $N$ for $N > N_c$ ($N_c$ is the critical chain length):

$$\eta \sim N^3 \tag{84}$$

However, the experiments showed that the exponent in the entangled regime is about 3.4 for all linear entangled polymers (experimental error analysis: taking $N = 200 \sim 1000$, $N^{3.4}/N^{3.3} \approx 1.7 \sim 2$; that means the theoretical value should be in the range of $3.4 \pm 0.03$). In order to explain and verify the mode of the multi-chain macromolecule system, it is important to find out the reason for this deviation from the exponent of 3.

Derivation of Eq. (84) is based on the assumption that the chain of length $N$ is a 'free chain', that is, a test long-chain can move *freely* in the melt. In the derivation, the number of degrees of freedom, $N^*$, of a test chain length $N$, is *implicitly* proportional to the chain length $N$.

It takes the co-operative motion energy of $n$ degrees of freedom for one 'free chain-particle with $n$ degrees of freedom' to diffuse one step. If $n$ is independent of $N$, the relationship between viscosity $\eta$ and chain-length $N$ is exactly shown in Eq. (84) deduced by de Gennes using reptation model. The complexity of the 3.4 power law results from the fact that $n$ depends $N$.

(i) In polymer physics, a more accurate concept and definition of the statistical chain-segment to describe the cooperative motion for chain-particles is proposed in Section 3.5. The mode at the minimum energy to excite motion in system shows that: according to extending Mott disorder theory (Section 2.2.), in a reference 3-*D* domain (local space), all the interface cross -coupled electron pairs only first occur in local z-space to maximally avoid the outside electrons each other in order to minimize the total dominant delocalized potential energy. The scale of a 3-*D* domain or local space is the scale of the disorder-induced localization. Accordingly, in a reference 3-*D* domain, when a chain-particle *walks* one step along z-axial, all the chain-particles in the domain must first be considered together with their cooperation motion in the z-space. In other words, within the entire temperature range from $T_g$ to $T_m$, on a reference chain of length $N$ that is randomly distributed in 3-*D* space, the cooperative motion to move one step for $N$ chain-particles can only be in a way that all the $N$ particles move along one direction, e.g. along $\pm z$-axial. Therefore, the cooperative motion of the $N$ chain-particles cannot be directly based on the diffusion motion of a chain-particle in any random direction. In order to represent the motion characteristic for a random long-chain macromolecule, $N_z$ is used to denote the long-chain connected by $N$ covalent bond z-components on a reference long-chain $N$ in 3-$D$ space.

The cooperative motion of $N_z$ particles on a long-chain $N$ is categorized into 8 orders, defined as 8 orders of chain-segments. The $i$-th order of chain-segment is marked as $l_i$. The number of chain particles, the relaxation time and the cooperatively $2\pi$-twisting direction of the $i$-th order of chain-segment are all determined as in Eq.(5).

(ii) From Section 3.5 and Eq.(5), the one-step-walk along +z-axial, of a reference chain-particle $a_0$ on a reference long-chain $N_z$, results from that the action of the 8 orders of chain-segments, cooperatively from short to long and from fast to slow, on the particle $a_0$ in $\pm z$-direction. Such action is *back* a minus $2\pi$-twisting and *forth* a plus $2\pi$-twisting *along the z-space topological direction* of the long-chain.

(iii) Only when the 8th order of chain-segment appears and *acts* on the particle $a_0$ (here the action refers to the compaction in z-direction of all the extra vacancy volumes in covalent bond z-components, transferred from interface excitations, in the 8th order of chain-segment, which is similar to the compacting clusters in Section 3.3, 3.19), should one 'particle-cavity' in +z-direction (the so-called 'defect' in de Gennes reptation model) appear in local field of $a_0$ and bring the particle $a_0$ to migrate (or hop, walk, move) one step along local +z-axial.

(iv) Statistically, the number of particles acting to a reference particle $a_0$ can not only be the number of particles, $\sigma_i$, in the 8 orders of hard-spheres in local excited field of $a_0$, as in Eq. (4); but also the number of particles, $l_i$, in the 8 orders of chain-segment on reference long-chain $N_z$, as in Eq. (5), i.e., $\sigma_i = l_i$.

(v) Each chain-segment is a $2\pi$-twisting vector with one chain-topological direction along which its entire chain-particles move (Section 3.5) on long-chain $N_z$. The number of chain-particles in the 8th order of chain-segment is actually the number of chain-particles in the critical entangled macromolecule, $N_c$, i.e. $N_c = l_8 = \sigma_8 = 200$, (Section 3.8).

The energy for cooperative motion of $N_c$ chain-particles along the z-direction is the intrinsic character energy in random system, $E_c(\tau_8)$. $E_c(\tau_8) = kT_g°(\tau_8)$, see Sections 3.10, 3.13, 3.14. Specifically, the character energy, $E_c(\tau_8)$, to excite a reference particle $a_0$ moving one step along z-axial, is the energy for inverse cascade and cascade of $\sigma_8$ excitation particles in the reference $a_0$ local excited field. Statistically, it is also the energy for inverse cascade and cascade of $l_8$ excitation chain-particles on the reference long-chain $N_z$ (when $N_z \geq N_c$).

(vi) The singularities of glass transition reveal that within the entire temperature range from $T_g$ to $T_m$, the four different concepts of energy in condensed matter physics, i.e., the energy of thermo-disorder induced localization, the energy of mobility edge, the critical energy of percolation and the transfer energy from inverse cascade to cascade, converge to *the same one concept*, a single constant energy, $E_c(\tau_8)$. $E_c(\tau_8)$ is actually an intrinsic character energy in random system independent of the temperature, and is also the localization energy in the 8th order of 2-$D$ (interface excitation) energy loop-flow in a local excited field.

(vii) When $N_z = N > N_c$, statistically, the cooperative energy $E_c(\tau_8)$ of the 200 excitation particles in a reference particle $a_0$ local excited field can be transferred to the local excited field of some other particle on the reference long-chain $N_z$ with equal probability. In order to minimize the number of the 8th order of 2-D loops, the $N_z$ chain-particles on the reference long-chain $N$ should tend to move along z-space together with the reference particle $a_0$. That is, the $N$ chain-particles should *share* the energy $E_c(\tau_8)$ of the 8th order of 2-D energy loop-flow in the reference $a_0$ local field. Therefore, the number of degrees of freedom of the reference particle $a_0$ is dependent on $N$.

(viii) Notice that the energy of the reference particle $a_0$ moving one step along z-axial is the energy of the 8th order potential well, $\varepsilon_0(\tau_8)$, with relaxation time $\tau_8$. $\varepsilon_0(\tau_8)$ is also the unit energy of degree of freedom with scale time $\tau_8$. The external degree of freedom of the 8th order of 2-D energy loop-flow is also one. Therefore, the 8th order of 2-D energy loop-flow is provided with the external motion energy of $\varepsilon_0(\tau_8)$, which is the non-integrable phase induced energy of the 8 orders of loop-flows to induce particle $a_0$ moving along z-axial (Section 7.1. (iii)).

(ix) The statistical effect in local excited field

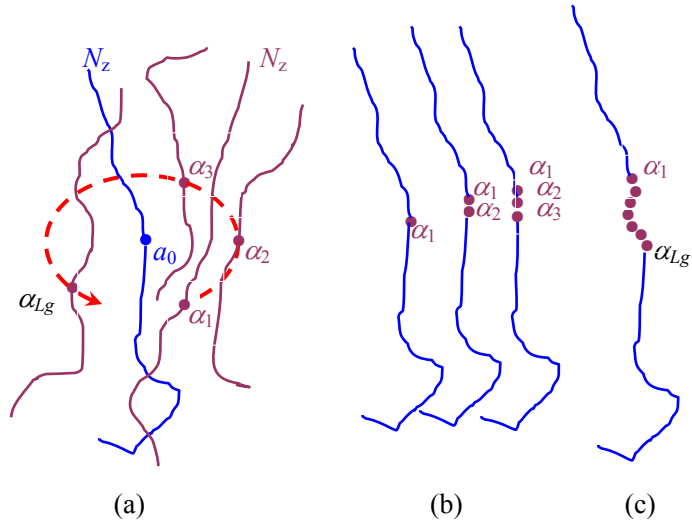

Fig.14. The equivalent particles for reference free particle $a_0$. (a) The one-by-one appearance of $L_g$ equivalent particles ($\alpha_1, \alpha_2, \alpha_3 \ldots \alpha_{Lg}$) in time; (b) the gradually increasing energy to assist $a_0$ moving along +z-axial. $L_g$ equivalent particles form an 'equivalent particle energy-flow' (the red loop-arrow), which is equivalent to an 8th order of 2-D loop-flow. (c) After the substitution of $a_0$ with $L_g$ equivalent particles (pansy dots), particle $a_0$ moves freely in local z-axial. The blue curve in (c) represents an unexcited reference long chain $N_z$. When chain $N_z$ is excited, the energy of the $L_g$ equivalent particles should be shared with the $N_z$ particles, and the number of degree of freedom and of equivalent particles for $a_0$ should be determined as $n_z$ (Eq. 93).

Statistically, the migrating motion along z-space of the 200 neighboring particles in reference $a_0$ local field should share the energy $E_c(\tau_8)$ with particle $a_0$. Due to cage effect, even if particle $a_0$ has obtained a cavity with one-step-walk, the motion of particle $a_0$ should still be correlated with the 200 particles in its local field. In other words, the neighboring particles still 'drag' the one-step-walk of the particle $a_0$.

In order to make particle $a_0$ move *freely* in z-axial in $a_0$ local field, let $E_c(\tau_8)/\varepsilon_0(\tau_8) = L_g$, where $L_g$ is the equivalent number of particles taking $\varepsilon_0(\tau_8)$ as the energy unit for degree of freedom. If the number of degree of freedom of the particle $a_0$ has taken into account the 'dragging action' of $L_g$, the motion of $a_0$ is free in z-axial in the local field. The reverse is also true. Let the equivalent particles with $L_g$ number of degrees of freedom, or, the 'equivalent particle-flow' (the pansy dots in Fig.14 (c)) formed from the $L_g$ equivalent particles, substitute the reference particle $a_0$ on the reference long-chain (the blue line in Fig.14 (a)), and the equivalent particle-flow will enable particle $a_0$ to move freely in z-axial in the local field on the new substituted long-chain.

The free motion in z-axial of a reference particle $a_0$ in de Gennes tube is equivalent to the motion of $L_g$ equivalent particles appearing one by one in time, as shown in Fig.14 (b). This shows statistically that the $L_g$ equivalent particles in $L_g$ tubes in multichain system also move one by one in time, which implies that all the equivalent particles share the same energy $E_c(\tau_8)$. The multichain reptation effect is reflected substituting the reference particle $a_0$ with the 'equivalent particle-flow' of chain-length $L_g$. However, such substitution only eliminates the statistical effect in local field.

(x) The statistical effect in z-space

The migration of each chain-particle is contributed by the 200 z-components in its local z-space. Therefore, each step of migration along z-direction for $N_z$ chain is the co-Brownian motion by its 200 $N_z$ chain-particles in z-space. In order to further eliminate the statistical effect in z-space, for the same reason in 7.1. (ix), the energy of the $L_g$ equivalent particles moving in z-space in Fig.14 (c) is still statistically shared with the $N_z$ z-components on a reference long-chain $N_z$. Substituting the long-chain $N_z$ with the '*equivalent particle-chain*' formed from $N_z^*$ equivalent particles, we will find out the number of degree of freedom $N_z^*$ for the long chain to move along in z-space.

(xi) The statistical effect in 3-D space

In the same way, in 3-*D* space, when a chain-particle *diffuses* one step randomly in any direction, the cooperative motion of all the chain-particles in system should respectively be taken into account in x-, y- and z-space. That is, substituting the long-chain $N$ in de Gennes tube with the '*equivalent particle-chain*' formed from $N^*$ equivalent particles, we will find out the number of degree of freedom $N^*$ ($N^* = N_z^* \cdot N_y^* \cdot N_x^*$) for its reptation in 3-*D* space.

Accordingly, only through substitution of the reference chain of length $N$ in de Gennes reptation model with an equivalent particle-chain of length $N^*$, can the diffuse motion of the reference chain $N$ be entirely free in tube.

$$\eta \sim N^{*3} \qquad (85)$$

Eq. (85) is the adjustment of Eq. (84), using the 'multi-chain reptation model' (equivalent particle-chain of length $N^*$) instead of the single chain (chain of length $N$) model.

## 7.2. The number of degrees of freedom $N^*$ for chain of length $N$ when $N > N_c$

In polymer physics, it has been an open issue to find the number of degrees of freedom for a chain with length $N$.  Using the mode of the inverse cascade – cascade, of the multi-macromolecular system presented in this paper, we can deduce the non-linear scaling relationship between $N^*$ and $N$, when $N > N_c$.  The energy of $N^*$ degrees of freedom, for chain of length $N$, is contributed by all neighboring chains in system.

(i) On the one hand, from the viewpoint of a reference particle $a_0$ on a reference long-chain $N_z$, the $N_z$ chain-particles on long-chain $N_z$ share the energy $E_c(\tau_8)$ (Sections 7.1. (vii), (viii)), thus, the probability that the reference particle $a_0$ is possessed of one unit degree of freedom on the reference long-chain is $1/N$.

(ii) On the other hand, from the viewpoint of the $L_g$ equivalent particles in the reference particle $a_0$ local field, statistically, each equivalent particle is located at its own long-chain with length $N_z$ (Fig.14 (a)), and the probability that each equivalent particle is possessed of one unit degree of freedom on its own long-chain is also $1/N$.

(iii) From Fig.14 (c), when the particle $a_0$ on long-chain $N_z$ is substituted by $L_g$ equivalent particles, the probability of the event that it obtains one unit degree of freedom and *freely* migrates one step along z-direction, denoted as $p_+(a_0)$, is namely the probability of the event that *all* the $L_g$ equivalent particles migrate one step along z-direction in the co-Brownian motion by its 200 $N_z$ chain-particles.  Therefore, $p_+(a_0)$ is given in Eq. (87):

$$p_+(a_0) = (\frac{1}{N})^{L_g} \qquad (87)$$

Where $\quad L_g = \dfrac{kT_g^\circ(\tau_8)}{\varepsilon_0(\tau_8)} \qquad (88)$

In Eq. (88), $kT_g^\circ(\tau_8) = E_c(\tau_8)$.  $kT_g^\circ(\tau_8)$ is exactly the localization energy (Section 3.10; 5.6), also the critical flow-percolation energy, or the mobility edge.

(iv) From 7.1. (ix), the energy of the $L_g$ equivalent particles is $E_c(\tau_8)$, which is also the localization energy in the 8th order of 2-D energy loop-flow excited in inverse cascade and cascade.  The one-by-one appearance of the $L_g$ equivalent particles in time can be regarded as the cooperative formation of an 'equivalent particle-flow'.  In other words, an 'equivalent particle-flow' is statistically equivalent to an 8th order of 2-D energy loop-flow.  Or, the appearance of one 'equivalent particle-flow' is equivalent to that of one 8th order of 2-D energy loop-flow (Fig.14 (a)).

In Section 2.18, we emphasized that the generation of the reference 8th order of 2-D energy loop-flow, in a reference $a_0$ local excited field, always comes from the contributions of its 4 neighboring 5-particle cooperative local excited fields. Therefore, the probability $p_+(a_0)$ can be also regarded as the probability that one 8th order of 2-D energy loop-flow *appears* in a reference local field among all the 8th order of 2-D energy loop-flows in co-Brownian motion by 200 $N_z$ chain-particles in z-space.

(v) Statistically, the probability of each particle freely migrating one step in z-space should all be equal to $p_+(a_0)$.  To maintain the balance of motion between inverse cascade and cascade in z-space, the probability that one 8th 2-D energy loop-flow *disappears* among all the 8th order of 2-D energy loop-flows in cascade in co-Brownian motion by 200 $N_z$ chain-particles in z-space is denoted as $p_-(a_0)$, and it is obtained as follows.

(vi) The statistical effect in conformational rearrangement of the long-chain $N_z$

Let the reference chain particle $a_0$ have $n_z$ degrees of freedom with $\varepsilon_0(\tau_8)$ as the unit in its cascade motion. $n_z$ is actually the number of degrees of freedom to make conformational rearrangement of the long-chain $N_z$ in z-space, which comes from the contribution of the 8th order 2-D energy loop-flow transferred from the reference particle $a_0$ to the other chain-particle on the long-chain.

Similar to Eq. (87),

$$p_-(a_0) = (n_z)^{L_m} \tag{89}$$

In Eq. (89), $L_m$ is the number of 'equivalent particles' to *eliminate* the reference 8th order of 2-D energy loop-flow, contributed by the reference particle $a_0$ with $n_z$ degrees of freedom, after the co-Brownian motion of the 200 $N_z$ z-component particles.

(vii) In melt, each of the $N_z$ local excited fields has 5 degrees of freedom with $\varepsilon_0(\tau_8)$ as unit degree of freedom in the event its 8th order cluster disappears, that is, when *its cascade motion starts*.

Similar to Section 7.1.(ix), since $N_z > N_c$ and due to the strong-weak bond resonance effect, the starting kinetic energy $kT_m(\tau_8)$ in the cascade motion in a reference particle $a_0$ local excited field can be entirely transferred to the covalent-bond long-chain, in the manner of 8 orders of kinetic energy $kT_m(\tau_i)$ (Section 3.11. Notice: numerically, $kT_m = kT_m(\tau_8) = kT_m(\tau_i)$, the energy of the macroscopic melting state, the kinetic energy of *i*-th order of chain-segment, $i = 1, 2\ldots 8$), in order to eliminate the 8th order of 2-D energy loop-flow with the mode of 8 orders of cascade motions from top to bottom. $kT_m(\tau_8)$ is independent of the experiment temperature. $T_m(\tau_8)$ is exactly the melting phase transition temperature described in Section 3.11., or in Eq. (62):

$$kT_m(\tau_8) = kT_g^*(\tau_8) + 4\varepsilon_0(\tau_8) \tag{90}$$

Similar to Eq. (88)

$$L_m = \frac{kT_m}{\varepsilon_0(\tau_8)} \tag{91}$$

(vii) In the complex inverse cascade-cascade motion, the equilibrium condition for inverse cascade - cascade motion is

$$p_+(a_0) = p_-(a_0) = p \tag{92}$$

As a result

$$(n_z)^{L_m} = (\frac{1}{N})^{L_g}, \text{ or}$$

$$n_z = (\frac{1}{N})^{\frac{T_g^\circ}{T_m}} \tag{93}$$

(viii) The number of degrees of freedom of $N_z$ chain-particles on chain of length $N_z$ to make conformational rearrangement in z-space is $N_z^*$, $N_z^* = n_z N_z$

$$N_z^* = N^{(1-\frac{T_g}{T_m})} \tag{94}$$

$N_z^*$ is namely the number of degrees of freedom of chain of length $N_z$ to have free motion in z-space. In Eq. (94), $T_g^\circ$ has been approximately replaced by its glass transition temperature $T_g$.

(ix) In each of the 8 orders of cascade motions from top to bottom, each local excited field has 5 degrees of freedom of $i$-th order with energy $\varepsilon_0(\tau_i)$ and relaxation time $\tau_i$ (Section 2.15). In melt state, in cascade, the local excited field of the reference particle $a_0$ has 5 degrees of freedom with energy $\varepsilon_0(\tau_8)$, but $a_0$ has only $n_z$ degrees with respect to the long-chain $N_z$.

(x) In Figure 6, after the reference chain-particle $a_0$ migrates one step along the local z-direction, it takes a long latency time for $a_0$ to rebuild another local excited field along an arbitrary direction. That is, in Figure 6, only after the 200-th particle has completed the local z-direction migration, is it possible for particle $a_0$ to obtain a new excitation interface. The reason is that the minimum excitation energy condition determines the priority of the first interface excitation (Section 2.4.). After the $a_0$ particle, the $b_0$ particle is excited, and then the $c_0$ particle … until the 200-th particle is excited. Thus, within the required walk time during which the reference long chain $N_z$ walks one step along z-space, denoted noted as $t_{mig}(N_z)$, the $a_0$ particle cannot obtain new local excitation energy from the $n_z$ degree-of-freedom energy, i.e. the particle $a_0$ cannot *inverse* cascade along another arbitrary direction during the time $t_{mig}(N_z)$.

The $N_z^*$ degrees of freedom for chain length $N_z$ is the balance condition to maintain the long-chain $N_z$ inverse cascade – cascade and reset configuration in the z-space.

(xi) Therefore, during the reptation of long-chain $N$ in the 3-$D$ space, the number of degrees of freedom $N_x^*$ in the x-axial and $N_y^*$ in the y-axial have $N_z^* = N_y^* = N_x^*$, or

$$N^* = N^{3(1-\frac{T_g}{T_m})} \tag{95}$$

In Eq. (95), $N^*$ is the number of degrees of freedom of long-chain $N$ when $N > N_c$

For flexible chain polymers, from Eqs (90) and (10),

$$N^* = N^{\frac{9}{8}} \tag{96}$$

**7.3. The 3.4 power law of entangled macromolecules**

Substituting Eq. (95) into Eq. (85), we obtain the general power law expression of viscosity in macromolecular entanglement

$$\eta \propto N^{9(1-\frac{T_g}{T_m})} \quad \text{(For general polymer system)} \tag{97}$$

or

$$\eta \propto M^{9(1-\frac{T_g}{T_m})} \quad \text{(For general polymer system)} \tag{98}$$

For flexible chain polymers, from Eqs (96), (98)

$$\eta \propto M^{\frac{27}{8}} = M^{3.375} \approx M^{3.4} \quad \text{(For flexible polymer system)} \tag{99}$$

This theoretical result conforms well to the experimental data. For non-flexible chain polymer, Eq. (97) can be verified. For example, for the polypropylene (PP), the glass transition temperature of PP is $-10\,°C = 263\,k$; the melt phase transition temperature of PP is $176\,°C = 449\,k$; therefore, the theoretical value given $\eta_{pp}(\text{theory}) \sim N^{3.73}$, conforms well to the experimental data[68]: $\eta_{pp}(\text{experimental}) \sim N^{3.72}$.

### 7.4. Nature of macromolecular entanglement

Eq. (96) can be used to estimate the number of degrees of freedom, $n$, of a chain-particle on macromolecular long-chain. When $N = 2^8$, $n = N^*/N = 2$; and $N = 3^8$, $n = N^*/N = 3$. The chain-length of the entangled macromolecules generally used in experiments is in the range from 250 to 6500. The average number of degree of freedom of each chain-particle is indeed as estimated in most literature, i.e., $n$ is about between $2 \sim 3$. However, only Eq. (96) reveals that $n = N^*/N = N^{1/8}$, and deduces the non-linear scaling relationship between $n$ and $N$.

The theoretical proof for the power law of viscosity of entangled macromolecules describes the entanglement structure as follows:

(i) When $N > N_c$, each one-step-walk of a chain-particle requires the cooperative 8 orders of inverse cascade and cascade in the walk-direction by 200 ($= N_c$) chains with chain-length $N$. Among all the 8 orders of inverse cascade and cascade formed by the 200 $N$ chain-particles, the appearing of the first 8th order of 2-$D$ energy loop-flow first produces a cavity for the chain-particle to walk one step. After that, the energy of this loop-flow is shared with the 200 $N$ chain-particles. Eentanglement is not the so-called topology restrictions or topology structure as 'clew'. The entanglement structure is exactly the appearance and the distribution in space-time of all the 8th order of 2-$D$ energy loop-flows formed by the 200 $N_z$ chain- particles, see following.

(ii) Esq. (93), (94) and Section 7.1. (ii), (iii) reveal a z-direction *solitary wave* moving $N_z$ steps on a random long-chain $N_z$ from one end to other end. The distinct character of the solitary wave is that the energy of each one-step-walk is only $n_z\varepsilon_0(\tau_i)$, ($i = 1, 2\ldots 8$), and the step-size of each one-step-walk is only the scale of $n_z$ particles. $n_z$ is a small numerical value, e.g., if taking $N = 256$, from Eqs (10), (90), (94), $n_z = 0.03125$. However, in Eq. (93), $n_z$ connects with the chain-length $N$ in each one-step-walk, the length-boundaries of solitary wave. The foremost character of solitary wave has been revealed that the wave-particle duality of solitary wave directly results from the two equal probabilities, $p_+(a_0)$ and $p_-(a_0)$ in Eq. (92). The determinate particle-energy of solitary wave is $n_z\varepsilon_0(\tau_i)$ and the number of step of its traveling wave is $N$.

(iii) Esq. (93), (94) also distinctly reveal the minimum energy manner of cooperative orient migration of the z-component of an entangled macromolecule $N_z$, is the model of solitary wave. For example, for $N = 256$, the delocalization energy of a chain-particle is only $0.03125\varepsilon_0 = 0.25\Delta\varepsilon$ which is smaller than far the interface excitation energy, $\Delta\varepsilon$, and the one-step migration energy of a whole long-chain $N_z$ is only $8\varepsilon_0$. Comparing with the delocalization energy of two-body in Section 4.15, the re-coupling delocalization energy of two $\sigma_1$-clusters (each $\sigma_1$ contains 17 particles) is $1/8\varepsilon_0$, the interface excitation energy.

(iv) The reptation in 3-$D$ space of an entangled macromolecule chain is assembled in time by the

three solitary waves on the chain in x-space, y-space and z-space, which comes from the characteristic of Brownian motion.

Accordingly, so-called entanglement structure is namely the localized 8 orders of transient 2-D mosaic geometric structures (thermo-excited ripplon) on x-y projection plane and the delocalized solitary wave in z-space.

**7.5. Chain- length corresponding to Reynolds number in hydrodynamics**

The theoretical proof further confirms that inverse cascade – cascade mode is the fundamental motion mode in the solid-to-liquid transition whether in macromolecular system or in small molecular system. Since cascade is also the fundamental mode in turbulence, Eq. (96) predicts that there is intrinsic relationship in scaling law between flexible macromolecular motion and homogeneous isotropic turbulence.

In order to relate the entangled macromolecular motion with turbulent flow, we discuss the total number of degree of freedom $N_L$ to excite multi-chain reptation, in the cube with edge length of $L$ ($L \geq N^*$) in entangled polymer melts. The unit for the length of $L$ here is the loop scale of the 8th order of 2-D energy loop-flow in co-Brownian motion

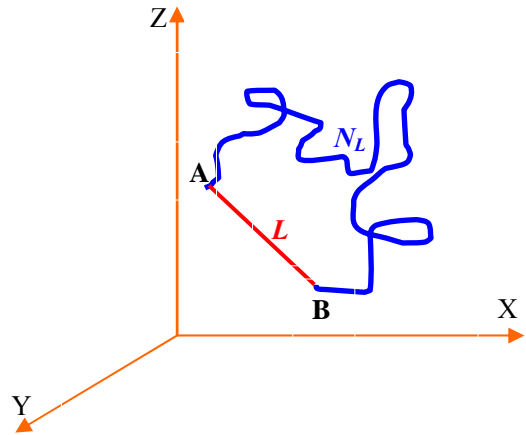

Fig.15. The unit of length of 'chain-length' $N_L$ of a 'random-walk chain' (the blue curve) is the loop scale of 8th order of 2-D energy loop-flow in 3-D space; the length of end-to-end distance (the red line) is $L$. $N_L = L^2$. $N_L$ is the total number of degree of freedom in cube with edge length $L$ in entanglement melts.

It is easy to find out the numerical value of $N_L$ using the properties of flow-percolation proposed in this paper (Sections 2.16., 4.19.). To determine the number of degree of freedom is namely to find the number of the appearance of the 8th order 2-D energy loop-flows, because each degree of freedom corresponds to the appearance of an 8th order of 2-D energy loop-flow. From Section 2.16., 4.19., flow-percolation is a 'random walk' energy flow connected by the 8th order of 2-D energy loop-flows which randomly appear one after another in 3-D space. In Fig.6, the connection between two neighboring 8th order loop-flows is a 'free-connection' between their excitation interfaces. Besides, each of the four neighboring 8th order loop-flows (each of 4 $V_7$–loops in Fig.6 evolves to its own $V_8$–loop) surrounding the center loop can be in any direction. Based on these analyses, simple way to obtain $N_L$ is as follows.

In Fig.15, in 3-*D* space, consider any one 'random-walk chain' (the blue curve in Fig.15) with 'chain-length' $N_L$ and with each 'chain-particle' of one unit length (the loop scale of 8th order of 2-*D* energy loop-flow). If the end-to-end distance (the length of line AB in Fig.15) of the random walk curve is always equal to *L*, *L* is then the root-mean-square of end-to-end 'chain-length' $N_L$ in polymer physics, that is, $N_L = L^2$.

When $L = N^*$, $N_L = (N^*)^2$. This means the probability density of degree of freedom in unit volume is $N_L / (N^*)^3 = 1/N^*$. The $N^*$ degrees of freedom acting on the long-chain *N* are indeed appearing and acting one by one. They form an 'equivalent particle-chain', which is determined by the character of Brownian motion.

And when $L \geq N^*$, $N_L \sim (N^*)^2$, from (96)

$$N_L \sim N^{9/4} \qquad (L \geq N^*) \tag{100}$$

Eq. (100) can be compared with the famous relationship [69] between the number of degrees of freedom and Reynolds numbers $R_e$ in turbulent flow in Eq (101).

$$N_R \sim R_e^{9/4} \tag{101}$$

It can be seen that the two terms in two different subjects: entanglement and turbulence both describe the same complex motion phenomena about inverse cascade – cascade. Both of them obey the same scaling law which indicates there is a universal theory behind them.

The macromolecular chain-length *N* corresponds to the Reynolds number $R_e$ in the universal entanglement / turbulence theory. Reynolds number is actually the ratio of the inertia force to the drag force, and the macromolecular chain-length also reflects the ratio of the number of the particles 'inertially' sharing the characteristic energy $E_c(\tau_i)$ to the one particle entangled with them in motion. The critical entanglement chain-length corresponds to the critical Reynolds number in the concept (rather than numerical value) of the critical point when entanglement / turbulence appears. The major difference between them is that: in randomly melt state (liquid state), with the length of entangled macromolecule expanding, the system slowly turns from the 'melt disorder state' to a new type of 'order in disorder state', during which the two characteristic energies in random system, $E_c(\tau_i)$ and $kT_m(\tau_i)$, also increase slowly.

## 8. The physical origin for turbulence
### 8.1. The universal theory of order to disorder transition for particle-cluster flowing

Notice that the theory of solid-to-liquid glass transition proposed in this paper, whether for small molecule or for macromolecule, does not need any presupposition and relevant parameter. The primary foundation is Mott disorder theory and de Gennes theory. If we consider the solid-to-liquid transition as the transition from order to disorder regarding particle-cluster *moving*, then the proposed theory can be considered as a universal theory for any liquid flow-transition 'from one disorder system to another more disorder system', that is, the self-similar multi-levels of the intrinsic 8 orders of transient 2-*D* mosaic geometric structures in disorder system. It also provides a new explanation of physical origin for turbulence.

The interface excitation state and the excitation interface introduced in Section 2 can be called as the *ground state* interface excitation state and the *ground state* excitation interface. The intrinsic 8 orders of transient 2-*D* mosaic geometric structures can be called as the *ground state* 8 orders of 2-*D* mosaic geometric structures, or the first level of 8 orders of 2-*D* mosaic geometric structure. The 8 orders of 2-*D* interface excitation energy loop-flow can be called as the *ground state* 8th order of 2-*D* energy loop-flow.

What is discussed in Section 2-7 has not involved the concept of energy level. However, for each energy level of small molecular, e.g. for the first excitation energy level, there is still a few Van der Waals repulsive cross-coupled electron pairs that can be excited one by one on the interfaces. Such interface excitation state is called as *the first energy level* interface excitation state, and such interface as *the first energy level* excitation interface. And in the first energy level excitation field, a new intrinsic 8 orders of transient 2-*D* mosaic geometric structure can be formed, which is called the *second level* of 8 orders of 2-*D* mosaic structure.

(i) For small molecule, in the liquid state after the glass-liquid transition, only about 1/3 ($\Delta E_{co}$ = 320$\Delta\varepsilon$; $8kT_m(\tau_8)/\Delta E_{co}$ = 1/3) interface cross-coupled electron pairs have been excited in system. As disorder increases, more ground state excitation interfaces of 8 orders 2-*D* loop-flows will be excited to accelerate liquid flow, which corresponds to the ordinary flow development.

(ii) As disorder keeps increasing and flow continues speedup, each interface on the ground state 8th order of 2-*D* energy loop-flow (i.e. each interface on the cluster $V_8(a_0)$), for a reference particle $a_0$ in z-space, can be in the *first energy level interface excitation state* which has a 'weaker interface bond energy' with new 8 orders of relaxation times, new 8 orders of additional restoring torques, of quantized energies and of extra volumes. All other interfaces (on the cluster $V_i(a_0)$, $i < 8$) on x-y projection plane are still in the ground state interface excitation state; besides, for the 6 interfaces of a small molecule, the two ±z-axial interfaces of each molecule for all the cooperative particles in $a_0$ local field are still in Van der Waals ground state (absent Van der Waals repulsion electron-electron) with 'a stronger interface bond energy' (taking the action similar to covalent bond, these two ground state interfaces are not excited in turbulence in order to minimize the totally dominant potential energy and to satisfy the condition of *n* = 0 in the *n* vector model in Section 2.3). These small molecules can form a *random* 'long-chain', whose 'chain-section' has 136 molecules, through the relatively stronger ground state 'z-axial Van der Waals interface bonds'. The mode of minimum energy for self-similar liquid-cluster flow is also regarded as the 'reptation' (or say, the solitary wave movement along z-direction on 3-*D* random 'long-chain') of the random 'long-chain' in the first energy level interface excitation field in z-space.

From Section 2, inverse cascade is the accumulation of the potential energy in loop-flows, and cascade is the transfer of energy from potential to kinetic to induce cooperative movement of liquid cluster. The larger loop is formed in inverse cascade, the bigger liquid cluster is driven in cascade, and the higher efficiency the system has to excite liquid flowing, and the faster the velocity of flow is. Therefore, when the flow needs speedup, the cluster $V_8(a_0)$ can only continue inverse cascade in the first energy level excited field of the reference particle $a_0$ to accumulate potential in loop-flow.

(iii) The number of interfaces of the cluster $V_8(a_0)$ continuing on inverse cascade is 60 (see Section 2.12.). 60 excitation interfaces in the first energy level can form the 0-th 'cluster -particle' in z-space in the second level of 8 orders of 2-*D* loop-flow, denoted as $V_0^{(1)}(a_0)$, (the figure in the superscript parentheses denotes the level of electron excitation state). The 0-th 'cluster-particle' is self-similar to the particle $V_0(a_0)$ in Fig.3 (b). The scale of the 0-th 'cluster-particle' is that of $V_8(a_0)$. The 0-th 'cluster-particle' has also 4 interfaces (the blue side in Fig.6) with the side length of 15 excitation interfaces in the first energy level. Each excitation interface has its new 8 orders of relaxation times and its new 8 orders of extra vacancy volumes.

Fig.6 is also valid in the first energy level excited field, as long as the unit length of lattice in Fig.6 is considered as the energy of the 15 excitation interfaces in first energy level.

(iv) Only 60 excitation interfaces in the first energy level are required for the 136 ground state particles to co-operatively move in the excited field of first energy level. In the solid-to-liquid transition, the process to break solid-lattice is the evolution of the 5-particle cooperative excited field in Section 2.14. Similarly, in the 'disorder to more disorder' transition, the process to break 'ground state disorder lattice' of liquid is the evolution of the 5 second-level zeroth 'cluster-particle' cooperative excited field, i.e., the reference particle field and 4 neighboring $V_7$-center fields in Fig. 6. They are self-similar to the 5-particle cooperative excited field. The inverse cascade of the 5-second-level zeroth 'cluster-particle' cooperative excited field can form the second level of 8 orders of 2-D mosaic geometric structure, cluster and 2-D loop-flow.

(vi) Similarity in Section 4.19, the second level of 8 orders of 2-D loop-flow is a loop-potential with larger loop-scale. This loop-scale is also the characteristic scale the scale of disorder induced in the more disorder system (in the first energy level excitation field). When it appears and cascades, the extra vacancy volumes on the first energy level excitation interfaces will be compacted and form a bigger cavity for the cluster $V_8(a_0)$ cluster-migrating in $V_0^{(1)}(a_0)$ field. Therefore, there are very few first energy level excitation interfaces, and these few interfaces are non-ergodic in system.

(vii) In the same way, when the flow keeps on speedup, the third, the forth…the $l$-th level of 8 orders 2-D mosaic geometric structures can be constructed one by one, in order to augment the loop scale and the loop potential so that the maximum loop potential reaches the value of the potential forced by environment to the liquid flow. After that, the cascade phenomenon of the maximum loop-flow appears. This is namely the homogeneous isotropic turbulence.

**8.2. Physical approach to turbulence**

As discussed thus far, the theory of the solid-to-liquid transition presents a universal physical picture of the particle-cluster flowing not only for small molecules but also for macromolecules and yields a series of significant results that pursue a physical approach to turbulence.

(i) The two intrinsic characteristic energies $E_c(\tau_i)$ and $kT_m(\tau_i)$, the disorder-induced localized energy and the renewed energy of microscopic cluster, independent of temperature in random system proposed in this paper, show that there should be two characteristic energies $E_c^{(l)}(\tau_i)$ and $kT_m^{(l)}(\tau_i)$ in the $l$-th level of 8 orders 2-D loop-flow. Same as non-flexible macromolecule, one of the complexities of turbulence is that each real system has its own $E_c^{(l)}(\tau_i)$ and $kT_m^{(l)}(\tau_i)$. One of the most important concepts in turbulence is that, the maximal loop potential $E_c^{(l)}(\tau_8)$ is exactly the localization energy in the disorder system.

(ii) The testification in Section 4 also indicates that any liquid always exists in the local cluster growth phase transition state in inverse cascade and in the structure rearrangement state in cascade. The increase of temperature or restricted disorder only increases the number of inverse cascade and cascade.

(iii) The density fluctuation in liquid results from the compacting clusters in inverse cascade and

cascade, or say, the generation and transfer of interface excitations.

(iv) The flowing path of particle-cluster in liquid is always along 8 orders of geodesic.

(v) The flowing mode of particle-cluster in liquid is the solitary wave traveling along topological 'strong-bond-direction', and the 'strong-bond' of small molecules is precisely the unexcited Van der Waals interfaces absent cross-coupled electron pairs. The wave-particle duality of solitary wave results from the balance of probability between inverse cascade and cascade.

(vi) Fig.6 is also the *truncated* figure of a famous ABC (Arnold-Beltrami-Childress) flow [70] in turbulence. The finite lattice in Fig.6 also shows the structure of *heteroclinic orbits* in turbulence. For example, the length of heteroclinic orbit is the length of 15 excitation interfaces in the case in Section 8.1. (iii).

(vii) The testification of WLF equation implies that the continuum of flow-percolation is only an infinity fractal subsystem in system. The equations of hydrodynamics and physical laws can only be applied in each subsystem in order to keep the enthalpy $H$ constant (Section 5.5) and the chemical potential zero in each subsystem. In other words, similar to the glass transition, the complicated turbulence phenomenon is also an emergent behavior or emergent property of the subsystems of system.

(viii) The 8 orders of 2-*D* transient mosaic geometric structure is the basic structure from disorder to more disorder. The testification of the 3.4 power law of viscosity of entangled macromolecules indicates that the self-similar multi-levels of the intrinsic 8 orders of transient 2-*D* mosaic geometric structures is likely to be the *coherent structure*, which has long been looked after in turbulence research.

The 8 orders of 2-*D* loop-flows mean that the acting physical quantity is the viscosity unit of [energy·time], $E_c(\tau_i)\cdot\tau_i$; and the energy dissipation rate unit of [energy/time], $E_c(\tau_i)/\tau_i$, when *i*-th order of loop-flow appears.

The method of statistical thermodynamics in Section 5 can also be applied to turbulence. The kinetic energy in cascade always keeps balance with the potential energy in inverse cascade. The phase difference of the two is always $\pi$. The chemical potentials are always zero in each subsystem, where the subsystem is the flow-percolation in the *l*-th level of *i*-th order of 2-*D* energy loop-flow.

(ix) In the solid-to-liquid transition, a 5-particle cooperative excited field can excite $f$ ($1 \leq f \leq 5$) 8th order 2-*D* loops in a local zone (Section 2.13). There is one 8th order 2-*D* loop in a local zone corresponding to the glass transition (cross-coupled electron pairs only occur in local z-space). There are $f$ ($1 < f < 5$) 8th order 2-*D* loops corresponding to the viscoelastic state, and $f = 5$ to the melt state (cross-coupled electron pairs occur in local 3-*D* space). Similarly, a 5-*l*-level-zeroth-'cluster-particle' cooperative excited field can also excite $f$ ($1 \leq f \leq 5$) maximal 2-*D* loops in a disorder-induced localization zone in the *l*-th energy level excited filed. The intermittency turbulence appears when $f < 5$.

## 9. Conclusions

Three famous problems in condensed matter physics are discussed in this paper, the glass transition, the macromolecular entanglement and the turbulence. The common root problem of the three is: what is the paradigm of order-to-disorder transition for particle-cluster moving, or disorder-to-more-disorder transition for liquid flow? The theoretical framework for solid-to-liquid transition of particle-cluster moving is presented in this paper highlighting the theoretic relevancy of the three problems.

The intrinsic 8 orders of transient 2-*D* mosaic geometric structures formed by exciting interfaces exist in any random system at any temperature. This paper deduces this result without any presupposition or relevant parameter, but through extending the interface electron in the Mott disorder theory in solid physics to the cross-coupled electron pairs. In solid-to-liquid transition, or in 'disorder to more disorder' transition, all accidentally arisen Van der Waals repulsion electron pairs, must change as cross-coupled pairs with 2-*D* self-avoiding closed loops, called interface excitation states, either in electronic ground state or in electronic excitation state in any random system, are all correlated and they form local 2-*D* lattice to maximally avoid each other in order to minimize the total potential energy to break solid-lattice in solid-to-liquid transition, or to break 'ground state liquid disorder lattice' in 'disorder to more disorder' transition.

This paper proves that an interface excitation is a vector with 8 orders of components. An action reference particle is defined by its 4 excitation interfaces. Each order of harmonic restoring torque gives rise to an additional position-asymmetry of Lindemann displacement; thus, in order to eliminate the additional position-asymmetry, the accompanied 8 orders of transient 2-*D* clusters with the 4 interface relaxations of the reference particle have been formed. Here, a 2-*D* cluster is strictly defined by an instant interface excitation $2\pi$ loop-flow on projection plane. In order to step by step eliminate the anharmonic interface tensions, the 8 orders of dynamical 3-*D* hard-spheres should be adopted. This model directly deduces the geometrical frustration using geometrical method and directly deduces the average energy of cooperative migration along one direction, which is also a critical energy $kT_2$ similar to the energy of Curie temperature in magnetism.

It is worth noting that the physical picture provides a unified mechanism to explain a series of current various theories and phenomena in glass transition, containing Boson peak, 'tunneling', and potential energy landscape; especially, Reynolds number, solitary wave, heteroclinic orbit and the intermittency turbulence in hydrodynamics.

This model provides a universal picture for small molecules and macromolecules, from $T_g$ to $T_m$, of particle-clusters cooperative migration in one direction in solid-to-liquid transition. The mode is solitary wave to thaw solid domain, rebuild local structure, induced particle-cluster one by one delocalization along geodesic in inverse cascade-cascade.

In the solid-to-liquid glass transition, two of the most striking findings are that the existence of fixed point for self-similar L-J potentials and the existence of fixed point for reduced second Virial coefficients. They can be proved by the model without other assumption.

The existences of the two fixed points moreover has proved that (a) the hard-sphere model can be directly deduced from the fixed point of self-similar L-J potentials; (b) the origin and transfer of interface excitation come of the balance effect between self-similar L-J potential fluctuation and geometric-phase-induced potential fluctuation; (c) a universal behavior: two orthogonal degenerate states, the fast reduced geometric phase factor 3/8 and the slow reduced geometric phase factor 5/8, is accompanied with the appearance of each order 2-*D* symmetrical interface excitation energy loop-flow in the glass transition and the sum of the two degenerate reduced phase factors is exactly equal to 1; (d) the reduced energy of thermo-disorder induced localization $T_g*$, which is also a fixed point and independent of temperature, to characterize the universal intrinsic property in the temperature range from $T_g$ to $T_m$: increase of temperature only increases the number of inverse cascade to thawing; (e) a universal Lindemann ratio: $d_L =$ 0.1047…; (f) $B_2^* \equiv 3/8$ and $B_3^* \equiv 5/8$, the potentials of 2body-3body coupling clusters, in critical local cluster growth phase transition, balance the kinetic energy in the glass transition; (g) the so-called tunneling turns out to be the generating and transferring of all quantized interface excitations should pass through the additional attractive potential center of $-17/16\varepsilon_0$, which is lower than potential well energy $\varepsilon_0$ and comes from the balance effect of reinforce-restraint of self-similar potential fluctuations; (h) the delocalization path (the re-coupling path) of particle-pairs is along 8 orders of geodesic and the delocalization energy only as $1/8\varepsilon_0$ of L-J potential well energy, which is exactly the quantized interface excitation energy; (i) the percolation in glass transition is only an infinity sub-system, which is connected, one by one, by local fields in interface excitation state; (j) the glass transition occurs if the chemical potentials of subsystems hold zero.

The theoretical proof of the WLF equation has been directly deduced based on the intrinsic 8 orders of 2-$D$ mosaic geometric structure. The derivation of WLF equation validates that (a) $kC_2$ in WLF equation is the potential well energy $\varepsilon_0$, and $C_1$, taking logarithm, is the non-dimension orientation activity energy $\Delta E_{co}/\varepsilon_0$; (b) the energy of all the 320 interface excitation states is also the orientation activity energy to break solid-lattice; (c) the Clapeyron equation governing the first order phase transition in thermodynamics still hold true in the glass transition, the singularity is that the Clapeyron equation only holds true in subsystems, instead of system; (d) the solid-liquid coexist state in a local zone evolves as the particle-flow through 8 orders of fast-slow states; and the solid-liquid coexist state is also the 8 orders of local relaxation states; (e) the quantized interface excitation energy, $\Delta\varepsilon \approx 6.45k$ ($\approx 5.6\times 10^{-4}$ eV), has also been deduced for flexible system in the glass transition.

The key of the theoretical proof for the 3.4 power law is to find out the number of degrees of freedom for a chain-length $N$. This proof furthermore confirms that (a) the mode of multi-chain motion is that of the inverse cascade-cascade, which is also the mode of universal hydrodynamics; (b) entanglement structure is namely the localized 8 orders of transient 2-$D$ mosaic geometric structures on x-y projection plane and the delocalized solitary wave in z-space; (c) the reptation is precisely assembled in time by the three solitary waves in x-space, y-space and z-space in 3-$D$ space; (d) macromolecular chain-length corresponds to the Reynolds number in inverse cascade – cascade.

Finally, it must be emphasized that the most important concept correlating glass transition, macromolecular entanglement and turbulence is the intrinsic localization energy induced by disorder, which is also the induced potential of a maximum 2-$D$ excitation interface energy loop-flow in random system. Once the loop-flow is formed, the glass transition appears, and the macromolecular critical entanglement phenomenon arises, and the cascade in turbulence begins. Accordingly, a new explanation of the physical origin for turbulence has also been proposed in this paper: the multi-levels of 8 orders of transient 2-$D$ interface excitation energy loop-flow formed on their intrinsic self-similar transient 2-$D$ coherent structure is the physical origin for turbulence.

## Acknowledgments

This paper covers the major contributions of the author's original research work in the past 20 years or so.  The author is grateful to all colleagues he had the pleasure to collaborate and interact, especially when he found the fundamental physics origin for the orientation activation energy obtained experimentally on melt high-speed spinning-line in 1986. In particular, the author would like to thank, in random order, Academician Yuan Tseh Lee and Academician Sheng Hsien Lin of the Academia Sinica (Taiwan), Professor Yun Huang of Beijing University, Professor Da-Cheng Wu of Sichuan University, and Professor Bo-Ren Liang and Professor Cheng-Xun Wu of Donghua University in China.

## References

[1] Ashburner M, 1995 *Science* **267** 1609

[2] Donth E. W, 2001 *The glass transition: relaxation dynamics in liquids and disordered materials*. (Springer, Berlin) p. 3

[3] Eckert T and Bartsch E, 2004 *J. Phys.: Condens. Matter* **16** S4937

[4] Peter J. Lu, Emanuela Z et al, 2008 *Nature* **453** 499

[5] Trappe V, Prasad V et al, 2001 *Nature* **411** 772

[6] Pham K N, Puertas A. M et al, 2002 *Science* **296** 104

[7] Flores H M, Naumis G G. 2009 *J. Chem. Phys.* **131** 154501

[8] Tanaka H, 2001 *J. Phys. Soci. Japan* **70** 1178

[9] Lubchenko V and Wolynes P. G, 2003 *Proc. Natl. Acad. Sci. USA* **100** 1515

[10] Angell C A, 2004 *J. Phys.: Condens. Matter* **16** S5153

[11] Chumakov A I, Sergueev I et al, 2004 *Phys. Rev. Lett*. **92** 245508

[12] Zaccarelli E, Foffi G et al, 2003 *J. Phys.: Condens. Matter* **15** S367

[13] Pontoni D, Narayanan T et al, 2003 *Phys. Rev. Lett*. **90** 188301

[14] Weigt M and Hartmann A K, 2003 *Europhys. Lett*. **62** 533

[15] Liu A J and Nagel S R, 1998 *Nature* **396** 21

[16] D'Anna G and Gremaud G, 2001 *Nature* **413** 407

[17] de Genne P G, 2002 *C. R. Physique* **3** 1263

[18] Yong Li, Peng Yu and Bai H Y, 2008 *J. Appl. Phys*. **104** 013520

[19] Wolynes P G, 1992 *Acc Chem. Res*. **25** 513

[20] Dzeroi M, Schmalian J and Wolynes P G, 2005 *Phys. Rev. Lett. B*. **72** 100201

[21] Glotzer S C, 2000 *Phys. World* **13** Apr. 22

[22] Weeks E, Crocker J C et al, 2000 *Science* **287** 627

[23] Laird B B and Bembenek S D, 1996 *J. Phys.: Condens. Matter* **8** 9569

[24] Barkai E, Naumov A V et al, 2003 *Phys. Rev. Lett*. **91** 075502

[25] Fredrickson G H and Andersen H C, 1984 *Phys. Rev. Lett*. **53** 1244

[26] Matyushov D V and Angell C A, 2005 *J. Chem. Phys*. **123** 034506

[27] Merolle M, Garrahan J P, Chandle D, 2005 *Proc. Natl. Acad. Sci. USA* **102** 10837

[28] Hetayothin B, Blum F D, 2008 *Polymer preprints* **49** (1) 4741

[29] Berthier L and Garrahan J P, 2005 *J. Phys. Chem. B*. **109** 3578

[30] Xia X Y and Wolynes P G, 2000 *Proc. Natl. Acad. Sci*. **97** 2990

[31] Lubchenko V and Wolynes P G, 2004 *J. Chem. Phys*. **121** 2852

[32] Tempere J, Klimin N, Silvera F and Devreese T, 2003 *Eur. Phys. J. B* **32** 329

[33] de Genners P G, 1979 *Scaling concepts in polymer physics* (Cornell University Press)

[34] Zhang J, Liang Yu, Jizhong Yan and Jianzhong Lou. 2007 *Polymer* **48** (16) 4900

[35] Garrahan J P, Chqndler D, 2003 *Proc. Natl. Acad. Sci. USA.* **100** 9710

[36] Berthier L, 2003 *Phys. Rev. Lett.* **91** 055701

[37] Lester O H, Robert L J, Juan P G, David C, 2009 *Science* **323** 1309

[38] Lubchenko V and Wolynes P G, 2003 *Preprint* cond-mat/ 0307089v1

[39] Straub J S and Keyes T, 1995 *J. Chem. Phys.* **103** 1235

[40] Lubchenko V and Wolynes P G 2001 *Phys. Rev. Lett.* **87** 195901

[41] Pahl S, Fleischer G, Fujara F and Geil B, 1997 *Macromolecules* **30** 1414

[42] Grest G., Cohen M, 1981 *Adv. Chem. Phys.* **48** 436

[43] Berker N and Kadanoff Leo P, 1980 *J. Phys. A* **13** L259

[44] Ferry J D, 1980 *Viscoelastic properties of polymers. 3rd. ed.* (Wiley, New York)

[45] Doi M and Edwards S F, 1986 *The theory of polymer dynamics* (Clarendon, Oxford) p. 157

[46] Zallen R, 1983 *The physics of amorphous solids* (A Wiley-Interscience Publication, New York)

[47] Jackle J, 1986 *Rep. Prog. Phys.* **49** 171

[48] Falconer K and Falconer K J, 2003 *Fractal Geometry- Mathematical Foundations and plications 2nd Edi.* (John Wiley and Sons) chapter 16

[49] Aharonov Y, Anandan, 1987 *J. Phys. Rev. Lett.* **58** 1953

[50] Wu J. L, Guan D and Quian B. Q, 1986 *Intern. Polymer Processing* **1** 25

[51] Mohazzabi P and Mansoori A G, 2005 *J. Comput. Theor. Nanosci.* **2** 138

[52] Kristóf T, Szalai I, 2003 *Phys. Rev.* **68** 041109

[53] Stewart M C and Evans R, 2005 *Phys. Rev. E* **71** 011602

[54] David R R and Patrick C B, 2005 *J. Stat. Mech.* P05013

[55] Fasolo M and Sollich P, 2003 *Phys. Rev. Lett.* **91** 068301

[56] Guckenheimer J, 1977 *Invent Math* **39** 165

[57] Shapere A and Wilczek F ed, 1989 *Geometric Phases in Physics Singapore* (World Scientific Pulb Co Ltd)

[58] Levi B G, 1993 *Phys. Today*, **March**17

[59] Samuel J, Bhandar, R, 1988 *Phys. Rev. Lett*, **60** 2339

[60] Physics Survey Committee 1986, *Physics through the 1990s: Condensed-Matter Physics* (Notional Academy Press)

[61] Zhou Y, Karplus M et al, 2002 *J. Chem. Phys.* **116** 2323

[62] David R R and Patrick C, 2005 *J. Stat. Mech.* P05013

[63] Reichl L E, 1980 *A Modern Course in Statistical Physics* (University of Texas Press)

[64] Katsura S, 1959 *Phys. Rev.* **115** 1417

[65] Pathria R K, 1977 *Statistical Mechanics* (Pergamon Press) chapter 9.4

[66] Williams M, Landel R, Ferry J, 1955 *J. Am Chem. Soc.* **77** 3701

[67] Kadanoff L P, 1991 *Today* **44** (3) 9

[68] Zhao D. 1981 Bulletin of Macromolecula (Chinese) 385

[69] Frisch U. 1995 *Turbulence* (Cambridge University Press) chapter 7.4

[70] Sapsis T. and Haller G. 2010 *Chaos* **20** 017515

Address correspondence to this author at the College of Material Science and Engineering, Donghua University, Shanghai 201620, P. R. China; Tel: +86-21-52180896, Fax: +86-21-68532821;
Email: jlwu@dhu.edu.cn